\documentclass[12pt]{iopart} 

\usepackage{amssymb}
\usepackage{latexsym}
\usepackage{amsfonts}
\usepackage{psfrag}
\usepackage{graphicx}
\usepackage{subfigure}
\usepackage{cite}

\newcommand{\sket}[1]{|#1\rangle}
\newcommand{\ket}[1]{\left|#1\right>}
\newcommand{\bra}[1]{\left<#1\right|}   
\newcommand{\sbraket}[2]{\langle#1|#2\rangle}
\newcommand{\f}[1]{\mbox{\boldmath$#1$}}
\newcommand{\fk}[1]{\mbox{\boldmath$\scriptstyle#1$}}
\newcommand{\bea}{\begin{eqnarray}}
\newcommand{\ea}{\end{eqnarray}}

\begin{document}

\title{Entangling photons via the quantum Zeno effect}

\author{Nicolai ten Brinke, Andreas Osterloh and Ralf Sch\"utzhold}

\address{Fakult\"at f\"ur Physik, Universit\"at Duisburg-Essen, 
Lotharstrasse 1, 47057 Duisburg, Germany}

\ead{ralf.schuetzhold@uni-due.de}

\begin{abstract}
The quantum Zeno effect describes the inhibition of quantum evolution by 
frequent measurements.
Here, we propose a scheme for entangling two given photons based on this 
effect. 
We consider a linear-optics set-up with an absorber medium whose two-photon 
absorption rate $\xi_{2\gamma}$ exceeds the one-photon loss rate 
$\xi_{1\gamma}$. 
In order to reach an error probability $P_{\rm error}$,
we need $\xi_{1\gamma}/\xi_{2\gamma}<2P_{\rm error}^2/\pi^2$, 
which is a factor of 64 better than previous approaches 
(e.g., by Franson {\em et al.}). 

Since typical media have $\xi_{2\gamma}<\xi_{1\gamma}$, we discuss three 
mechanisms for enhancing two-photon absorption as compared to one-photon loss.
The first mechanism again employs the quantum Zeno effect via 
self-interference effects when sending two photons repeatedly 
through the same absorber.
The second mechanism is based on coherent excitations of many atoms and 
exploits the fact that $\xi_{2\gamma}$ scales with the number of excitations 
but $\xi_{1\gamma}$ does not. 
The third mechanism envisages three-level systems where the middle level 
is meta-stable ($\Lambda$-system). 
In this case, $\xi_{1\gamma}$ is more strongly reduced than $\xi_{2\gamma}$ 
and thus it should be possible to achieve $\xi_{2\gamma}/\xi_{1\gamma}\gg1$. 

In conclusion, although our scheme poses challenges regarding the density 
of active atoms/molecules in the absorber medium, their coupling constants 
and the detuning, etc., we find that a two-photon gate with an error 
probability $P_{\rm error}$ below $25\%$ might be feasible using present-day 
technology.
\end{abstract}

\submitto{New Journal of Physics}

\maketitle

\section{Introduction}
\label{sec:introduction}

The idea that quantum mechanics or, more precisely, quantum systems could be 
used for (quantum) computation 
\cite{Benioff:1980kx,Feynman:1982fk,Feynman:1986uq}, 
led to the advent of a whole new research area known as quantum information 
processing. Important theoretical progress has been made in this field, 
showing that quantum information processing exceeds classical information 
processing in a fundamental sense 
\cite{Shor:1997vn,Grover:1997ys,Feynman:1982fk,Lloyd:1996vn}. 

Accordingly, strong efforts are undertaken to physically realize quantum 
computers, following such different approaches as ion traps, quantum dots, 
Josephson junctions, nuclear spins and quantum optics, all with their 
respective assets and drawbacks \cite{Exp:2000kx}. 
Picking up one of the promising routes to quantum information processing, 
this work is connected with a quantum optics approach. 
In this approach, 
the information is encoded in the polarization or spatial degree of freedom 
of optical modes, i.e. qubits are physically implemented as single photons.

It is well known that photons possess many properties which make them very 
suitable candidates for quantum information processing. 
They can be well controlled and manipulated, as well as created and measured, 
even down to the single-photon level 
\cite{DiVincenzo:2000fk,Alleaume:2004uq,Chen:2006kx,Varnava:2008fk}. 
Furthermore, because of their long decoherence time, photons can propagate 
over relatively long distances without significantly coupling to the 
environment.
Moreover, when thinking of the quantum circuit model, see e.g. 
\cite{Nielsen:2000uq}, the necessary single-qubit quantum gates 
(such as Hadamard and phase shift gates) exist for photons and are 
easily implemented. If an adequate two-photon entangling gate 
(such as a CNOT-gate) with low enough error-probability existed, 
one would have a universal set of gates, from which every quantum 
algorithm could be constructed.
Unfortunately, the weak coupling of photons to the environment is also 
related to their main drawback -- 
it is very hard to make two photons interact, and thus hard to build a 
two-photon entangling gate, as the CNOT-gate.
Typically, before two photons interact by means of some non-linear 
medium, at least one of them is absorbed. 
This motivates the idea to turn the problem around and to actually exploit 
the absorption in order to make photons interact. 

An indirect way of doing this is realized by the Knill-Laflamme-Milburn
(KLM) proposal \cite{Knill:2001vn}, which initiated the field of {\em linear} 
optical quantum computation. In the KLM paper it was shown that scalable 
quantum computing is theoretically possible using only linear optical 
elements (like beam-splitters and phase shifters), single photon sources 
and detectors. In this approach, interactions are induced 
(probabilistically) by means of entangled ancilla photons on which 
measurements are performed.
Linear optical quantum computation is being studied in great detail 
theoretically 
(e.g., \cite{Raussendorf:2001ve,Pittman:2001zr,Ralph:2001fk,
Franson:2002kx,Knill:2002uq,Yoran:2003fk,Nielsen:2004uq,al:2004kx,
Browne:2005vn,Kok:2007nx,Gilchrist:2007bs,Hayes:2008bh,Gong:2010ly,
Hayes:2010dq,Jennewein:2011vn,Berry:2011fk}) 
and experimentally 
(e.g., \cite{Pittman:2002fk,Pittman:2003uq,OBrien:2003kx,Sanaka:2004ys,
Gasparoni:2004vn,Zhao:2005zr,Pittman:2005ly,Chen:2008ys,Schmid:2009qf,
Lemr:2010ly,Barz:2010ve,Mikova:2012zr}), 
and is currently one of the most elaborate schemes for optical 
quantum information processing. 
A major drawback is the vast amount of resources 
(ancilla photons, entanglement) which is 
-- despite all theoretical progress -- 
still needed when scaling up this scheme.

A more direct way of exploiting the absorption of photons is to employ the 
quantum Zeno effect which describes the slow-down or even inhibition of 
quantum evolution by repeated measurements 
\cite{Misra:1977ul,Itano:1990pd,Kofman:1996cr,Schulman:1998nx,
Kwiat:1998tg,Kofman:2000kl,Kofman:2001oq}. 
Imagine a quantum particle in a double-well potential initially confined
to the right well, for example.
After some time $T$, it would tunnel to the left well (and then back, etc).
However, if we measure the position of the particle frequently, i.e., after 
very short time intervals $\Delta t\ll T$, it does not tunnel since each 
measurement projects the quantum state back to the right well. 
This is the basic picture of the quantum Zeno effect.
Since the absorption of a photon in some medium is equivalent to measuring
its position, a strong enough absorption probability can actually 
prevent the photon from tunnelling/propagating into this medium \cite{Horodecki:2001fk}. 

The idea to utilize the quantum Zeno effect to build a two-photon entangling 
gate was first brought up by Franson and co-workers \cite{Franson:2004fj} 
and further developed in \cite{Leung:2006jo,Franson:2007yo,Myers:2007fc,
Leung:2007dw,Huang:2008xq}. In their set-up, the two photons (qubits) 
are guided within two fibre cores, which are coupled by their evanescent 
fields. When doping the fibre cores with two-photon absorbing atoms/molecules, 
unwanted two-photon amplitudes are suppressed due to the quantum Zeno effect. 
One advantage of this approach is that it requires much less resources when 
compared to linear optics quantum computation (KLM-style) schemes. 
However, an important issue of the Zeno gate is that while strong two-photon 
absorption is strictly required, one-photon loss leads to its failure. 
That is, the rate of two-photon absorption needs to be several orders of 
magnitude larger than the rate of one-photon loss for successful gate 
operation.

In this paper, after a brief introduction on the quantum Zeno effect 
(section \ref{ssec:the_quantum_zeno_effect}),
a new optical set-up for an entangling quantum Zeno gate is developed 
(section \ref{sec:two_photon_entangling_zeno_gate}). 
In contrast to earlier proposals for photon gates based on the quantum 
Zeno effect \cite{Franson:2004fj,Leung:2006jo,Franson:2007yo,Myers:2007fc,
Leung:2007dw,Huang:2008xq}, the set-up presented in this paper is modified 
and so yields a significantly reduced error probability, 
see (\ref{eq:constraint}), which was derived considering finite two-photon 
absorption and non-negligible one-photon loss.

Furthermore, a realization based on free-space propagation is proposed, 
i.e., without waveguides, resonators, or optical fibres -- which may induce 
additional decoherence. 
Thus in section \ref{sec:single_three_level_atom}, 
an analysis of two-photon absorption and one-photon scattering is given 
for a single three-level atom in free-space.
As our gate requires an absorbing material which features strong two-photon 
absorption compared to one-photon loss, we present three mechanisms 
(sections \ref{sec:repeated_inducing}, 
\ref{sec:coherent_interaction_with_Sgg1_atoms}, 
and \ref{sec:alternative_level_scheme}) 
to enhance two-photon absorption compared to one-photon loss, which could 
be applied in free-space. Additionally, by inserting example values 
(sections \ref{sec:example_values} 
and \ref{sec:alternative_level_scheme}), 
we show that our scheme is not out of reach experimentally. 
We summarize and conclude in section \ref{sec:summary_and_conclusion}.

\subsection{The quantum Zeno effect}
\label{ssec:the_quantum_zeno_effect}
\begin{figure}[h]
\begin{center}
\includegraphics[width=0.2\columnwidth]{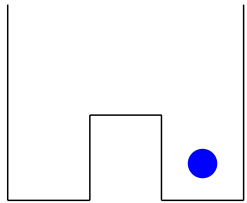}
\caption{Double-well potential with particle in the right well initially.}
\label{fig:double-well}
\end{center}
\end{figure}

In our approach to entangle photons, we employ the quantum Zeno effect.
The quantum Zeno effect describes the suppression of the time evolution of a quantum state by frequent measurements. It is based on a subtle application of the quantum theory of measurement processes \cite{Sakurai:1994uq}.
Imagine a quantum particle in a double-well potential as depicted in Figure~\ref{fig:double-well}.
\noindent
Let us assume that the quantum particle initially is confined in the right well,
\bea
\label{eq:double-well_initial}
\ket{\psi \left( 0 \right)}
=
\ket{\rm R}
\,,
\ea
where a ket-vector $\ket{\rm R}$ denotes the basis state where the particle is in the right well and $\ket{\rm L}$ denotes the basis state where the particle occupies the left well.
As the two wells of the double-well potential are coupled, according to the rules of quantum mechanics, the quantum particle starts tunnelling from the right well into the left well. After a certain time $T$, the quantum state will have evolved to
\bea
\label{eq:double-well_final}
\ket{\psi \left( T \right)}
=
\ket{\rm L}
\,,
\ea
i.e., the particle will have tunneled entirely into the left well.
Then the particle starts tunnelling back such that it is entirely in the right well at $t = 2 T$ etc. 

However, when dividing the tunnelling time $T$ into $N \gg 1$ equal-spaced time steps $\Delta t = T/N$, the quantum state after the first time step would be
\bea
\label{eq:double-well_firststep}
\ket{\psi \left( \Delta t \right) }
=
\cos \left( \frac{\pi}{2 N} \right) \ket{\rm R} + \sin \left( \frac{\pi}{2 N} \right) \ket{\rm L}
\,,
\ea
and after the $n$-th timestep, the state would evolve into
\bea
\label{eq:double-well_nthstep}
\ket{\psi \left( n \Delta t \right)}
=
\cos \left( n \frac{\pi}{2 N} \right) \ket{\rm R} + \sin \left( n \frac{\pi}{2 N} \right) \ket{\rm L}
\,.
\ea
So far, there were no measurements made on the system. Now imagine we measure the position of the particle frequently, i.e. after each time step $\Delta t$. This means that the quantum state (\ref{eq:double-well_firststep}) is projected back onto the initial state $\ket{\psi \left( 0 \right)} = \ket{\rm R}$ with probability $P_{\rm R} \left( \Delta t \right) = \cos^2 \pi / \left( 2 N \right)$ or is projected onto the state $\ket{\rm L}$ with probability $P_{\rm L} \left( \Delta t \right) = \sin^2 \pi / \left( 2 N \right)$.
Furthermore, when measuring the position of the particle after each time step, the probability for the particle remaining in the right well is given by
\bea
\label{eq:double-well_finalprob}
P_{\rm R} \left( T \right)
=
P_{\rm R}^N \left( \Delta t \right)
=
\cos^{2N} \left(  \frac{\pi}{2 N} \right) 
=
1 - \frac{\pi^2}{4 N} + \Or \left( N^{-2} \right)
\,,
\ea
which approaches unity in the limit of large $N \gg 1$. With this probability (\ref{eq:double-well_finalprob}), the time evolution of the quantum state is completely supressed, i.e. the quantum particle stays in the right well instead of the left well, where it would have arrived without measurements.
\begin{figure}[h]
\begin{center}
\subfigure[]{\includegraphics[width=0.5\columnwidth]{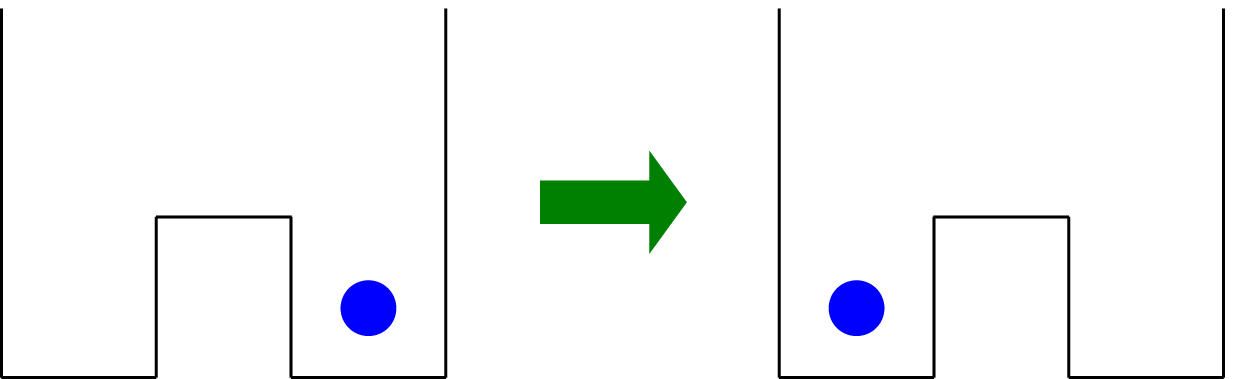}}
\hspace{.5cm}
\subfigure[]{\includegraphics[width=0.5\columnwidth]{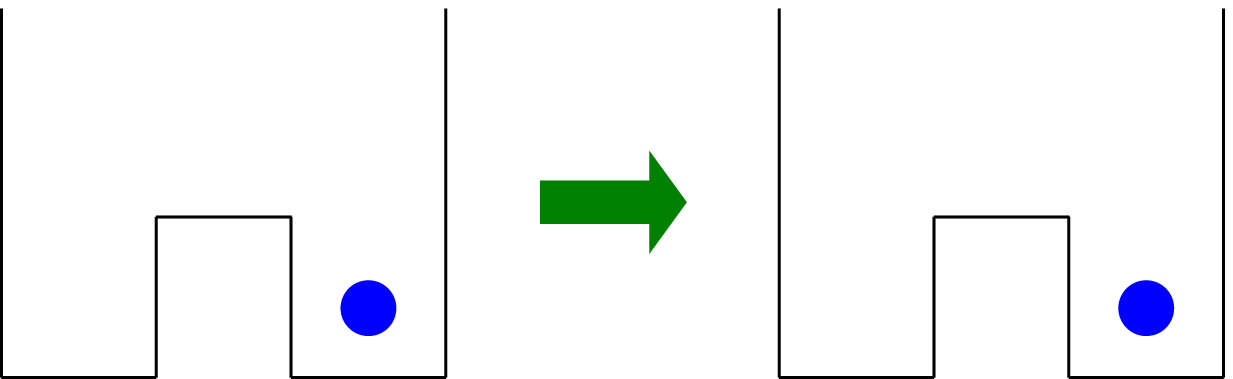}}
\caption{Comparison of the dynamics of the particle with (b) and without (a) frequent measurements. (a) Without quantum Zeno effect, the quantum particle tunnels from the right well ($t = 0$) into the left well ($t = T$). (b) When measuring the position of the quantum particle after each time step $\Delta t$, the particle is still found in the right well for $t = T$. This is a simple example of the quantum Zeno effect.}
\label{fig:double-well-comp}
\end{center}
\end{figure}
\section{Two-photon entangling Zeno gate}
\label{sec:two_photon_entangling_zeno_gate}
\subsection{Two-branch gate}
\label{ssec:two_branch_gate}
In section \ref{ssec:the_quantum_zeno_effect}, the basic principle of the quantum Zeno effect was briefly discussed. In this section we will demonstrate how the quantum Zeno effect can be employed in an otherwise linear optics set-up for building a two-photon entangling gate.
For this purpose we will first study the underlying concept via a reduced version of our entangling gate, called the ``Two-branch gate''. On this basis we will be able to understand the functionality of the ``Three-branch gate'' (section \ref{ssec:three_branch_gate}) and to derive the dependency $\Or(P_{\rm error}) = \sqrt{\xi_{1\gamma}/\xi_{2\gamma}}$.
\begin{figure}[h]
\begin{center}
\includegraphics[width=1.0\columnwidth]{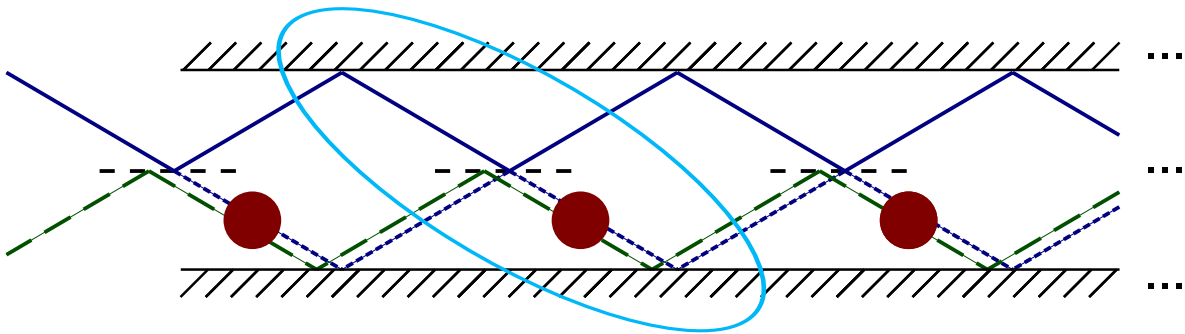}
\caption{Sketch of the macro-structure of the two-branch gate.
Horizontal solid lines (top and bottom) denote perfect mirrors and 
horizontal dashed lines indicate beam splitters.
The target photon (slanting dark blue line) enters the upper branch. 
The control photon (slanting dark green dashed line) enters the lower 
branch (if present). 
Then both, the control photon and the part of the target photon which 
tunneled through the beam splitter into the lower branch, pass the 
two-photon absorbing medium (brown circle).   
Each segment (tilted light blue oval) consists of an absorbing medium 
and a beam splitter.}
\label{fig:2branch}
\end{center}
\end{figure}

The proposed quantum Zeno gate consists of perfect mirrors at the top and bottom, weak beam splitters in the middle, and a two-photon absorbing medium in the lower branch, see Figure~\ref{fig:2branch}.
The quantum state of the target photon is represented by a 2-vector $\vec\psi=(x,y)^T$ 
where $x$ and $y$ are the amplitudes for the target photon being in the upper and lower branch, respectively.
Additionally, the control photon is either incident in the lower branch, $\ket{1}_C$, or not, $\ket{0}_C$. The polarization and/or frequency of the control photon is chosen such that it is perfectly reflected by the beam splitters and thus maintains its state in the lower branch for the whole gate. The target photon, in contrast, is affected by weak beam splitters with reflectivity $\cos\epsilon$ and transmittivity $\sin\epsilon$, where $\epsilon \ll 1$. Their effect on the state $\vec\psi$ can be represented by a 2$\times$2 rotation matrix
\bea
\label{eq:beamsplitter}
\left[
\begin{array}{cc}
\cos\epsilon  & \sin\epsilon\\
-\sin\epsilon & \cos\epsilon\\
\end{array}
\right]
\,.
\ea
The absorbers in the lower branch reduce the amplitude by a factor $\rme^{-\xi}$, where $\xi$ takes different values depending on whether the control photon is incident, $\xi_{2\gamma}$, or not, $\xi_{1\gamma}$. Therefore the whole operation of one segment of the gate on the target photon state reads
\bea
\label{eq:2branchunit}
\left[
\begin{array}{cc}
 \cos\epsilon & \sin\epsilon \\
 -\rme^{-\xi} \sin\epsilon & \rme^{-\xi} \cos\epsilon
\end{array}
\right]
=
\left[
\begin{array}{cc}
 1 & 0 \\
 0 & \rme^{-\xi} 
\end{array}
\right]
\cdot
\left[
\begin{array}{cc}
\cos\epsilon  & \sin\epsilon\\
-\sin\epsilon & \cos\epsilon\\
\end{array}
\right]
\,.
\ea
As our gate consists of $N$ of such segments, the output state is given by
\bea
\label{eq:outputstate}
\vec\psi_{\rm out} 
=
\left[
\begin{array}{cc}
 \cos\epsilon & \sin\epsilon \\
 -\rme^{-\xi} \sin\epsilon & \rme^{-\xi} \cos\epsilon
\end{array}
\right]^N
\cdot
\vec\psi_{\rm in}
\,.
\ea
To explain how the gate works, we first consider the ideal case where the absorber exhibits perfect two-photon absorption, $\xi_{2\gamma} = \infty$, and non-existent one-photon loss, $\xi_{1\gamma} = 0$. 
If we assume an input state where the target photon enters the upper branch, $\vec\psi_{\rm in}=(1,0)^T$, and there is no control photon (thus $\xi = \xi_{1\gamma} = 0$), then the small rotations (\ref{eq:beamsplitter}) of the target photon state simply add up
\bea
\label{eq:outputstate_without}
\vec\psi_{\rm out}^{1\gamma}
=
\left[
\begin{array}{cc}
 \cos\epsilon & \sin\epsilon \\
 - \sin\epsilon & \cos\epsilon
\end{array}
\right]^N
\cdot
\left[
\begin{array}{cc}
 1 \\
 0
\end{array}
\right]
=
\left[
\begin{array}{cc}
 \cos N\epsilon \\
 -\sin N\epsilon
\end{array}
\right]
\,.
\ea
Physically, the target photon in the upper branch gradually tunnels into the lower branch (and would continue to tunnel back and forth afterwards) via the weak beam splitters. Choosing $\epsilon$ to fulfill $N \epsilon = \pi / 2$, the target photon will have completely tunneled into the lower branch at the end of the gate, $\vec\psi_{\rm out}^{1\gamma}=(0,-1)^T$, as then $N$ rotations by a small angle $\epsilon$ add up to a $\pi / 2$-rotation.

However, if there is a control photon incident and thus $\xi = \xi_{2\gamma} = \infty$, the second row of the matrix (\ref{eq:2branchunit}) is zero and therefore the output state (\ref{eq:outputstate}) is given by 
\bea
\label{eq:outputstate_with}
\vec\psi_{\rm out}^{2\gamma} 
=
\left[
\begin{array}{cc}
 \cos\epsilon & \sin\epsilon \\
 0 & 0
\end{array}
\right]^N
\cdot
\left[
\begin{array}{cc}
 1 \\
 0
\end{array}
\right]
=
\left[
\begin{array}{cc}
 \cos^N \epsilon \\
 0
\end{array}
\right]
\,.
\ea
In this case, due to the quantum Zeno effect, the target photon stays in the upper branch. Physically, each beam splitter transmits a small fraction $\sin\epsilon$ of the amplitude from the upper branch into the lower branch, which immediately gets canceled by the two-photon absorbing medium. Hence, the amplitude in the upper branch is reduced by a factor $\cos\epsilon$ for each beam splitter. In the limit of large $N$, $\vec\psi_{\rm out}^{2\gamma}=(\cos^N \epsilon,0)^T$ approaches $\vec\psi_{\rm out}^{2\gamma}=(1,0)^T$, where the error due to finite $N$ (discretization error) scales with $\Or \left( N^{-1} \right)$
\bea
\label{eq:discretization_error}
\cos^N \frac{\pi}{2 N}
=
1 - \frac{\pi^2}{8 N} + \Or \left( N^{-2} \right)
\,.
\ea
For ideal absorbers ($\xi_{2\gamma} = \infty$, $\xi_{1\gamma} = 0$) and $N \rightarrow \infty$, the illustrated two-branch-gate would deterministically entangle photons, as the branch where the target photon ends up is totally determined by the presence or absence of the control photon.
These requirements, however, are clearly unrealistic when it comes to real, physical absorbers. In the following, we therefore investigate how the gate operation is affected when considering finite two-photon absorption ($\xi_{2\gamma} < \infty$) and non-negligible one-photon loss ($\xi_{1\gamma} > 0$). For arbitrary parameters $\xi$ and $N$, (\ref{eq:outputstate}) can be evaluated to 
\bea
\label{eq:outputstate_general}
\vec\psi_{\rm out} 
=
\frac{1}{2^N r}\left[
\begin{array}{cc}
 \left( \beta_+ \alpha_-^N - \beta_- \alpha_+^N \right)/2 & \left( \alpha_+^N - \alpha_-^N \right) \sin\epsilon \\
 \left( \alpha_-^N - \alpha_+^N \right) \rme^{-\xi} \sin\epsilon & \left( \beta_+ \alpha_+^N - \beta_- \alpha_-^N \right)/2
\end{array}
\right]
\cdot
\vec\psi_{\rm in}
\,,
\ea
with abbreviations
\bea
&&r = \sqrt{ \left(\rme^{-\xi} + 1\right)^2 \cos^2 \epsilon - 4 \rme^{-\xi}}\nonumber\\
&&\alpha_\pm = \left(\rme^{-\xi} + 1\right) \cos\epsilon \pm r\nonumber\\
&&\beta_\pm = \left(\rme^{-\xi} - 1\right) \cos\epsilon \pm r
\,.
\ea
It is insightful to throw a glance at the output state (\ref{eq:outputstate_general}) for the one-photon case, expanded for small $\xi = \xi_{1\gamma} \ll 1$ in first order, where $\vec\psi_{\rm in}=(1,0)^T$. It reads
\bea
\label{eq:outputstate_firstorder}
\vec\psi_{\rm out}^{1\gamma}
=
\left[
\begin{array}{cc}
 \cos N\epsilon \left( 1 - \xi_{1\gamma} \left( N - \cot\epsilon \tan N\epsilon \right) / 2 \right)\\
 -\sin N\epsilon \left( 1 - \xi_{1\gamma} \left( N + 1 \right) / 2  \right)
\end{array}
\right] + \Or \left( \xi_{1\gamma}^2 \right)
\,.
\ea
As was shown before (discussion below (\ref{eq:outputstate_without})), the absolute value of the amplitude in the lower branch at the end of the gate is $1$ in the ideal case, where $\xi_{1\gamma} = 0$. Also in the general case, $\xi_{1\gamma} > 0$, the absolute value of the second line in (\ref{eq:outputstate_firstorder}) needs to be to close to $1$ in order to approximate the desired gate operation. This, however, is only possible when $\xi_{1\gamma}$ scales with $1/N$, such that $N \xi_{1\gamma} \ll 1$. Moreover, (\ref{eq:outputstate_firstorder}) shows that $N \epsilon = \pi / 2$ is still the optimal choice for $\epsilon$, even if the one-photon loss is non-zero.
\subsubsection{Error probabilities}
\label{sssec:error_probabilites_2branch}
In both operating modes (with and without control photon), errors occur when considering non-ideal values for $\xi_{1\gamma}$, $\xi_{2\gamma}$ and finite $N$. We are going to derive the respective error probabilities and discuss the requirements on the absorber when the effective error probability of the gate should be below a certain error threshold $P_{\rm error}$.

When the control photon is absent, the target photon could get lost (e.g. scattered) for $\xi = \xi_{1\gamma} > 0$. This error corresponds to a situation where the target photon, which is initially in the upper branch, $\vec\psi_{\rm in}=(1,0)^T$, is not found in the lower branch at the end, thus
\bea
\label{eq:error_without}
P_{\rm error}^{1\gamma} = 1 - \left|(0,1) \cdot \vec\psi_{\rm out}^{1\gamma}\right|^2
\,.
\ea
With control photon, finite $N$ leads to a non-zero probability that the two photons actually get absorbed (discretization error, see discussion below (\ref{eq:outputstate_with})) and imperfect two-photon absorption, $\xi = \xi_{2\gamma} < \infty$, allows a non-vanishing amplitude in the ideally forbidden lower branch. In summary, the gate fails when the target photon is not found in the upper branch at the end
\bea
\label{eq:error_with}
P_{\rm error}^{2\gamma} = 1 - \left|(1,0) \cdot \vec\psi_{\rm out}^{2\gamma}\right|^2
\,.
\ea
Both error probabilities (\ref{eq:error_without}), (\ref{eq:error_with}) are analytically well defined by (\ref{eq:outputstate_general}), but the exact expressions will not be quoted here. Instead we consider the limit $N \gg 1$, assuming that $\xi_{1\gamma}$ and $\xi_{2\gamma}$ scale with $1/N$. For $\xi_{1\gamma}$ this is necessary for proper gate operation, see discussion after (\ref{eq:outputstate_firstorder}), and for $\xi_{2\gamma}$ we need to choose the same dependence to obtain a quotient $\xi_{2\gamma} / \xi_{1\gamma}$ which is independent of $N$ (see below). Additionally, we demand that the one-photon loss rate is small even when multiplied with $N$, and that the two-photon absorption rate multiplied with $N$ is much larger than one ($N \xi_{2\gamma} \gg 1 \gg N \xi_{1\gamma}$). These limits are essential in order to obtain reasonable success probabilities. We then get for the error probability $P_{\rm error}^{1\gamma}$ in case without control photon (for the first non-vanishing order in $1/N$, see \ref{ap:ssec:success_probabilities_of_the_two_branch_gate})
\bea
\label{eq:error_nocontrol}
P_{\rm error}^{1\gamma} = N \xi_{1\gamma} + \Or \left( N^{-1} \right)
\,.
\ea
The error probability $P_{\rm error}^{2\gamma}$ in case with control photon reads
\bea
\label{eq:error_control}
P_{\rm error}^{2\gamma} = \frac{\pi^2}{2 N \xi_{2\gamma}} + \Or \left( N^{-2} \right)
\,.
\ea
Usually both $\xi_{1\gamma}$ and $\xi_{2\gamma}$ scale linearly with the absorber length and therefore we can trade-off one error probability against the other. If we make the absorber longer, for example, both $\xi_{1\gamma}$ and $\xi_{2\gamma}$ increase by the same factor, which results in a lower $P_{\rm error}^{2\gamma}$ but in a higher $P_{\rm error}^{1\gamma}$. 
\begin{figure}[h]
\begin{center}
\psfrag{Perror}{\Large{$P_{\rm error}$}}
\psfrag{Perror1gamma}{\Large{$P_{\rm error}^{1\gamma}$}}
\psfrag{Perror2gamma}{\Large{$P_{\rm error}^{2\gamma}$}}
\psfrag{Xi2gamma}{\Large{$\xi_{2\gamma}$}}
\psfrag{1.0}{$1.0$}
\psfrag{0.8}{$0.8$}
\psfrag{0.6}{$0.6$}
\psfrag{0.4}{$0.4$}
\psfrag{0.2}{$0.2$}
\psfrag{0.00}{\,$0.00$}
\psfrag{0.02}{\,$0.02$}
\psfrag{0.04}{\,$0.04$}
\psfrag{0.06}{\,$0.06$}
\psfrag{0.08}{\,$0.08$}
\psfrag{0.10}{\,$0.10$}
\psfrag{0.12}{\,$0.12$}
\psfrag{0.14}{\,$0.14$}
\includegraphics[width=1.0\columnwidth]{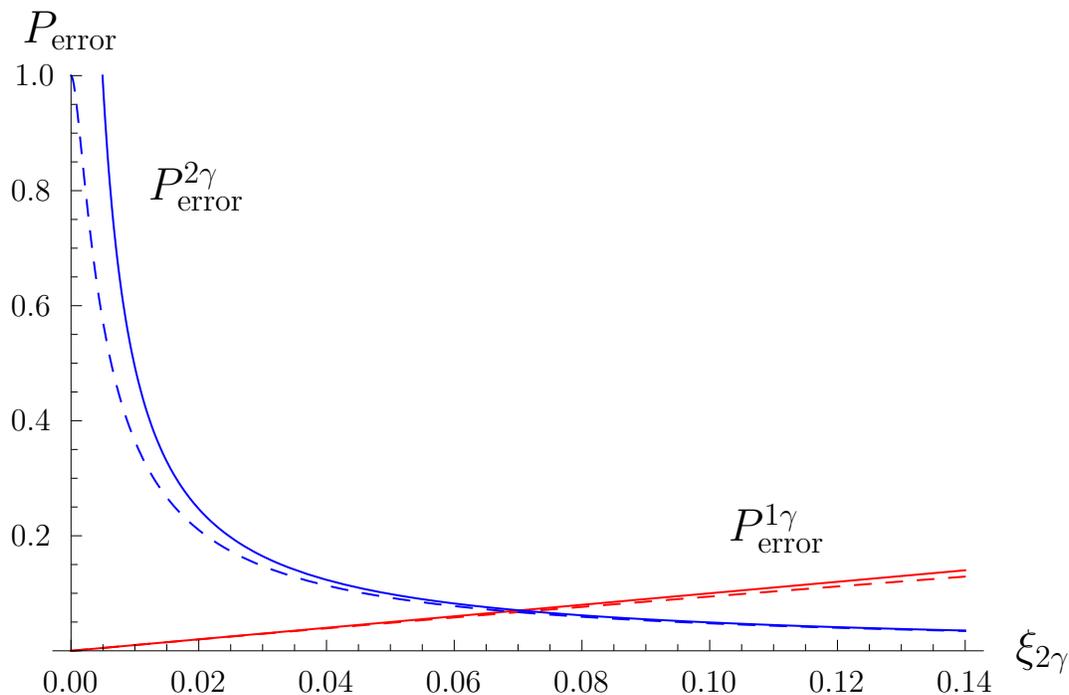}
\caption{Plot of the error probabilities for the one-photon case (red) and for the two-photon case (blue) with $\kappa = 10^3$ and $N = 10^3$.
The approximated error probabilities (\ref{eq:error_nocontrol}), (\ref{eq:error_control}) are plotted as solid lines, while the dashed lines indicate the exact error probabilities (\ref{eq:error_without}), (\ref{eq:error_with}).
$\xi_{2\gamma} = \xi_{1\gamma} \kappa$ is proportional to the absorber length. The optimal absorber length is found at the intersection between the blue and the red graph.}
\label{fig:error_plot}
\end{center}
\end{figure}
If we define the effective error probability as the maximum of the error probabilities for both cases, the optimal absorber length is reached when $P_{\rm error}^{1\gamma} = P_{\rm error}^{2\gamma}$, as $P_{\rm error}^{1\gamma}$ is monotonically increasing with the absorber length, while $P_{\rm error}^{2\gamma}$ is monotonically decreasing with the absorber length, see Figure~\ref{fig:error_plot}. Assuming that the quotient $\kappa := \xi_{2\gamma} / \xi_{1\gamma}$ is fixed as a material property of the absorber, we arrive at optimal rates
\bea
\label{eq:two_branch_optimal_rates}
\xi_{1\gamma} = \frac{1}{\sqrt{\kappa}} \frac{\pi}{\sqrt{2} N}\,, \quad\quad \xi_{2\gamma} = \sqrt{\kappa} \frac{\pi}{\sqrt{2} N}
\,.
\ea
Reinserting this result into (\ref{eq:error_nocontrol}), (\ref{eq:error_control}) yields the effective error probability for given $\kappa$, optimized absorber length, and large $N$
\bea
\label{eq:overall_error_prob}
P_{\rm error} = \frac{\pi}{\sqrt{2 \kappa}} + \Or \left( N^{-1} \right)
\,.
\ea
This relation can also be understood as a requirement on $\kappa$ in order to reach a certain error threshold $P_{\rm error}$
\bea
\label{eq:constraint_2branch}
\kappa
=
\frac{\xi_{2\gamma}}{\xi_{1\gamma}}
=
\frac{\pi^2}{2P_{\rm error}^2}
\gg1
\,.
\ea
Before we examine (\ref{eq:constraint_2branch}) in detail, let us discuss the relation to earlier proposals in \cite{Franson:2004fj,Leung:2006jo}.
The proposed quantum Zeno gate in Figure~\ref{fig:2branch} has substantial similarities to the quantum Zeno CSIGN-gate proposed by Leung and Ralph \cite{Leung:2006jo} based on the work of Franson et. al. \cite{Franson:2004fj}. However, the main difference here is that our set-up is not symmetric, i.e. the control photon always enters in the lower branch and stays there, and there are no absorbers in the upper branch. With the set-up described in this section, it is difficult to design a quantum CNOT-gate.

\subsection{Three-branch gate}
\label{ssec:three_branch_gate}
\begin{figure}[h]
\begin{center}
\includegraphics[width=1.0\columnwidth]{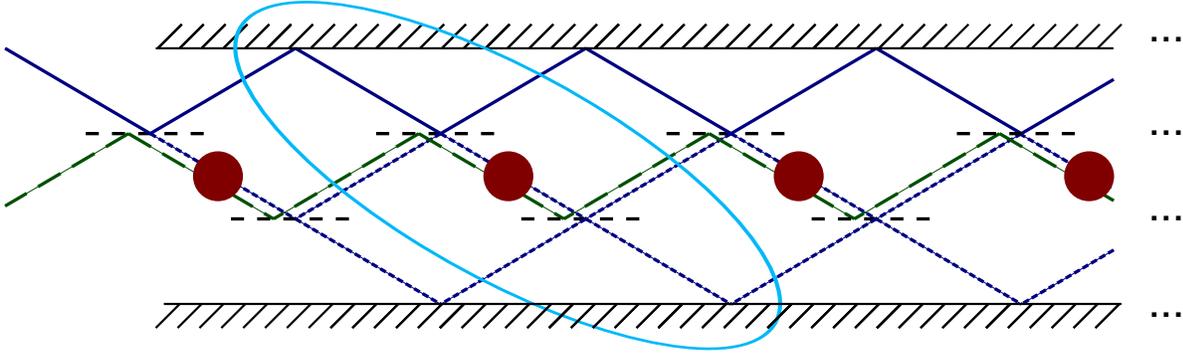}
\caption{Sketch of the macro-structure of the three-branch gate.
Again, horizontal solid lines (top and bottom) denote perfect mirrors and 
horizontal dashed lines indicate beam splitters.
The target photon (slanting dark blue line) enters the upper branch 
(initial state $\ket{0}$) or the lower branch (initial state $\ket{1}$). 
The control photon (slanting dark green dashed line) enters the middle 
branch (if present, i.e, if in the state $\ket{1}_C$). 
Then both, the control photon and the part of the target photon which 
tunneled through the beam splitter into the middle branch, pass the 
two-photon absorbing medium (brown circle).   
If the target photon is not absorbed, it may continue to tunnel to 
the lower branch through the second beam splitter. 
Now, each segment (tilted light blue oval) consists of an absorbing medium 
and two beam splitters.}
\label{fig:3branch}
\end{center}
\end{figure}
\noindent
This obstacle can be overcome by making our set-up symmetric, i.e. adding a third branch at the bottom, such that the control photon always stays in the middle branch while the target photon can enter the gate in the top- or bottom-branch and tunnel to the opposite side (without control photon) or stay there (with control photon). 
Now the quantum state of the target photon is represented by a 3-vector $\vec\psi=(x,y,z)^T$, where $x$ stands for the amplitude in the upper branch, $y$ for the amplitude in the middle branch and $z$ for the amplitude in the lower branch. That is $\vec\psi_{\rm in}=(1,0,0)^T$ or $\vec\psi_{\rm in}=(0,0,1)^T$. The second line of beamsplitters is identical to the first and thus the operation of one segment of the gate on the target photon state reads
\bea
\label{eq:3branchunit}
&&\left[
\begin{array}{ccc}
\cos\epsilon & -\sin\epsilon & 0 
\\
\rme^{-\xi}\cos\epsilon\sin\epsilon & \rme^{-\xi}\cos^2\epsilon & -\sin\epsilon
\\
\rme^{-\xi}\sin^2\epsilon & \rme^{-\xi}\cos\epsilon\sin\epsilon & \cos\epsilon
\end{array}
\right]\nonumber\\
=
&&\left[
\begin{array}{ccc}
1 & 0 & 0 
\\
0 & \cos\epsilon & -\sin\epsilon
\\
0 & \sin\epsilon & \cos\epsilon
\end{array}
\right]
\cdot
\left[
\begin{array}{ccc}
1 & 0 & 0 
\\
0 & \rme^{-\xi} & 0
\\
0 & 0 & 1
\end{array}
\right]
\cdot
\left[
\begin{array}{ccc}
\cos\epsilon & -\sin\epsilon & 0 
\\
\sin\epsilon & \cos\epsilon & 0
\\
0 & 0 & 1
\end{array}
\right]
\,.
\ea
Therefore the output state of the three-branch gate is given by
\bea
\label{eq:3branch_outputstate}
\vec\psi_{\rm out}
=
\left[
\begin{array}{ccc}
\cos\epsilon & -\sin\epsilon & 0 
\\
\rme^{-\xi}\cos\epsilon\sin\epsilon & \rme^{-\xi}\cos^2\epsilon & -\sin\epsilon
\\
\rme^{-\xi}\sin^2\epsilon & \rme^{-\xi}\cos\epsilon\sin\epsilon & \cos\epsilon
\end{array}
\right]^N
\cdot
\vec\psi_{\rm in}
\,.
\ea
The relation between $\epsilon$ and $N$ in order to have the target photon completely tunneled into the opposite branch at the end of the gate slightly changes into $\epsilon = \pi / \left( \sqrt{2} N \right)$ due to the additional beam splitter barrier. 
\subsubsection{Error probabilities}
\label{sssec:error_probabilites_3branch}
Now, without control photon ($\xi = \xi_{1\gamma}$), the error probability is given by the probability to not find the target photon, which is initially in the upper branch, $\vec\psi_{\rm in}=(1,0,0)^T$, in the lower branch at the end of the gate
\bea
\label{eq:3branch_error1gamma}
P_{\rm error}^{1\gamma} = 1 - \left|(0,0,1) \cdot \vec\psi_{\rm out}^{1\gamma}\right|^2
\,.
\ea
With control photon ($\xi = \xi_{2\gamma}$), as for the two-branch gate, the operation fails when the target photon is not in the upper branch at the end 
\bea
\label{eq:3branch_error2gamma}
P_{\rm error}^{2\gamma} = 1 - \left|(1,0,0) \cdot \vec\psi_{\rm out}^{2\gamma}\right|^2
\,.
\ea
Note that in both cases the error probabilities would be the same if the target photon alternatively starts in the lower branch and error probability definitions (\ref{eq:3branch_error1gamma}), (\ref{eq:3branch_error2gamma}) change accordingly (setup is symmetric).

$\vec\psi_{\rm out}$ in (\ref{eq:3branch_outputstate}) was calculated using \texttt{Mathematica}, but we will not quote the exact result here due to excessive length. Instead, we give the approximated results for the error probabilities (\ref{eq:3branch_error1gamma}), (\ref{eq:3branch_error2gamma}), which were derived using the same assumptions as for the two-branch gate ($N \gg 1$ and $N \xi_{2\gamma} \gg 1 \gg N \xi_{1\gamma}$). Interestingly, the results differ only slightly from the results described in section \ref{sssec:error_probabilites_2branch}
 (again, for first non-vanishing order in $1/N$, see \ref{ap:ssec:success_probabilities_of_the_three_branch_gate})
\bea
\label{eq:3branch_error_prob}
P_{\rm error}^{1\gamma} &=& \frac{N \xi_{1\gamma}}{2} + \Or \left( N^{-1} \right)\,,\nonumber\\
P_{\rm error}^{2\gamma} &=& \frac{\pi^2}{N \xi_{2\gamma}} + \Or \left( N^{-2} \right) 
\,.
\ea
In analogy to section \ref{sssec:error_probabilites_2branch}, we are looking for the optimal two-photon absorption ($\xi_{2\gamma}$) and one-photon loss ($\xi_{1\gamma}$) rates, given a certain quotient $\kappa := \xi_{2\gamma} / \xi_{1\gamma}$. We arrive at 
\bea
\label{eq:3branch_optimal_rates}
\xi_{1\gamma} = \frac{1}{\sqrt{\kappa}} \frac{\sqrt{2} \pi}{N}\,, \quad\quad \xi_{2\gamma} = \sqrt{\kappa} \frac{\sqrt{2} \pi}{N}
\,.
\ea
When reinserting (\ref{eq:3branch_optimal_rates}) in (\ref{eq:3branch_error_prob}), we find that we achieve the same effective error probability (\ref{eq:overall_error_prob}) and the same constraint on the quotient between two-photon absorption and one-photon loss as for the two-branch gate,
\bea
\label{eq:constraint}
\kappa
=
\frac{\xi_{2\gamma}}{\xi_{1\gamma}}
=
\frac{\pi^2}{2P_{\rm error}^2}
\gg1
\,.
\ea
The above requirement (\ref{eq:constraint}) represents a major experimental challenge 
since two-photon processes are typically much weaker than one-photon 
effects. 
There have been proposals to induce strong two-photon absorption in resonators, cavities, or fibres \cite{Franson:2006mw,You:2008kn,You:2009jw,Hendrickson:2010pi}. 
Of course, these ideas could be applied to our set-up in Figure~\ref{fig:3branch} as well.
But in the following sections, we shall focus on free-space propagation which offers some advantages as compared to wave-guides 
(but also has drawbacks, of course), and we are going to discuss the feasibility of the requirement (\ref{eq:constraint}) on a concrete example.
For an error threshold of $P_{\rm error} = 0.1$, i.e. error probability below 10\%, our gate would need a two-photon absorbing medium which features a ratio of $\kappa \approx 500$.
\subsubsection{Comparison with Franson-Gate}
\label{sssec:comparison_with_franson_gate}
Let us briefly compare the error probabilities of our gate with the error probabilities of the set-up discussed in \cite{Franson:2004fj,Leung:2006jo}. When using the same definitions for the error probabilities and applying the same limits, we obtain the following expressions for the Franson et. al. gate
\bea
\label{eq:franson_gate_p1gamma}
P_{\rm error}^{1\gamma} &=& 2 N \xi_{1\gamma}\,,\\
\label{eq:franson_gate_p2gamma}
P_{\rm error}^{2\gamma} &=& 4 N\xi_{1\gamma} + \frac{2\pi^2}{N \xi_{2\gamma}} + \Or \left( N^{-2} \right) 
\,.
\ea
It can be seen that the error probability for the Franson gate is higher for each case individually. Furthermore, the error probability for the two-photon case is always higher than the error probability for the one-photon case, which means that the optimal absorber length is given by minimizing $(\ref{eq:franson_gate_p2gamma})$ alone. Doing this (as always replacing $\kappa = \xi_{2\gamma} / \xi_{1\gamma}$), we arrive at
\bea
\label{eq:franson_gate_optimal_rates}
\xi_{1\gamma} = \frac{1}{\sqrt{\kappa}} \frac{\pi}{\sqrt{2} N}\,, \quad\quad \xi_{2\gamma} = \sqrt{\kappa} \frac{\pi}{\sqrt{2} N}
\,,
\ea
and thus we get an effective overall error probability of
\bea
\label{eq:franson_gate_overall_error_prob}
P_{\rm error} &=& \frac{4 \sqrt{2} \pi}{\sqrt{\kappa}} + \Or \left( N^{-1} \right)
\,.
\ea
Looking at (\ref{eq:franson_gate_overall_error_prob}) and (\ref{eq:constraint}), it is clear that the required value for $\kappa$ is a factor of 64 times smaller in our proposal compared to the Franson et. al. gate. This is mainly due to the fact that the absorbing medium is only in the middle branch in our set-up.
\subsubsection{Control photon loss}
\label{sssec:control_photon_loss}
Note that one-photon loss of the control photon was not taken into account so far. For a simple estimate of the additional error, assume $\xi_{\rm c}$ as the control photon loss rate which should be small, $N \xi_{\rm c} \ll 1$. The error probability due to this effect is roughly $1 - \left( \rme^{-\xi_{\rm c}} \right)^{2N} \approx 2 N \xi_{\rm c}$. Adding this to our existing error probability (\ref{eq:overall_error_prob}), we get
\bea
\label{eq:overall_error_including_control}
P_{\rm error} =&& \frac{\pi}{\sqrt{2 \kappa}} + 2 N \xi_{\rm c} + \Or \left( N^{-1} \right)
\,.
\ea
Assuming the worst case that the control photon loss rate is as high as the one-photon loss rate of the target photon, $\xi_{\rm c} = \xi_{1\gamma}$, we arrive at
\bea
\label{eq:overall_error_including_control_2}
P_{\rm error} =&& \frac{\pi}{\sqrt{2 \kappa}} \left( 1 + 4 \right) + \Or \left( N^{-1} \right)
\,,
\ea
and conclude that the error probability scales up by a factor of $5$, and therefore the necessary ratio $\kappa$ in (\ref{eq:constraint}) increases by a factor of $25$ (which is still less than the factor 64 then Franson et. al. gate has in {\em any} case). For useful gate operation it is thus a necessity to make the one-photon loss of the control photon as small as possible. We will show later that it is justified to neglect the one-photon loss of the control photon, as we can make it much smaller than the one-photon loss of the target photon, $\xi_{\rm c} \ll \xi_{1\gamma}$. 

Summing up, the three-branch gate features the same probability of success as the two-branch gate, while offering the possibility to serve as a full-fledged quantum CNOT-gate.
\section{Single three-level atom}
\label{sec:single_three_level_atom}
\begin{figure}[h]
\begin{center}
\psfrag{e1}{$E_1$}
\psfrag{e2}{$E_2$}
\psfrag{e3}{$E_3$}
\psfrag{1s}{1s}
\psfrag{2p}{2p}
\psfrag{3s}{3s}
\psfrag{delta}{$\Delta$}
\psfrag{omega_1}{$\omega_1$}
\psfrag{omega_2}{$\omega_2$}
\includegraphics[width=0.3\columnwidth]{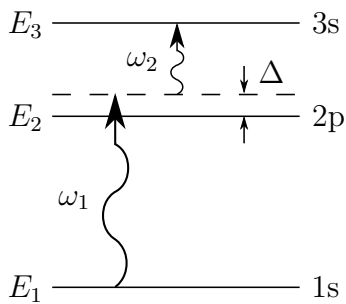}
\caption{Sketch (not to scale) of the level scheme. Both photons together are in resonance with the transition between 1s and 3s, $E_3 - E_1 = \omega_1 + \omega_2$, but one photon absorption is suppressed by the detuning $\Delta$.} 
\label{fig:niveau}
\end{center}
\end{figure}

As outlined in the previous section \ref{sec:two_photon_entangling_zeno_gate},
the proposed quantum CNOT-gate requires a two-photon absorption rate which is 
several orders of magnitudes larger than the one-photon loss rate. 
In order to investigate if this requirement (\ref{eq:constraint}) is 
satisfiable, we will now derive the two-photon absorption probability and 
one-photon scattering probability in the case of a simple three-level atom. 
The level scheme of the three-level atom is shown in Figure~\ref{fig:niveau}, 
it consists of three electron-eigenstates 1s, 2p and 3s with their respective 
energy levels $E_1$, $E_2$ and $E_3$. 
An alternative level scheme is investigated in section 
\ref{sec:alternative_level_scheme}.

\noindent
Mathematically speaking, we assume that the Hilbert space of the atom is spanned by the three orthonormal basis states
\bea
\label{eq:hilbert_space}
\mathcal{H}_{\rm atom} = \rm{lin}\{ \ket{1s}, \ket{2p}, \ket{3s} \}
\,.
\ea
When considering the well-established Hamiltonian for an electron bound by a potential,
\bea
\label{eq:Hamiltonian_0}
\hat{H}_0
&=&
\frac{\hat{\f{p}}^2}{2m} + V(\hat{\f{r}})
\,,
\ea
$E_1$, $E_2$ and $E_3$ are defined by the following characteristic equations
\bea
\label{eq:char_equation}
\hat{H}_0 \ket{\rm 1s} = E_1 \ket{\rm 1s}\,,\quad
\hat{H}_0 \ket{\rm 2p} = E_2 \ket{\rm 2p}\,,\quad
\hat{H}_0 \ket{\rm 3s} = E_3 \ket{\rm 3s}
\,.
\ea
The level spacings are shortly denoted by $E_{12} = E_2 - E_1$, $E_{23} = E_3 - E_2$ and $E_{13} = E_3 - E_1$. In order to be able to observe two-photon absorption, we assume two incoming photons, e.g. the target photon with frequency $\omega_1$ and the control photon with frequency $\omega_2$. Both photons together are in resonance with the transition between 1s and 3s, i.e. $E_{13} = \omega_1 + \omega_2$. Both photons by themselves however are detuned from the transition between 1s and 2p, such that a one-photon absorption is forbidden by energy conservation. The same applies for one-photon transitions between 1s and 3s, which are moreover forbidden by angular momentum conservation, as a photon always carries an angular momentum, but the involved states have the same angular momentum ($l=0$).
To sum up, the level scheme supports two-photon absorption but inhibits one-photon absorption. However, one-photon scattering will remain as the dominant one-photon loss mechanism.
\subsection{Hamiltonian and initial state}
\label{ssec:hamiltonian_and_initial_state}
We start most generally from the non-relativistic Hamiltonian for atom-field interactions \cite{Scully:1997fk} ($\hbar=c=\varepsilon_0=\mu_0=1$), dividing it into the non-interacting part $\hat{H}_0$ and the interaction part $\hat{H}_1$
\bea
\label{eq:Hamiltonian}
\hat H 
&=&
\frac{1}{2m}\left(
\hat{\f{p}}-q\hat{\f{A}\,}( \f{r}_0 + \f{\delta}\hat{\f{r}}, t )
\right)^2
+V(\hat{\f{r}}) + \hat{H}_{\rm field}
= 
\hat{H}_0 + \hat{H}_1
\,.
\ea
The non-interacting part $\hat{H}_0$ carries the energy of the electronic states as well as the field energy and is given by
\bea
\hat{H}_0 = \sum_{\fk{k},\lambda} \hat{a}_{\vphantom{\fk{\tilde{k}}}\fk{k}, \lambda}^\dagger \hat{a}_{\vphantom{\fk{\tilde{k}}}\fk{k}, \lambda}^{\phantom{\dagger}} \, \omega_k + E_1 \ket{\rm 1s} \bra{\rm 1s} + E_2 \ket{\rm 2p} \bra{\rm 2p} + E_3 \ket{\rm 3s} \bra {\rm 3s}
\,.
\ea
The interaction Hamiltonian,
\bea
\label{eq:Hamiltonian_interaction}
\hat{H}_1 = -\frac{q}{m} \hat{\f{p}} \cdot \hat{\f{A}\,}( \f{r}_0 + \f{\delta}\hat{\f{r}}, t ) + \frac{q^2}{2m} \hat{\f{A}\,}^2(\f{r}_0 + \f{\delta}\hat{\f{r}}, t )
\,,
\ea
contains the rest mass $m$ and the charge $q = -\left| e \right|$ of an electron, together with its momentum (vector) operator $\hat{\f{p}}$ and the vector potential operator for the quantized electromagnetic field (here written in the interaction picture)
\bea
\label{eq:vec_pot}
\hat{\f{A}\,}( \f{r}_0 + \f{\delta}\hat{\f{r}}, t ) = \sum_{\fk{k}, \lambda} \f{g}^{\lambda}_{\fk{k}} \hat{a}_{\vphantom{\fk{\tilde{k}}}\fk{k}, \lambda}^{\phantom{\dagger}} \rme^{-\rmi \omega_k t + \rmi\fk{k}\cdot( \fk{r}_0 + \fk{\delta}\hat{\fk{r}})} + \rm{H.c.}
\,,
\ea
at the position $\f{r}_0 + \f{\delta}\hat{\f{r}}$. Here $\f{r}_0$ is the position of the atomic nucleus and $\f{\delta}\hat{\f{r}}$ is the position of the electron relative to the nucleus position. The amplitude $\f{g}^{\lambda}_{\fk{k}}$ is given by 
\bea
\f{g}^{\lambda}_{\fk{k}} = -\rmi \sqrt{\frac{1}{2 \left( 2 \pi \right)^3 \omega_k}} \, \hat{\f{\epsilon}}^{\lambda}_{\fk{k}}
\,.
\ea
Hence the interaction Hamiltonian (\ref{eq:Hamiltonian_interaction}) describes one-photon transitions between adjacent levels by its mixed term $\propto \hat{\f{p}}\cdot\hat{\f{A}\,}( \f{r}_0 + \f{\delta}\hat{\f{r}}, t )$, and possible direct two-photon absorption or a one-photon scattering by its quadratic term $\propto \hat{\f{A}\,}^2(\f{r}_0 + \f{\delta}\hat{\f{r}}, t )$.

The Hamiltonian (\ref{eq:Hamiltonian_interaction}) can be rewritten in terms of atomic transition operators $\hat{\sigma}^{12} = \ket{\rm 1s}\bra{\rm 2p}$, $\hat{\sigma}^{23} = \ket{\rm 2p}\bra{\rm 3s}$ and $\hat{\sigma}^{13} = \ket{\rm 1s}\bra{\rm 3s}$ and photonic annihilation/creation operators $\hat{a}_{\vphantom{\fk{\tilde{k}}}\fk{k}, \lambda}^{\phantom{\dagger}} / \hat{a}_{\vphantom{\fk{\tilde{k}}}\fk{k}, \lambda}^\dagger$.
\bea
\label{eq:ham_inter_mod}
\hat{H}_{1, {\rm I}} ( \f{r}_0, t )
&=& \sum_{\fk{k}, \lambda} \left( g_{\vphantom{\fk{\tilde{k}}}\fk{k}, \lambda}^{12} \hat{\sigma}^{12} \rme^{-\rmi \Delta_k^{12} t} + g_{\vphantom{\fk{\tilde{k}}}\fk{k}, \lambda}^{23} \hat{\sigma}^{23} \rme^{-\rmi \Delta_k^{23} t} \right) \hat{a}_{\vphantom{\fk{\tilde{k}}}\fk{k}, \lambda}^\dagger \rme^{-\rmi\fk{k}\cdot\fk{r}_0} + \rm{H.c.}\nonumber\\
&-& \sum_{\fk{k}, \lambda} \left( g_{\vphantom{\fk{\tilde{k}}}\fk{k}, \lambda}^{12} \hat{\sigma}^{12} \rme^{-\rmi(E_{12} + \omega_k)t} + g_{\vphantom{\fk{\tilde{k}}}\fk{k}, \lambda}^{23} \hat{\sigma}^{23} \rme^{-\rmi(E_{23} + \omega_k)t} \right) \hat{a}_{\vphantom{\fk{\tilde{k}}}\fk{k}, \lambda}^{\phantom{\dagger}} \rme^{\rmi\fk{k}\cdot\fk{r}_0} + \rm{H.c.}\nonumber\\
&+& \sum_{\fk{k}, \fk{\tilde{k}}, \lambda, \tilde{\lambda}} g_{\vphantom{\fk{\tilde{k}}}\fk{k} \fk{\tilde{k}}, \lambda \tilde{\lambda}}^{13} \rme^{\rmi(\omega_k + \omega_{\tilde{k}} - E_{13})t} \hat{a}_{\vphantom{\fk{\tilde{k}}}\fk{k}, \lambda}^\dagger \hat{a}_{\fk{\tilde{k}}, \tilde{\lambda}}^\dagger \hat{\sigma}^{13} \rme^{-\rmi(\fk{k} + \fk{\tilde{k}})\cdot\fk{r}_0} + \rm{H.c.}\nonumber\\
&+& \sum_{\fk{k}, \fk{\tilde{k}}, \lambda, \tilde{\lambda}} g_{\vphantom{\fk{\tilde{k}}}\fk{k} \fk{\tilde{k}}, \lambda \tilde{\lambda}}^{11} \rme^{\rmi(\omega_k - \omega_{\tilde{k}})t} \hat{a}_{\vphantom{\fk{\tilde{k}}}\fk{k}, \lambda}^\dagger \hat{a}_{\fk{\tilde{k}}, \tilde{\lambda}}^{\phantom{\dagger}} \rme^{\rmi(\fk{\tilde{k}} - \fk{k})\cdot\fk{r}_0}\nonumber\\
&+& \hat{H}_{1, {\rm I}}^{\rm rem} ( \f{r}_0, t )
\,.
\ea
The derivation of the associated coupling constants $g_{\vphantom{\fk{\tilde{k}}}\fk{k}, \lambda}^{12}$ etc. can be found in \ref{ap:ssec:derivation_of_the_coupling_constants}.
Note that the Hamiltonian (\ref{eq:ham_inter_mod}) is written in the interaction picture and therefore now carries phases due to the atomic energies. The remaining part $\hat{H}_{1, {\rm I}}^{\rm rem} ( \f{r}_0, t )$ is given in \ref{ap:ssec:derivation_of_the_coupling_constants}, because the terms it contains are unimportant for the upcoming analysis. Abbrevations $\Delta_k^{12} := E_{12} - \omega_k$ and $\Delta_k^{23} := E_{23} - \omega_k$ are introduced to indicate the detunings between the photonic energies and the atomic transition energies.
The interaction Hamiltonian (\ref{eq:ham_inter_mod}) works on an incoming state which consists of the atom in the ground state $\ket{\rm 1s}$ and two incoming photons $\hat{a}_1^\dagger$ and $\hat{a}_2^\dagger$
\bea
\label{eq:input_state}
\ket{\psi_{\rm in}} = \hat{a}_1^\dagger \hat{a}_2^\dagger \ket{0} \ket{\rm 1s}
\,.
\ea
The two incident photons are generally smeared out in $\f{k}$-space,
\bea
\label{eq:anni_def}
\hat{a}_{1/2}^{\phantom{\dagger}} = \sum_{\fk{k}} \, f_{1/2}\left( \f{k} \right) \hat{a}_{\vphantom{\fk{\tilde{k}}}\fk{k}, \lambda_{1/2}}^{\phantom{\dagger}}
\,,
\ea
in such a way that they respect the usual commutation relations
\bea
\left[\hat{a}_1^{\phantom{\dagger}}, \hat{a}_{1}^\dagger \right] = 1,
\quad\quad
\left[\hat{a}_2^{\phantom{\dagger}}, \hat{a}_{2}^\dagger \right] = 1
\quad\quad {\rm and} \quad\quad
\left[ \hat{a}_1^{(\dagger)}, \hat{a}_2^{( \dagger)} \right] = 0
\,.
\ea
In order to do so, $f_{1/2}$ need to fulfill the conditions
\bea
\label{eq:f_conditions}
\sum_{\fk{k}} \left|f_{1/2} \left( \f{k} \right)\right|^2 = 1
\quad\quad {\rm and} \quad\quad
\sum_{\fk{k}} f_1^{\left( * \right)} \left( \f{k} \right) f_2^{\left( * \right)} \left( \f{k} \right) = 0
\,.
\ea
This can be seen by employing the fundamental commutation relation for creation and annihilation operators,
\bea
\label{eq:fund_comm_rel}
\left[ \hat{a}_{\vphantom{\fk{\tilde{k}}}\fk{k}, \lambda}^{\phantom{\dagger}}, \hat{a}_{\fk{\tilde{k}}, \tilde{\lambda}}^\dagger \right] = \delta^3 \left( \f{k} - \f{\tilde{k}} \right) \delta_{\vphantom{\tilde{\lambda}}\lambda \tilde{\lambda}}
\,.
\ea
\subsection{Perturbation theory}
\label{ssec:perturbation_theory}
\begin{figure}[h]
\begin{center}
\psfrag{e1}{$E_1$}
\psfrag{e2}{$E_2$}
\psfrag{e3}{$E_3$}
\psfrag{omega_1}{$\omega_1$}
\psfrag{omega_2}{$\omega_2$}
\psfrag{omega}{$\omega_{1/2}$}
\psfrag{omega_s}{$\omega_k$}
\subfigure[]{\includegraphics[height=0.25\columnwidth]{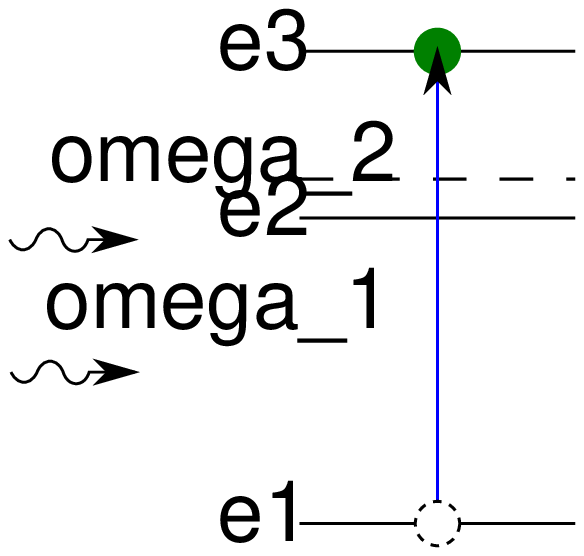}}
\hspace{.6cm}
\subfigure[]{\includegraphics[height=0.25\columnwidth]{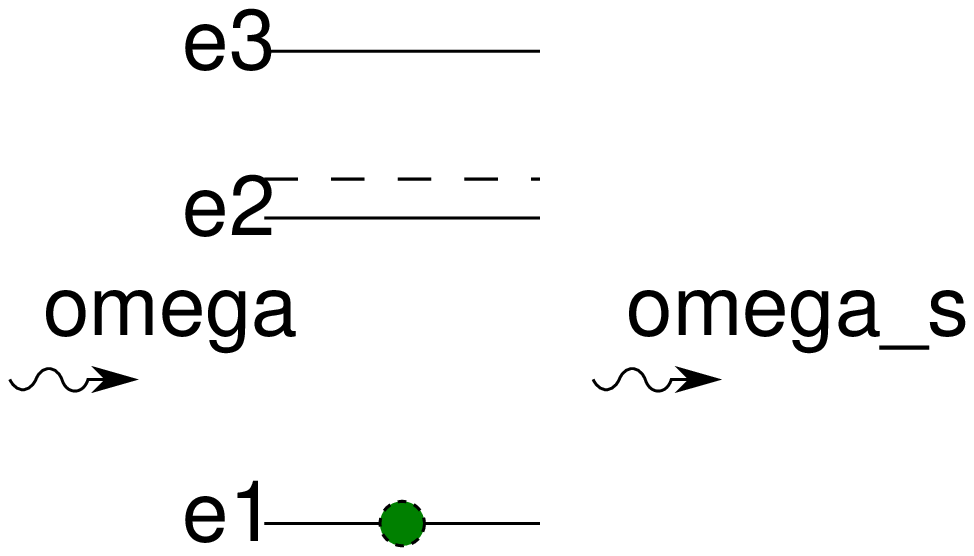}}
\caption{First-order processes of the Hamiltonian (\ref{eq:ham_inter_mod}) acting on the input state (\ref{eq:input_state}). The absorption (a) or scattering (b) works via the $\propto \hat{\f{A}\,}^2(\f{r}_0 + \f{\delta}\hat{\f{r}}, t )$-term of the Hamiltonian (\ref{eq:Hamiltonian_interaction}), thus the the 2p-level is irrelevant here. A circle with black dashed border symbolizes the initial state of the atom, while a green circle symbolizes the final state. For scattering, (b), initial and final state of the atom are the same.} 
\label{fig:niveau_proc_qA2}
\end{center}
\end{figure}

The time evolution of the system for infinite interaction time is governed by
\bea
\label{eq:time_evo_oper}
\hat{U} ( \f{r}_0, \infty ) = \mathcal{T} \exp\left( -\rmi \int_{-\infty}^{+\infty} dt' \hat{H}_{1, {\rm I}} ( \f{r}_0, t' ) \right)
\,,
\ea
with time ordering operator $\mathcal{T}$ and atomic (nucleus) position $\f{r}_0$. We calculate the outgoing state $\hat{U} ( \f{r}_0, \infty ) \ket{\psi_{\rm in}}$ via second-order standard perturbation theory \cite{Sakurai:1994uq}, dropping all terms of order $\hat{\f{A}\,}^3$ or higher. $\ket{\psi_1}$ and $\ket{\psi_2}$ represent the first-order term and the second-order term, respectively, such that
\bea
\hat{U} ( \f{r}_0, \infty ) \ket{\psi_{\rm in}} \approx \ket{\psi_{\rm in}} + \ket{\psi_1} + \ket{\psi_2}
\,.
\ea
For infinite interaction time, only terms matter which satisfy energy conservation, because other terms have a non-vanishing temporal phase which makes them cancel out in the infinite time integral. Considering this (using also $\omega_{1/2} < E_{13}$) and performing some commutations of the annihilation- and creation operators (see \ref{ap:ssec:commutation_relations}), we arrive at
\bea
\label{eq:first_order_term}
\ket{\psi_1} &=& -\rmi (2 \pi) \sum_{\fk{k}, \fk{\tilde{k}}, \lambda, \tilde{\lambda}} 2 \big(g_{\vphantom{\fk{\tilde{k}}}\fk{k} \fk{\tilde{k}}, \lambda \tilde{\lambda}}^{13}\big)^* \delta \left( \omega_k + \omega_{\tilde{k}} - E_{13} \right)\times\nonumber\\
&&\hspace{1.0cm}\times f_1^* ( \f{k} ) f_2^* ( \f{\tilde{k}} ) \delta_{\vphantom{\tilde{\lambda}}\lambda_1 \lambda} \delta_{\vphantom{\tilde{\lambda}}\lambda_2 \tilde{\lambda}} \rme^{\rmi( \fk{k} + \fk{\tilde{k}})\cdot\fk{r}_0} \ket{0} \ket{\rm 3s}-\nonumber\\
&&-\rmi (2 \pi) \sum_{\fk{k}, \fk{\tilde{k}}, \lambda, \tilde{\lambda}} g_{\vphantom{\fk{\tilde{k}}}\fk{k} \fk{\tilde{k}}, \lambda \tilde{\lambda}}^{11} \delta \left( \omega_k - \omega_{\tilde{k}} \right)\times\nonumber\\
&&\hspace{1.0cm}\times\left( f_2^* ( \f{\tilde{k}} ) \delta_{\vphantom{\tilde{\lambda}}\lambda_2 \tilde{\lambda}} \hat{a}_1^\dagger + f_1^* ( \f{\tilde{k}} ) \delta_{\vphantom{\tilde{\lambda}}\lambda_1 \tilde{\lambda}} \hat{a}_2^\dagger \right) \hat{a}_{\vphantom{\fk{\tilde{k}}}\fk{k}, \lambda}^\dagger \rme^{\rmi(\fk{\tilde{k}} - \fk{k})\cdot\fk{r}_0} \ket{0} \ket{\rm 1s}
\,.
\ea
That is, the first-order term $\ket{\psi_1}$ contains one-photon scattering and two-photon absorption via the $\propto \hat{\f{A}\,}^2( \f{r}_0 + \f{\delta}\hat{\f{r}}, t )$-term of the original Hamiltonian (\ref{eq:Hamiltonian_interaction}), see also Figure~\ref{fig:niveau_proc_qA2}.

\begin{figure}[h]
\begin{center}
\psfrag{e1}{$E_1$}
\psfrag{e2}{$E_2$}
\psfrag{e3}{$E_3$}
\psfrag{omega_1}{$\omega_1$}
\psfrag{omega_2}{$\omega_2$}
\psfrag{omega}{$\omega_1$}
\psfrag{omega_s}{$\omega_k$}
\psfrag{and}{\&}
\psfrag{delta_1}{$|\Delta_{k_1}^{12}|$}
\psfrag{delta_2}{$|\Delta_{k_2}^{12}|$}
\psfrag{delta_3}{$|\Delta_{k_1}^{12}|$}
\subfigure[]{\includegraphics[height=0.25\columnwidth]{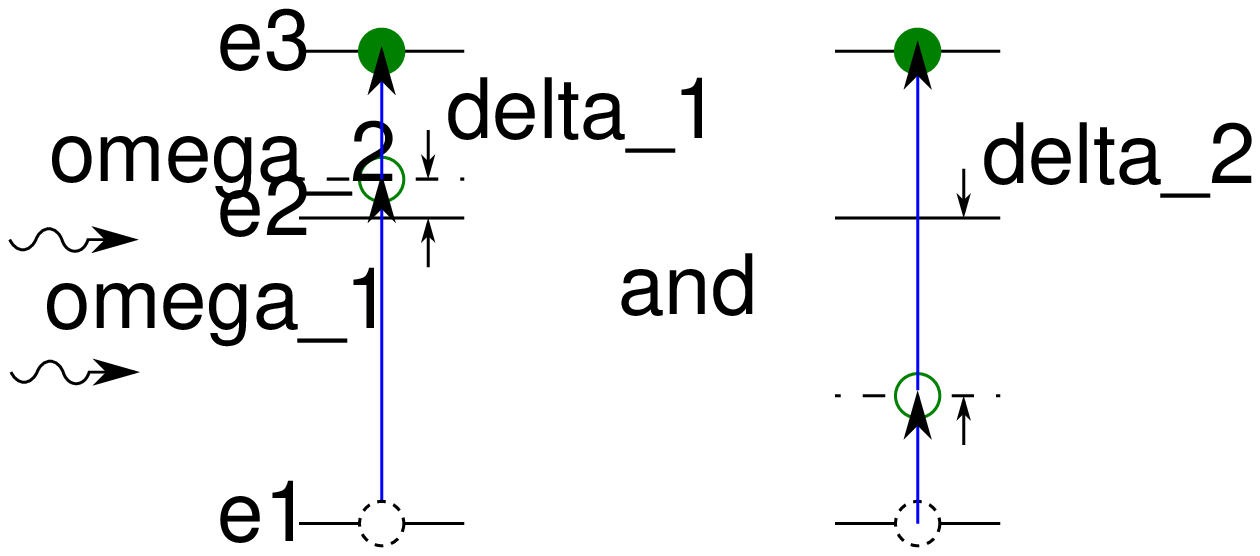}}
\hrule
\subfigure[]{\includegraphics[height=0.25\columnwidth]{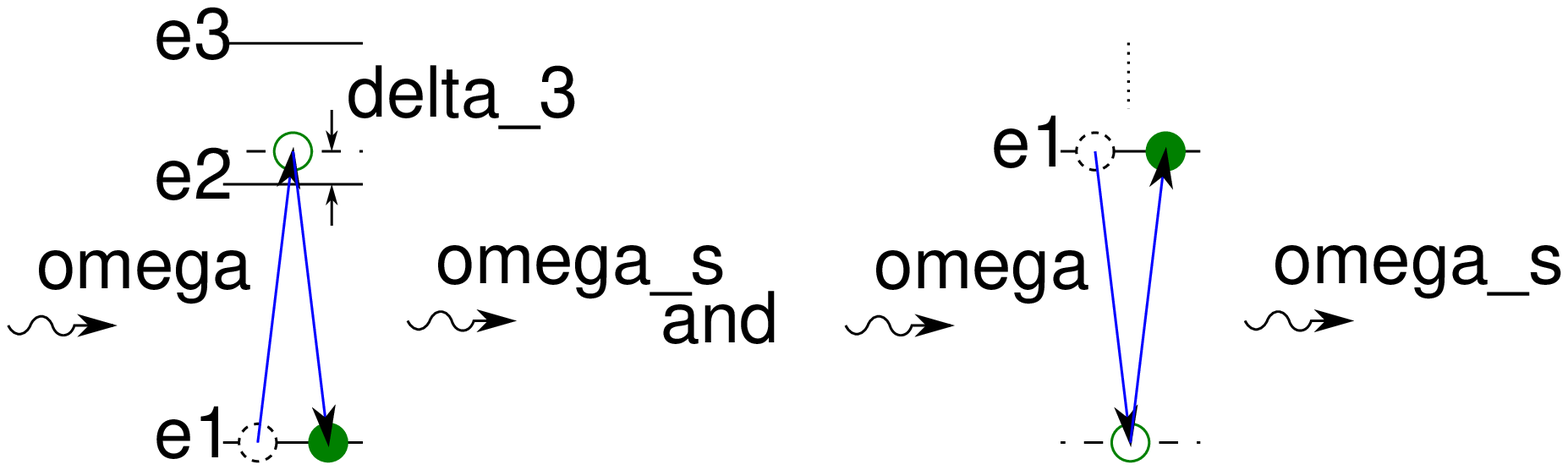}}
\caption{Second-order processes of the Hamiltonian (\ref{eq:ham_inter_mod}) working on the input state (\ref{eq:input_state}). The absorption (a) or scattering (b), exemplarily shown for $\omega_1$ getting scattered) works via the $\propto \hat{\f{p}}\cdot\hat{\f{A}\,}$-term of the Hamiltonian (\ref{eq:Hamiltonian_interaction}). A white circle with black dashed border symbolizes the initial state of the atom, while a white circle with green solid border symbolizes the occupation of a virtual level (black dashed line) during the transition. A filled green circle denotes the final state. The detuning of the respective process is given by the distance between the 2p-level (black solid line) and the virtual level (black dashed line).} 
\label{fig:niveau_proc_pA}
\end{center}
\end{figure}

\noindent
The second-order term is also given by (\ref{eq:time_evo_oper}), where we additionally have to attend for normal ordering of the photonic creation and annihilation operators \cite{Ryder:1996fk}.
As stated above, we leave out all terms of order $\hat{\f{A}\,}^3$ or higher, so for second-order perturbation theory, only the first two lines of the Hamiltonian (\ref{eq:ham_inter_mod}) contribute. Calculating the time-integrals, using the commutation relations (\ref{ap:ssec:commutation_relations}) and renaming indices in the first term we arrive at
\bea
\label{eq:sec_order_term}
\ket{\psi_2} &=& \rmi \left( 2 \pi \right) \sum_{\fk{k}, \fk{\tilde{k}}, \lambda, \tilde{\lambda}} \left( \frac{ g_{\fk{\tilde{k}}, \tilde{\lambda}}^{12} g_{\vphantom{\fk{\tilde{k}}}\fk{k}, \lambda}^{23} }{\Delta_{\tilde{k}}^{12}} + \frac{ g_{\vphantom{\fk{\tilde{k}}}\fk{k}, \lambda}^{12} g_{\fk{\tilde{k}}, \tilde{\lambda}}^{23} }{\Delta_k^{12}} \right)^* \delta \left( \omega_k + \omega_{\tilde{k}} - E_{13} \right)\times\nonumber\\
&&\hspace{1.0cm}\times f_1^* ( \f{k} ) f_2^* ( \f{\tilde{k}} ) \delta_{\vphantom{\tilde{\lambda}}\lambda_1 \lambda} \delta_{\vphantom{\tilde{\lambda}}\lambda_2 \tilde{\lambda}} \rme^{\rmi(\fk{k} + \fk{\tilde{k}})\cdot\fk{r}_0} \ket{0} \ket{\rm 3s}+ \nonumber\\
&+& \rmi \left( 2 \pi \right) \sum_{\fk{k}, \fk{\tilde{k}}, \lambda, \tilde{\lambda}} \left( \frac{g_{\vphantom{\fk{\tilde{k}}}\fk{k}, \lambda}^{12} \big(g_{\fk{\tilde{k}}, \tilde{\lambda}}^{12}\big)^* }{\Delta_{\tilde{k}}^{12}} 
+ \frac{ g_{\fk{\tilde{k}}, \tilde{\lambda}}^{12} \big(g_{\vphantom{\fk{\tilde{k}}}\fk{k}, \lambda}^{12}\big)^* }{E_{12} + \omega_k} \right) \delta \left( \omega_k - \omega_{\tilde{k}} \right)\times\nonumber\\
&&\hspace{1.0cm}\times \left( f_2^* ( \f{\tilde{k}} ) \delta_{\vphantom{\tilde{\lambda}}\lambda_2 \tilde{\lambda}} \hat{a}_1^\dagger + f_1^* ( \f{\tilde{k}} ) \delta_{\vphantom{\tilde{\lambda}}\lambda_1 \tilde{\lambda}} \hat{a}_2^\dagger \right) \hat{a}_{\vphantom{\fk{\tilde{k}}}\fk{k}, \lambda}^\dagger \rme^{\rmi(\fk{\tilde{k}} - \fk{k})\cdot\fk{r}_0} \ket{0} \ket{\rm 1s}
\,.
\ea
In the second-order term, we also have two-photon absorption and one-photon scattering, but this time going over the intermediate level 2p via the $\propto \hat{\f{p}}\cdot\hat{\f{A}\,}$-term of the Hamiltonian (\ref{eq:Hamiltonian_interaction}). Therefore the detuning of the two photons from the intermediate level, e.g. $\Delta_{\tilde{k}}^{12} = E_{12} - \omega_{\tilde{k}}$, also plays a role here. Note that there are two terms for two-photon absorption (bracket in the first line), as there are two possible orders of absorption. There are also two scattering mechanisms (bracket in the third line), whereof the second one has a detuning comparable to optical energies and thus will be neglected. All four processes are shown in Figure~\ref{fig:niveau_proc_pA}.
\subsection{Two-photon absorption probability}
\label{ssec:two_photon_absorption_probability}
Using both the first-order (\ref{eq:first_order_term}) and the second-order (\ref{eq:sec_order_term}) term, we can calculate the total two-photon absorption probability $P_{2\gamma}$
\bea
\label{eq:abs_prob}
P_{2\gamma} &=& \left|\bra{\rm 3s} \hat{U} ( \f{r}_0, \infty ) \ket{\psi_{\rm in}}\right|^2
\,.
\ea
In order to obtain a concrete result we now need to specify the distribution of the incoming photons in $\f{k}$-space. For the sake of simplicity, we assume rectangular functions around $\f{k_{1/2}} = k_{1/2} \cdot \f{\hat{e}_z}$, i.e. the incoming photons propagate almost parallel to the $z$-axis, with slight deviations in $x$- and $y$-direction. The side lengths are denoted $2 \Delta k_x$, $2 \Delta k_y$ and $2 \Delta k_z$, and of course $f_{1/2}$ are normalized such that (\ref{eq:f_conditions}) is fulfilled.

As the side lengths in $\f{k}$-space should be small, the integrand in (\ref{eq:abs_prob}) can be regarded as constant over the small integration volume at $\f{k} \approx \f{k_1}$ and $\f{\tilde{k}} \approx \f{k_2}$. The level scheme of the atom is designed in such a way that the detuning of the target photon ($\Delta_{k_1}^{12} =: \Delta$), is much smaller than the detuning of the control photon ($\Delta_{k_2}^{12}$), whose term is therefore omitted. For small detuning, i.e. in the optical regime, the direct transition $\big(g_{\vphantom{\fk{\tilde{k}}}\fk{k} \fk{\tilde{k}}, \lambda_1 \lambda_2}^{13}\big)^*$, produced via the quadratic $\propto \hat{\f{A}\,}^2$-term, can also be neglected (\ref{ap:ssec:derivation_of_the_coupling_constants}). The result is
\bea
\label{eq:2gamme_pre}
P_{2\gamma} = (2 \pi)^2 \frac{16 \left|g_{\vphantom{\fk{\tilde{k}}}\fk{k_1},\lambda_1}^{12}\right|^2 \left|g_{\vphantom{\fk{\tilde{k}}}\fk{k_2},\lambda_2}^{23}\right|^2}{\Delta^2} \Delta k_x^2 \Delta k_y^2
\,.
\ea
\noindent
Inserting the expressions for the coupling constants (notably the typical coupling length $\ell_{\rm atom} = |\bra{\psi_{\rm out}} \hat{\f{r}} \ket{\psi_{\rm in}}|$, see \ref{ap:ssec:derivation_of_the_coupling_constants} for details) and introducing the cross section area of the two photon wave-packets $A := 1/ \left( \Delta k_x \Delta k_y \right)$ finally yields 
\bea
\label{eq:2gamma}
P_{2\gamma}
=
\frac{4\alpha^2_{\rm QED}}{\pi^2\omega_1\omega_2}\,
\frac{E^2_{12}E^2_{23}}{\Delta^2}\,
\frac{\ell_{\rm atom}^4}{A^2}
\,,
\ea
where, for simplicity, we assumed that the polarization vectors are along the same axis as the absorbing orbitals. Otherwise, the result is still a reasonable estimate for the magnitude of the two-photon absorption. Note that the probability (\ref{eq:2gamma}) does not depend on the length of the photon wave-packets; only the transversal dimensions (i.e., $A$) enter.
\subsection{One-photon scattering probability}
\label{ssec:one_photon_scattering_probability}
From the final amplitudes (\ref{eq:first_order_term}) and (\ref{eq:sec_order_term}), we also may calculate the one-photon scattering probability $P_{1\gamma}$. 
\bea
P_{1\gamma} &=& \left|\bra{\rm 1s} \hat{U} ( \f{r}_0, \infty ) \ket{\psi_{\rm in}}\right|^2
\,.
\ea
Using the fundamental commutation relation, one can verify that $P_{1\gamma}$ can be simplified to the following form due to the orthogonality in $\f{k}$- and $\lambda$-space. The index $\zeta = 1,2$ sums over the two possibly scattered photons.
From the commutation relations there also arises a term of order $f^4$ which will be neglected, as it would be about a factor $\Or \left( \Delta k_x \Delta k_y / \left( k_1 k_2 \right) \right)$ smaller than the first term. As stated above, the second scattering mechanism was ignored due to its very large detuning $E_{12} + \omega_k$.
\bea
\label{eq:p1gamma_app}
P_{1\gamma} &=& (2 \pi)^2 \sum_{\fk{k}, \lambda, \zeta = 1,2} \left| \sum_{\fk{\tilde{k}}} \left( \frac{g_{\vphantom{\fk{\tilde{k}}}\fk{k}, \lambda}^{12} \big(g_{\fk{\tilde{k}}, \lambda_\zeta}^{12}\big)^* }{\Delta_{\tilde{k}}^{12}} - g_{\vphantom{\fk{\tilde{k}}}\fk{k} \fk{\tilde{k}}, \lambda \lambda_\zeta}^{11} \right)\times\right.\nonumber\\
&&\hspace{1.0cm}\left.\times\,\delta \left( \omega_k - \omega_{\tilde{k}} \right) f_\zeta^* ( \f{\tilde{k}} ) \rme^{\rmi(\fk{\tilde{k}} - \fk{k})\cdot\fk{r}_0} \vphantom{\frac{g_{\vphantom{\fk{\tilde{k}}}\fk{k}, \lambda}^{12} \big(g_{\fk{\tilde{k}}, \lambda_\zeta}^{12}\big)^* }{\Delta_{\tilde{k}}^{12}}}\right|^2 + \Or \left( f^4 \right)
\,.
\ea
After carrying out the inner integration (by assuming the integrand to be constant at $\f{\tilde{k}} \approx \f{k_\zeta}$), the outer integration can be done in spherical coordinates where the radial integration is restricted due to the Dirac delta function in the inner integration (for further details, see \ref{ap:ssec:scattering_probability_integration}). After inserting the expressions for the coupling constants and executing the outer integration and the sum over the two polarization modes, we arrive at the final result
\bea
\label{eq:scat_prob}
P_{1\gamma} = \frac{8 \alpha_{\rm QED}^2}{3 \pi} \sum_{\zeta = 1,2} \left[\frac{E_{12}^2}{\Delta_{k_\zeta}^{12}} - \frac{1}{m \ell_{\rm atom}^2}\right]^2 \frac{\ell_{\rm atom}^4}{A}
\,.
\ea
It can be seen, that the one-photon scattering probability is based on two mechanisms. The first term in the bracket corresponds to a scattering via a virtual occupation of the intermediate level 2p, while the second term corresponds to a direct scattering via the $\propto \hat{\f{A}\,}^2$-term of the Hamiltonian (\ref{eq:Hamiltonian_interaction}). Usually, for optical energies $E_{12}$ and very small detunings (but still much larger than the natural line width of the middle 2p-level), the first term is some orders of magnitude larger than the second term, which therefore can be neglected. The first term for $\zeta = 2$ describes the scattering probability of the control photon. Choosing the detuning of the control photon to be much larger than that of the target photon, i.e. $\Delta_{k_2}^{12} \gg \Delta_{k_1}^{12} = \Delta$, the scattering of the control photon can be neglected as required in section \ref{sssec:control_photon_loss}.
\subsection{Destructive interference}
\label{ssec:destructive_interference}
In principle, it is possible to suppress the scattering by destructive interference between the two contributions (from the $\propto \hat{\f{p}}\cdot\hat{\f{A}\,}$- and the $\propto \hat{\f{A}\,}^2$-term, respectively) to the scattering amplitude, leading to $P_{1\gamma}\ll P_{2\gamma}$. To this end, we need to go to the infrared regime and to large detunings, such that the first term in (\ref{eq:scat_prob}) gets small. When dealing with large detunings, neglecting the second scattering term in (\ref{eq:sec_order_term}) is not accurate and therefore we keep it for this consideration. For the interference to be destructive we obtain
\bea
\label{eq:destr_cond}
\left| \frac{g_{\vphantom{\fk{\tilde{k}}}\fk{k}, \lambda}^{12} \big(g_{\vphantom{\fk{\tilde{k}}}\fk{k_\zeta}, \lambda_\zeta}^{12}\big)^* }{E_{12} - \omega_{k_\zeta}} + \frac{ g_{\vphantom{\fk{\tilde{k}}}\fk{k_\zeta}, \lambda_\zeta}^{12} \big(g_{\vphantom{\fk{\tilde{k}}}\fk{k}, \lambda}^{12}\big)^* }{E_{12} + \omega_{k_\zeta}} - g_{\vphantom{\fk{\tilde{k}}}\fk{k} \fk{k_\zeta}, \lambda \lambda_\zeta}^{11} \right|^2 \stackrel{!}{=} 0
\,.
\ea
Whereas in the previous sections it was convenient to assume that the polarization vector of the incoming photon $\hat{\f{\epsilon}}_{\fk{k_\zeta}}^{\lambda_\zeta}$ is along the same axis as the absorbing orbital $\f{\ell}_{\rm atom}^{12} = \bra{1s} \hat{\f{r}} \ket{2p}$, it is necessary for destructive interference to happen. So when we require this and additionally $|\f{\ell}_{\rm atom}^{12}| = \ell_{\rm atom}$, our condition (\ref{eq:destr_cond}) simplifies to
\bea
\left| \left[ E_{12}^2 \ell_{\rm atom}^2 \left( \frac{1}{E_{12} - \omega_{k_\zeta}} + \frac{1}{E_{12} + \omega_{k_\zeta}} \right) \frac{\f{\ell}_{\rm atom}^{12}}{\ell_{\rm atom}} - \frac{1}{m} \hat{\f{\epsilon}}^{\lambda_\zeta}_{\fk{k_\zeta}} \right] \cdot \hat{\f{\epsilon}}^{\lambda}_{\fk{k}} \right|^2 \stackrel{!}{=} 0
\,.
\ea
The expression inside the square brackets is zero when 
\bea
E_{12}^2 \ell_{\rm atom}^2 \left( \frac{1}{E_{12} - \omega_{k_\zeta}} + \frac{1}{E_{12} + \omega_{k_\zeta}} \right) - \frac{1}{m} = 0
\,,
\ea
or written as a condition on the frequencies of the target photon and the control photon
\bea
\label{eq:destr_interf_cond}
\omega_1 = \omega_2 = E_{12} \sqrt{1 - 2 m \ell_{\rm atom}^2 E_{12}}
\,.
\ea
Together with the resonance condition $\omega_1 + \omega_2 = E_{12} + E_{23}$, there is only one parameter left. For example, choosing $E_{12}$ determines $\omega_1 = \omega_2$ by (\ref{eq:destr_interf_cond}). Due to the resonance condition, $E_{23}$ is also determined then.
\subsection{Comparison and conclusion}
\label{ssec:comparison_and_conclusion}
In the general case without destructive interference, one can neglect the second term inside the bracket of (\ref{eq:scat_prob}), as well as the first term for $\zeta = 2$ due to the large detuning of the control photon. What remains is the scattering probability of the target photon due to the $\propto \hat{\f{p}}\cdot\hat{\f{A}\,}$-term of the Hamiltonian (\ref{eq:Hamiltonian_interaction}), which is
\bea
\label{eq:scat_prob_simp}
P_{1\gamma} = \frac{8 \alpha_{\rm QED}^2}{3 \pi} \frac{E_{12}^4}{\Delta^2} \frac{\ell_{\rm atom}^4}{A}
\,.
\ea
We arrive at an expression for the quotient of the two-photon absorption probability and the one-photon scattering probability
\bea
\label{eq:ratio}
\frac{P_{2\gamma}}{P_{1\gamma}}= \frac{1}{\pi \omega_1 \omega_2 A} \frac{3 E_{23}^2}{2 E_{12}^2} = \frac{1}{\Or (\pi\omega^2_{1,2}A )}
\,.
\ea
To get an impression what this ratio is in view of the diffraction limit, we can insert $\omega \approx \omega_1 \approx \omega_2 \approx E_{12} \approx E_{23}$ (for simplicity) and $A = (\lambda/2)^2$ where $\lambda = 2 \pi / \omega$, which results in
\bea
\label{eq:ratio_example}
\frac{P_{2\gamma}}{P_{1\gamma}} \approx \frac{3}{2 \pi^3} \approx 0.05
\,.
\ea
So even at the diffraction limit, this ratio is smaller than one and thus 
the requirement (\ref{eq:constraint}) does not seem to be satisfiable 
without additional efforts. 
In the next sections \ref{sec:repeated_inducing}, 
\ref{sec:coherent_interaction_with_Sgg1_atoms},
and \ref{sec:alternative_level_scheme}, 
we present possible mechanisms to boost two-photon absorption in order to 
overcome this difficulty.
\subsection{Effective treatment via the ``slow variables'' formalism}
\label{ssec:effective_treatment_via_slow_variables_formalism}
In our analysis of two-photon absorption, section \ref{ssec:two_photon_absorption_probability}, we found that the two-step process via the $\propto \hat{\f{p}}\cdot\hat{\f{A}\,}$-term is the dominant process. In what follows, we will thus repeatedly resort to an effective treatment of two-photon absorption and emission, where we omit the middle (2p) level from the Hamiltonian and model the two-step process effectively as direct transition between the 1s- and the 3s-level, i.e.\ as a first-order process.

The ``slow variables'' formalism, which we explain below, provides an elegant way to pass into the effective treatment and to evaluate the associated coupling constants. Considering our three-level system, we start from a reduced Hamiltonian which treats the electromagnetic field classically, supporting only the modes $A_1$ and $A_2$ with frequencies $\omega_1$ and $\omega_2$, respectively. Moreover it supports only transitions between adjacent levels via the appropriate mode $A_1$ or $A_2$, i.e. the $\propto \hat{\f{A}\,}^2$-term is being completely disregarded
\bea
\label{eq:reduced_hamiltonian}
\hat{H}_{\rm red}
&=& E_1 \ket{\rm 1s}\bra{\rm 1s} + E_2 \ket{\rm 2p}\bra{\rm 2p} + E_3 \ket{\rm 3s}\bra{\rm 3s}+\nonumber\\
&&+ \left( g^{12} A_1^* \hat{\sigma}^{12} \rme^{\rmi \omega_1 t} + g^{23} A_2^* \hat{\sigma}^{23} \rme^{\rmi \omega_2 t} + \rm{H.c.}\right)
\,.
\ea
From this reduced Hamiltonian, we can construct an associated Lagrangian which treats the state-vectors as if they were complex numbers (which are of course normalized)
\bea
\label{eq:lagrangian}
L &=& E_1 \Psi_{\rm 1s}^* \Psi_{\rm 1s} + E_2 \Psi_{\rm 2p}^* \Psi_{\rm 2p} + E_3 \Psi_{\rm 3s}^* \Psi_{\rm 3s}-\nonumber\\
&& -\rmi \Psi_{\rm 1s}^* \dot{\Psi}_{\rm 1s} -\rmi \Psi_{\rm 2p}^* \dot{\Psi}_{\rm 2p} -\rmi \Psi_{\rm 3s}^* \dot{\Psi}_{\rm 3s}+\nonumber\\
&& + \left( g^{12} A_1^* \Psi_{\rm 1s}^* \Psi_{\rm 2p} \rme^{\rmi \omega_1 t} + g^{23} A_2^* \Psi_{\rm 2p}^* \Psi_{\rm 3s} \rme^{\rmi \omega_2 t} + \rm{H.c.} \right)
\,.
\ea
Redefining the phases of the complex variables in order to eliminate explicit phases in the interaction terms and to cancel out two of the three energy terms
\bea
\psi_{\rm 1s} = \Psi_{\rm 1s} \rme^{\rmi E_1 t}, \quad\quad\quad \psi_{\rm 2p} = \Psi_{\rm 2p} \rme^{\rmi(E_1 + \omega_1)t}, \quad\quad\quad \psi_{\rm 3s} = \Psi_{\rm 3s} \rme^{\rmi E_3 t}
\,,
\ea
yields a shorter Lagrangian
\bea
\label{eq:redefined_lagrangian}
L &=& \Delta \psi_{\rm 2p}^* \psi_{\rm 2p} -\rmi \left( \psi_{\rm 1s}^* \dot{\psi}_{\rm 1s} + \psi_{\rm 2p}^* \dot{\psi}_{\rm 2p} + \psi_{\rm 3s}^* \dot{\psi}_{\rm 3s} \right)+ \nonumber\\
&& + \left( g^{12} A_1^* \psi_{\rm 1s}^* \psi_{\rm 2p} + g^{23} A_2^* \psi_{\rm 2p}^* \psi_{\rm 3s} + \rm{H.c.} \right)
\,,
\ea
which involves the detuning $\Delta = E_{12} - \omega_1$. From (\ref{eq:redefined_lagrangian}) the Euler-Lagrange equations for the variables $\psi_{\rm 2p}^*$, $\psi_{\rm 1s}^*$ and $\psi_{\rm 3p}^*$ read
\bea
\label{eq:euler_lagrange_eqs}
\Delta \psi_{\rm 2p} -\rmi \dot{\psi}_{\rm 2p} + {g^{12}}^* A_1 \psi_{\rm 1s} + g^{23} A_2^* \psi_{\rm 3s} &=& 0,\\
-\rmi \dot{\psi}_{\rm 1s} + g^{12} A_1^* \psi_{\rm 2p} &=& 0,\\
-\rmi \dot{\psi}_{\rm 3s} + {g^{23}}^* A_2 \psi_{\rm 2p} &=& 0
\,.
\ea
(\ref{eq:euler_lagrange_eqs}) can be solved approximately for $\psi_{\rm 2p}$ in the following way
\bea
\label{eq:psi_2p}
\psi_{\rm 2p} &=& - \left( \Delta -\rmi \partial_t \right)^{-1} \left( {g^{12}}^* A_1 \psi_{\rm 1s} + g^{23} A_2^* \psi_{\rm 3s}\right)\nonumber\\
&=& -\frac{1}{\Delta} \left( {g^{12}}^* A_1 \psi_{\rm 1s} + g^{23} A_2^* \psi_{\rm 3s}\right)-\nonumber\\
&& - \frac{1}{\Delta^2} \left( | {g^{12}}^* A_1 |^2 + | g^{23} A_2^* |^2 \right) \psi_{\rm 2p} + \Or \left( | g A |^3 / \Delta^3 \right)
\,,
\ea
where the Euler-Lagrange equations for $\psi_{\rm 1s}^*$ and $\psi_{\rm 3p}^*$ have been inserted. We assume that $| g A | \ll \Delta$ and thus
\bea
\label{eq:psi_2p_approx}
\psi_{\rm 2p} &\approx& -\frac{1}{\Delta} \left( {g^{12}}^* A_1 \psi_{\rm 1s} + g^{23} A_2^* \psi_{\rm 3s}\right)
\,.
\ea
Reinserting (\ref{eq:psi_2p_approx}) into the Lagrangian (\ref{eq:redefined_lagrangian}), where the $\dot{\psi}_{\rm 2p}$-term was neglected as well for the same reason, yields an effective Lagrangian in which the 2p-level was integrated out
\bea
\label{eq:effective_lagrangian}
L_{\rm eff} &=& -\rmi \left( \psi_{\rm 1s}^* \dot{\psi}_{\rm 1s} + \psi_{\rm 3s}^* \dot{\psi}_{\rm 3s} \right) - \frac{\big| {g^{12}}^* A_1 \psi_{\rm 1s} + g^{23} A_2^* \psi_{\rm 3s} \big|^2}{\Delta}\nonumber\\
&=& -\rmi \left( \psi_{\rm 1s}^* \dot{\psi}_{\rm 1s} + \psi_{\rm 3s}^* \dot{\psi}_{\rm 3s} \right) - \frac{\big| {g^{12}}^* A_1 \big|^2}{\Delta}  \psi_{\rm 1s}^* \psi_{\rm 1s} - \frac{\big| g^{23} A_2^* \big|^2}{\Delta} \psi_{\rm 3s}^* \psi_{\rm 3s}+\nonumber\\
&& + \left( - \frac{{g^{12}}^* {g^{23}}^*}{\Delta} A_1 A_2 \psi_{\rm 3s}^* \psi_{\rm 1s} + {\rm H.c.} \right)
\,.
\ea
We find the two-photon absorption and emission processes in the last line. The coupling constant of these processes is thus given by
\bea
\label{eq:effective_coupling}
g_{\rm eff} = - \frac{{g^{12}}^* {g^{23}}^*}{\Delta}
\,.
\ea
\section{Repeated inducing}
\label{sec:repeated_inducing}
In section \ref{sec:two_photon_entangling_zeno_gate}, we presented our two-photon entangling scheme which requires the two-photon absorption rate of the absorbers to be much larger than the one-photon loss rate. We subsequently investigated the situation in case of a simple three-level atom (section \ref{sec:single_three_level_atom}), and obtained the result that the two-photon absorption probability is at best still about $20$-times smaller than the one-photon scattering probability. Therefore, additional enhancement of two-photon absorption is necessary for our scheme to work. Thus in this section, a mechanism is presented which could enhance the two-photon absorption probability compared to the one-photon scattering probability.
\begin{figure}[h]
\begin{center}
\includegraphics[width=0.3\columnwidth]{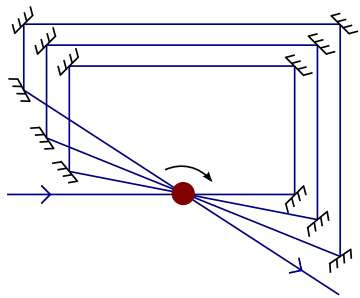}
\caption{Internal structure of the absorbers where two photons are sent through the same absorbing medium repeatedly (in this case, $n = 4$ times).
The photons always travel the same length $L$ between two successive passages of the absorbing medium, which was disregarded in this picture.
In order to ensure that the incident angle of the photons remains the same, the medium could be rotated.} 
\label{fig:repeated}
\end{center}
\end{figure}
\noindent
The mechanism is again based on the quantum Zeno effect. When sending the two photons $n$ times through the {\em same} absorbing medium (see Figure~\ref{fig:repeated}) with the correct optical path length $L$, the two-photon absorption amplitudes add up coherently, while for scattering processes, only the probabilities add up. To see this, let us first introduce an effective Hamiltonian
\bea
\label{eq:effective_hamiltonian_0}
\hat{V}_{\rm eff}^{(0)}&=& g_{13} \hat{\sigma}^{13} \hat{a}_1^\dagger \hat{a}_2^\dagger + g_{12} \hat{\sigma}^{12} \hat{a}_{1}^\dagger + g_{23} \hat{\sigma}^{23} \hat{a}_{2}^\dagger + g_{11} \sket{\psi_{\rm scatter}^{(0)}} \bra{\psi_{\rm in}} + \rm{H.c.} 
\,.
\ea
In this effective Hamiltonian, for the sake of completeness, we also included one-photon absorption and emission (the case of zero detuning). The index $ ^{(0)}$ denotes that this is the Hamiltonian for the first passage of the absorbing medium. The incoming state is (\ref{eq:input_state}) as before. In the outgoing state, there are small amplitudes for two-photon absorption, one-photon absorption and one-photon scattering according to (\ref{eq:effective_hamiltonian_0}). We assume small interaction time $\tau \ll 1$, i.e. we regard only up to first order in $\tau$
\bea
\label{eq:first_passage}
&&\sket{\Psi^{(1)}} = \left( 1 -\rmi \tau \hat{V}_{\rm eff}^{(0)}  + \Or \left( \tau \right)^2 \right) \ket{\psi_{\rm in}}\nonumber\\
&&= \ket{\psi_{\rm in}} -\rmi \tau \left[ g_{13}^* \ket{0} \ket{3s} + g_{12}^* \hat{a}_{2}^\dagger \ket{0} \ket{2p} + g_{11} \sket{\psi_{\rm scatter}^{(0)}}\right] + \Or \left( \tau \right)^2
\,.
\ea
Afterwards, the two photons are guided into the material again, crossing a certain optical path length $L$, which should be much larger than the size of the photon wave-packets. The Hamiltonian for the second passage, $\hat{V}_{\rm eff}^{(1)}$, therefore carries different phase factors depending on which photons are involved in the respective processes. That is because the Hamiltonian for the second passage couples to the vector potential at some position $\f{A} ( \f{r}_0 + \f{r}_L , t )$ instead of $\f{A} ( \f{r}_0 , t)$, and the spatial phase factors $\exp \{\rmi \f{k}\cdot\f{r}_0\}$ change accordingly.
In contrast to the two-photon absorption, the scattering process behaves incoherently, i.e., like repeated measurements, when $L$ is larger than the size of the photon wave-packets. There only the probabilities add up and the scattering amplitudes of the different passages are orthogonal. Hence they are labeled with indices $\sbraket{\psi_{\rm scatter}^{(i)}}{\psi_{\rm scatter}^{(j)}} = \delta_{i j}$
\bea
\label{eq:effective_hamiltonian_1}
\hat{V}_{\rm eff}^{(1)}&=& g_{13} \rme^{-\rmi(k_1 + k_2)L} \hat{\sigma}^{13} \hat{a}_1^\dagger \hat{a}_2^\dagger+\nonumber\\
&&+ g_{12} \rme^{-\rmi k_1 L} \hat{\sigma}^{12} \hat{a}_{1}^\dagger + g_{23} \rme^{-\rmi k_2 L} \hat{\sigma}^{23} \hat{a}_{2}^\dagger + g_{11} \sket{\psi_{\rm scatter}^{(1)}} \bra{\psi_{\rm in}} + \rm{H.c.}
\,.
\ea
So in the second passage, the state gathers additional amplitudes with their respective phase-factors
\bea
\label{eq:second_passage}
\sket{\Psi^{(2)}} &=& \left( 1 -\rmi \tau \hat{V}_{\rm eff}^{(1)}  + \Or \left( \tau \right)^2 \right) \sket{\Psi^{(1)}}\nonumber\\
&=& \ket{\psi_{\rm in}} -\rmi \tau \left[ g_{13}^* \left( 1 + \rme^{\rmi(k_1 + k_2)L} \right) \ket{0} \ket{3s} + g_{12}^* \left( 1 + \rme^{\rmi k_1 L} \right) \hat{a}_{2}^\dagger \ket{0} \ket{2p}+\right.\nonumber\\
&&\left.+ g_{11} \sket{\psi_{\rm scatter}^{(0)}} + g_{11} \sket{\psi_{\rm scatter}^{(1)}}\right] + \Or \left( \tau \right)^2
\,.
\ea
When the two photons are repeatedly guided into the absorbing material for $n$ times, each time passing the same optical path length $L$, the state evolves as
\bea
\label{eq:nth_passage}
\sket{\Psi^{(n)}} &=& \prod_{\mu=0}^{n-1} \left( 1 -\rmi \tau \hat{V}_{\rm eff}^{(\mu)} + \Or \left( \tau \right)^2 \right) \ket{\psi_{\rm in}}\nonumber\\
&=& \ket{\psi_{\rm in}} -\rmi \tau \sum_{\mu=0}^{n-1} \left[ g_{13}^* \rme^{\rmi(k_1 + k_2)\mu L} \ket{0} \ket{3s}+\right.\nonumber\\
&&+\left. g_{12}^* \rme^{\rmi k_1 \mu L} \hat{a}_{2}^\dagger \ket{0} \ket{2p} + g_{11} \sket{\psi_{\rm scatter}^{(\mu)}} \right] + \Or \left( \tau \right)^2
\,.
\ea
Moreover, when calculating the overall two-photon absorption probability, we readily see that it is enhanced by a factor $n^2$, assumed we choose the optical path length such that is related to the sum of the two photon wave-numbers via $(k_1+k_2)L\in 2\pi\mathbb N$
\bea
P_{2\gamma} = \left|\sbraket{3s}{\Psi^{(n)}}\right|^2 &=& \tau^2 \left| g_{13}^* \right|^2 \left| \sum_{\mu=0}^{n-1} \rme^{\rmi(k_1 + k_2)\mu L} \right|^2 = \tau^2 \left| g_{13}^* \right|^2 n^2
\,.
\ea
A possible local one-photon absorption effect, in contrast, would violate the phase matching requirements when choosing $k_1 L\notin 2\pi\mathbb N$. Therefore the one-photon absorption probability does not even scale with $n$. It is heavily suppressed
\bea
P_{1\gamma}^{\rm abs} = \left|\sbraket{2p}{\Psi^{(n)}}\right|^2 &=& \tau^2 \left| g_{12}^* \right|^2 \left| \sum_{\mu=0}^{n-1} \rme^{\rmi k_1 \mu L} \right|^2\nonumber\\
&=& \tau^2 \left| g_{12}^* \right|^2 \frac{\cos n k_1 L - 1}{\cos k_1 L - 1}\nonumber\\
&=& \tau^2 \left| g_{12}^* \right|^2 \Or \left( 1 \right)
\,.
\ea
Most important, for one-photon scattering only the probabilities add up
\bea
P_{1\gamma}^{\rm scatter} &=& \sum_{\mu=0}^{n-1} \left|\sbraket{\psi_{\rm scatter}^{(\mu)}}{\Psi^{(n)}}\right|^2  =
\tau^2 \left| g_{11} \right|^2 n 
\,.
\ea
In this way, we can enhance the total two-photon absorption probability
by a factor of $n^2$ as compared to the one-photon scattering losses, which scale with $n$.
Concluding this section, we see that the requirement (\ref{eq:constraint}) could be achieved 
by sufficiently large $n$
\bea
\label{eq:ratio-n}
\kappa
=
\frac{\xi_{2\gamma}}{\xi_{1\gamma}}
=
\frac{n}{\Or (\pi\omega^2_{1,2}A )}
\,.
\ea
\section{Coherent excitation of many atoms}
\label{sec:coherent_interaction_with_Sgg1_atoms}
It was already pointed out that additional efforts are needed to further enhance two-photon absorption as compared to one-photon scattering. One mechanism to achieve such an enhancement was already given in section \ref{sec:repeated_inducing}. In this section, we present an alternative approach based on coherent excitation of a large number $S$ of atoms/molecules. We will show that the two-photon absorption probability scales with the number of excitations, whereas the one-photon processes do not. In addition we discuss how an excited state of many atoms/molecules could be sustained by pump lasers.
Coherently excited states of many atoms are often referred to as ``Dicke-states'' \cite{Dicke:1954fk}, which are typically investigated regarding collective spontaneous emission also known as ``Dicke super-radiance'' \cite{Rehler:1971uq,Scully:2006fk,Eberly:2006ly,Scully:2007fk,Svidzinsky:2008qf,Porras:2008kx,Scully:2009fk,Sete:2010fu,Wiegner:2011zr}.
As we now deal with many three-level atoms instead of one, our Hamiltonian is now the sum of the individual single-atom Hamiltonians in (\ref{eq:ham_inter_mod}), i.e.
\bea
\label{interaction_hamiltonian_many}
\hat{H}_{\rm Cluster}( \{ \f{r}_\ell \}, t )
=
\sum_{\ell=1}^S \hat{H}_{1, {\rm I}}( \f{r}_\ell, t )
\,.
\ea
The single atomic transition operators inside the individual single-atom Hamiltonians $\hat{H}_{1, {\rm I}}( \f{r}_\ell, t )$ are now denoted with an index $\ell$, e.g. $\hat{\sigma}^{13}_\ell$, which means that they act on the $\ell$-th atom. The initial state changes in so far, that we start from an $S$-atom state $\ket{s}$ instead of the ground state of a single atom $\ket{\rm 1s}$,
\bea
\label{eq:input_state_s}
\ket{\psi_{\rm in}^s} = \hat{a}_1^\dagger \hat{a}_2^\dagger \ket{0} \ket{s}
\,.
\ea
The $s$ in $\ket{s}$ means that out of $S$ atoms, $s$ atoms should be coherently excited to the 3s-level in order to enhance two-photon absorption, while $S-s$ atoms are still in the ground state $\ket{\rm 1s}$. None of the atoms in $\ket{s}$ is in the state $\ket{\rm 2p}$, which corresponds to the middle (2p) level.
\subsection{Effective Hamiltonian and quasispin operators}
\label{ssec:effective_hamiltonian_and_quasispin_operators}
To get the idea behind coherent excitation, it is sufficient to look at the following effective (interaction) Hamiltonian, considering $S$ atoms/molecules with their respective positions $\f{r}_\ell$
\bea
\label{eq:Dicke}
\hat H_{\rm Dicke}
=
g\hat a_1\hat a_2\sum_{\ell=1}^S
\hat{\sigma}^{13\dagger}_\ell \exp\{\rmi\f{r}_\ell\cdot(\f{k_1}+\f{k_2})\}
+{\rm H.c.}
\,.
\ea
It can be understood as reduced version of (\ref{interaction_hamiltonian_many}), with the middle (2p) level integrated out (resulting in an effective coupling constant $g$, see \ref{ap:ssec:integrating_out_the_(2p)_level}) and all effects disregarded except (coherent) two-photon absorption and emission. As we only have the 1s- and the 3s-level left in this effective treatment, the Hamiltonian (\ref{eq:Dicke}) describes $S$ \emph{two-level} systems. These $S$ two-level systems can also be regarded as $S$ spin-$\frac{1}{2}$ systems, identifying the $\ket{\rm 1s}$-state with spin down, $s_z = -\frac{1}{2}$, and the $\ket{\rm 3s}$ state with spin up, $s_z = +\frac{1}{2}$. Accordingly, the atomic transition operators correspond to ladder operators from Pauli matrices $\hat{\sigma}_\ell^\pm = ( \hat{\sigma}_\ell^x \pm \rmi \hat{\sigma}_\ell^y )/2$, such that $\hat{\sigma}_\ell^+ = \hat{\sigma}^{13\dagger}_\ell$ and $\hat{\sigma}_\ell^- = \hat{\sigma}^{13}_\ell$. 
With this in mind, we can define quasispin-$S$ operators which sum over the $S$ individual spin-$\frac{1}{2}$ systems as
\bea
\label{QuasispinXY}
\hat{\Sigma}^x\pm \rmi\hat{\Sigma}^y
=
\hat{\Sigma}^\pm
:=
\sum_{\ell=1}^S
\hat{\sigma}_\ell^\pm\exp\{\pm \rmi\f{r}_\ell\cdot(\f{k_1}+\f{k_2})\}
\,.
\ea
Moreover, for the $S$ spin-$\frac{1}{2}$ systems, we also have the $\hat{\sigma}_\ell^z$-operator from the Pauli-$z$-matrix, with eigenvalues $\pm 1$, such that $\hat{\sigma}_\ell^z \ket{\rm 1s}_\ell = -\ket{\rm 1s}_\ell$ and $\hat{\sigma}_\ell^z \ket{\rm 3s}_\ell = +\ket{\rm 3s}_\ell$. We define the quasispin-$S$ operator 
\bea
\label{QuasispinZ}
\hat{\Sigma}^z
:=
\frac{1}{2} \sum_{\ell=1}^S
\hat{\sigma}_\ell^z
\,,
\ea
which essentially sums up the energy due to excitations, $\hat{H}_0 = E_{13} ( \hat{\Sigma}^z + S/2 )$.
The Hamiltonian (\ref{eq:Dicke}) can then be written shortly as
\bea
\label{Dicke_short}
\hat H_{\rm Dicke} = g\hat a_1\hat a_2\hat{\Sigma}^++{\rm H.c.}
\,.
\ea
The operators $\hat{\Sigma}^x$, $\hat{\Sigma}^y$ and $\hat{\Sigma}^z$ form an $SU(2)$ algebra $\left[ \hat{\Sigma}^\mu, \hat{\Sigma}^\nu \right] = \sum_\rho \rmi \varepsilon_{\mu \nu \rho} \hat{\Sigma}^\rho$ \cite{Lipkin:2002fk}. Therefore they have the same characteristics as the usual spin operators and especially they have the same normalized eigenstates.
As is well-known \cite{Sakurai:1994uq}, the following relations for the ladder operators arise solely from the angular momentum operator algebra $\left[ \hat{J}_\mu, \hat{J}_\nu \right] = \sum_\rho \rmi \varepsilon_{\mu \nu \rho} \hat{J}_\rho$. The $| j, m \rangle$ form a set of normalized eigenstates of $\hat{\vec{J}}^{\,2}$ and $\hat{J}_z$
\bea
&&\hat{J}_+ |j,m\rangle = \sqrt{( j - m )( j + m + 1)} |j,m+1\rangle\,,\nonumber\\
&&\hat{J}_- |j,m\rangle = \sqrt{( j + m )( j - m + 1)} |j,m-1\rangle
\,,
\ea
where the quantum number $m$ can take values from $-j$ to $j$ with unit steps. 

The same is true for our quasispin operators $\hat{\Sigma}^x$, $\hat{\Sigma}^y$ and $\hat{\Sigma}^z$. The variables however are defined slightly different. That is, the total number of the individual spin-$\frac{1}{2}$ systems $S$ corresponds to the maximum value of $m$, i.e. $S = 2j$, and the number of excitons $s$ is related to the quantum number $m$ by $s = m + j$, $m$ ranging in $-S/2$ to $+S/2$, $s$ takes values from $0$ to $S$. This results in the following normalization factors
\bea
\label{TransitionMatrixQuasispin}
\hat{\Sigma}^+\ket{s} &=& \sqrt{(S-s)(s+1)}\ket{s+1}\,,\\
\label{TransitionMatrixQuasispin2}
\hat{\Sigma}^-\ket{s} &=& \sqrt{(S-s+1)s}\ket{s-1}
\,,
\ea
where the state $\ket{s+1}$ denotes an entangled state of $s+1$ coherently excited atoms and $S-s-1$ atoms in the ground state. Analogously, $\ket{s-1}$ represents a state where $s-1$ atoms are coherently excited to the 3s-level and $S-s+1$ atoms are in the ground state.
\subsection{Enhanced two-photon absorption probability}
Regarding particularly (\ref{TransitionMatrixQuasispin}), one sees that the absorption probability scales with $Ss$ in the limit $s \ll S$
\bea
P_{2\gamma}^{\rm s} = (S-s)(s+1) P_{2\gamma} &\approx& S s P_{2\gamma}
\,.
\ea
This means that the two-photon absorption probability is enhanced by a factor $s$, as it would only scale with $S$ in the usual case without coherent excitation, i.e. $s=0$.
Please note that the corresponding directed two-photon emission (see \cite{Scully:2007fk,Scully:2006fk,Sete:2010fu}) is amplified too, (\ref{TransitionMatrixQuasispin2}), but this means no harm to our  desired two-photon absorption, as long as the atoms can be kept in the excited state, e.g. by constantly pumping (see below).

\begin{figure}[h]
\begin{center}
\psfrag{e1}{$E_1$}
\psfrag{e2}{$E_2$}
\psfrag{e3}{$E_3$}
\psfrag{omega_1}{$\omega_1$}
\psfrag{omega_2}{$\omega_2$}
\psfrag{omega}{$\omega_1$}
\psfrag{omega_ex}{$\omega_2$}
\psfrag{omega_s}{$\omega_k$}
\psfrag{and}{\&}
\subfigure[]{\includegraphics[height=0.20\columnwidth]{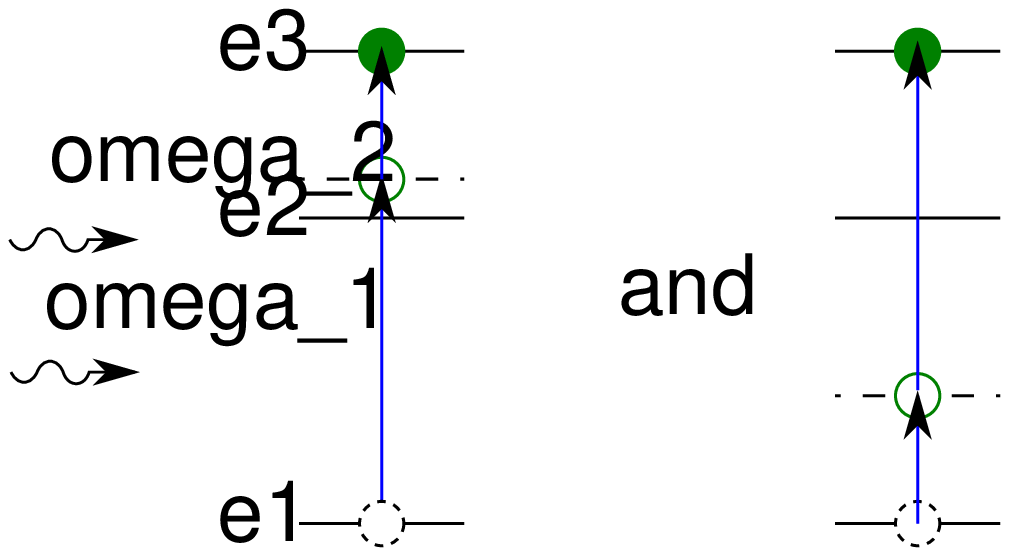}}
\hrule
\subfigure[]{\includegraphics[height=0.20\columnwidth]{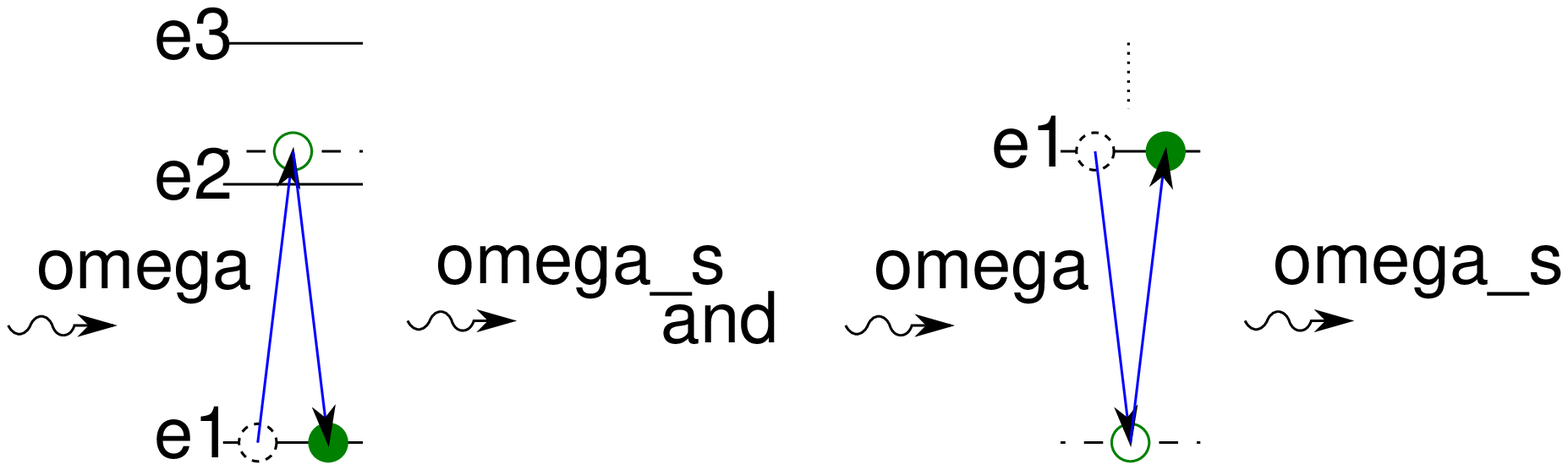}}
\hrule
\subfigure[]{\includegraphics[height=0.20\columnwidth]{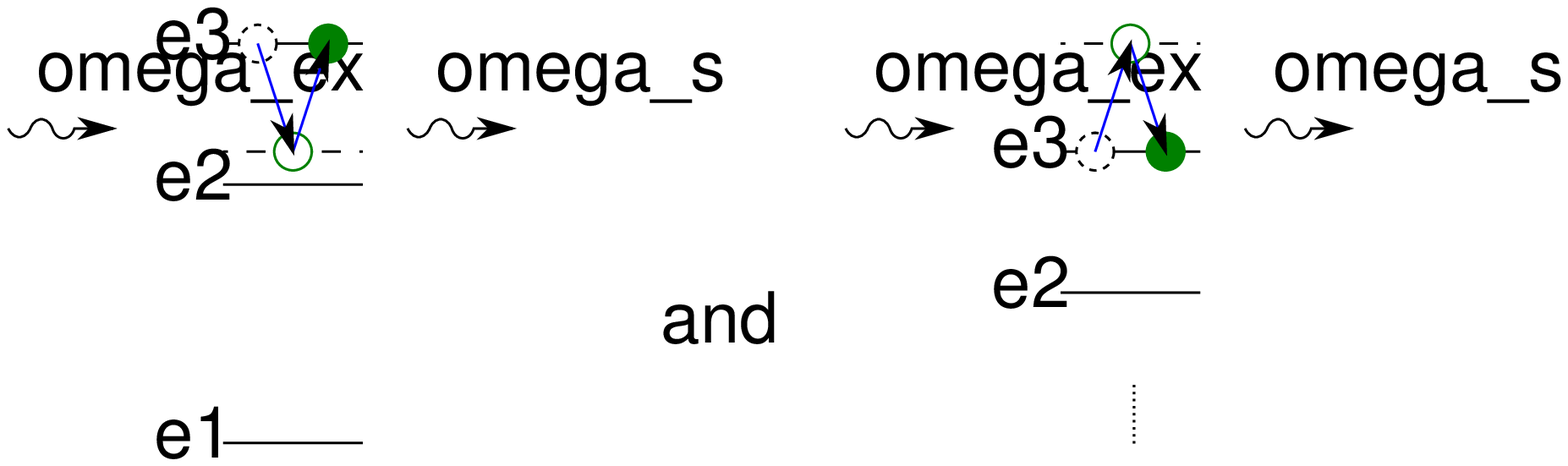}}
\hrule
\subfigure[]{\includegraphics[height=0.35\columnwidth]{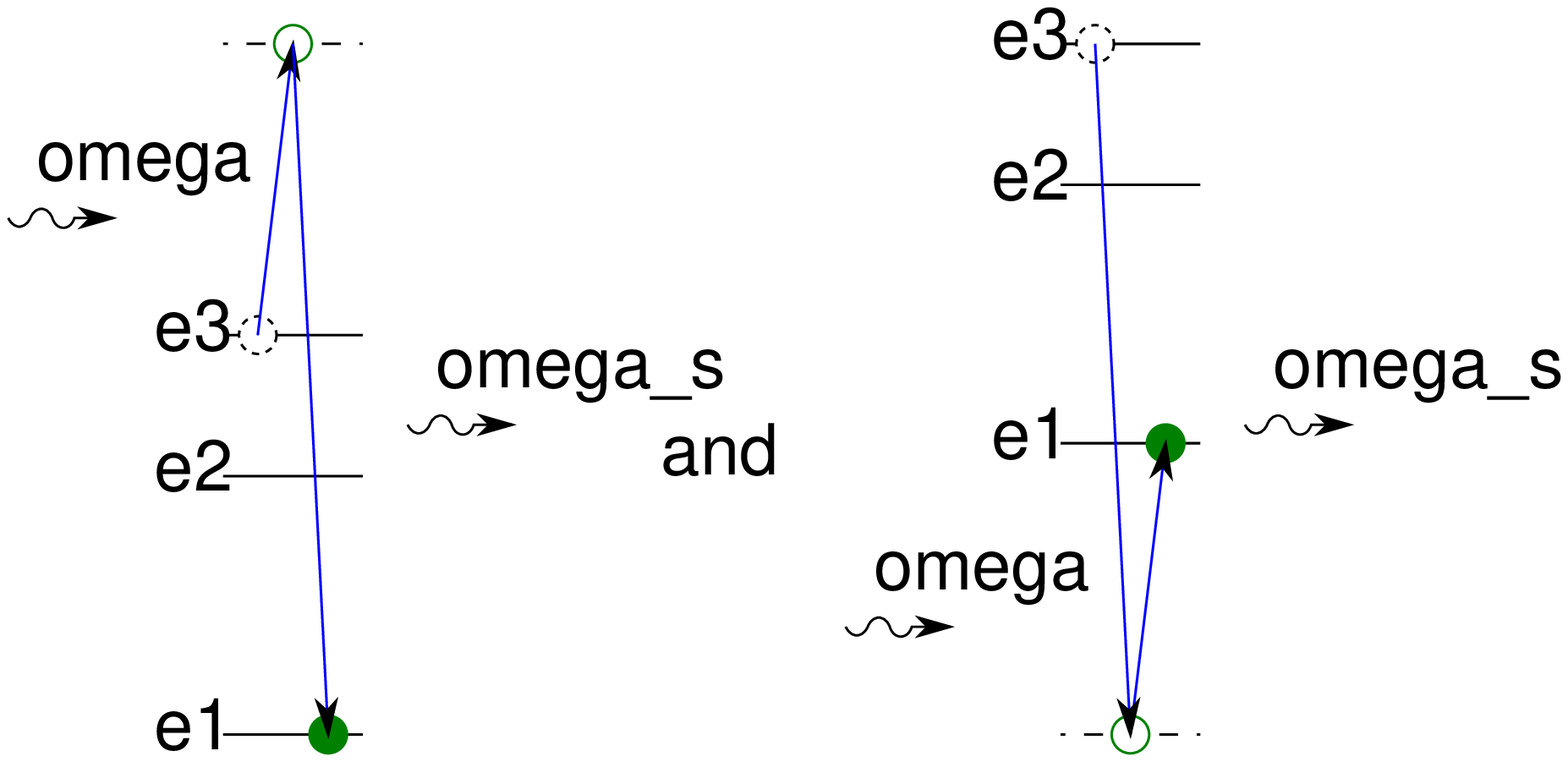}}
\caption{All important processes (coming from the $\hat{\f{p}}\cdot\hat{\f{A}\,}$-term in the Hamiltonian) in the situation of a cluster of coherently excited atoms $\ket{s}$ are shown. (a) Two-photon absorption, probability $\propto Ss$. (b) One-photon scattering on the ground state (exemplarily shown for $\omega_1$ getting scattered), probability $\propto \Or(S)$. (c) One-photon scattering on the excited state (for $\omega_2$), probability $\propto \Or(s)$. (d) One-photon upconversion (for $\omega_1$), probability $\propto Ss$, but very large detuning.}
\label{fig:all_processes}
\end{center}
\end{figure}

Aside from two-photon absorption and emission and one-photon emission processes (see above), we find one-photon scattering on the ground state as well as on the excited state, and a special upconversion process where one photon is absorbed, one atom relaxes into the ground state, and a higher energetic photon is emitted again. As for the single-atom case, full perturbation theory calculations have been performed for (\ref{interaction_hamiltonian_many}) in order to maintain the probability for any possible process. All important processes are shown in Figure~\ref{fig:all_processes}.
\subsection{One-photon scattering probability}
Let us now take a closer look at the one-photon scattering processes. The amplitude for the one-photon scattering on the ground state is proportional (taking only the atomic part, the rest is comparable to the scattering amplitude in section \ref{sec:single_three_level_atom}) to
\bea
\label{scattering_amplitude}
\ket{\psi_{1\gamma}^{\rm s}} \propto \sum_{\ell=1}^S \sigma^{12}_\ell \sigma^{12\dagger}_\ell \exp\{\rmi\f{r}_\ell\cdot(\f{\tilde{k}} - \f{k})\} \ket{s}
\,,
\ea
where $\f{\tilde{k}}$ is the $\f{k}$-vector of the scattered photon before the scattering, e.g. $\f{\tilde{k}} = \f{k}_1$, and $\f{k}$ is the wave vector after the scattering. For $\f{k} \neq \f{k}_1$, the phases inside the sum are randomly distributed because of the random atomic positions $\f{r}_\ell$. Therefore the phases do not add up coherently but rather statistically (like in a two-dimensional random walk)
\bea
\label{phases_squared}
\left| \sum_{\ell \in I} \exp\{\rmi\f{r}_\ell\cdot(\f{k}_1 - \f{k})\} \right|^2 = \Or \left( S \right)
\,.
\ea
$I$ denotes an arbitrary index set of $\left( S - s \right)$ atoms in the ground state. Thus for the scattering probability in case of $S$ atoms we simply arrive at
\bea
P_{1\gamma}^{\rm s} = \Or \left( S \right) \; P_{1\gamma}
\,,
\ea
where $P_{1\gamma}$ was the scattering probability for one atom. The same argument applies for scattering on the excited state, except that there are only $s$ atoms in $I$ and thus the scattering probability is only $\propto \Or \left( s \right) \; P_{1\gamma}$.

If, however, $\f{k} = \f{k}_1$, (\ref{phases_squared}) evaluates to $\left( S - s \right)^2$ and the scattering seems to be heavily enhanced. But in this case the photon is exactly in the same state as before, i.e. it {\em actually has not been scattered} but just {\em passed through} the atomic cluster. In other words, nothing happened.

To summarize, the one-photon scattering processes are at most enhanced by a factor $\Or \left( S \right)$ compared with the single-atom case (section \ref{sec:single_three_level_atom}). This result is also intuitively plausible, as the photon can now be scattered by $S$ different atoms instead of just one. Notably, the scattering probability does not scale with the number of excitations $s$, as the two-photon absorption, since it does not satisfy the phase matching condition.
\subsection{One-photon upconversion probability}
In this process, a single photon gets collectively (virtually) absorbed by the excited atoms, followed by the emission of a higher-energetic photon via relaxation into the next lower coherent state $\ket{s-1}$. The emission is also directed due to the collective character of the transition, i.e. the phase matching condition. For example the target photon $\f{k}_1$ is (virtually) absorbed, followed by the emission of a photon with wave vector $2\f{k}_1+\f{k}_2$. This can be seen by looking at the atomic part of the amplitude of this upconversion process
\bea
\label{conversion_amplitude}
\ket{\psi_{\rm conv}^{\rm s}} \propto \sum_{\ell=1}^S
\sigma_\ell^- \exp\{-\rmi\f{r}_\ell\cdot(\f{k} - \f{\tilde{k}})\} \ket{s}
\,.
\ea
Again, $\f{\tilde{k}}$ is the $\f{k}$-vector of the photon before the absorption. Sticking to our example, $\f{\tilde{k}} = \f{k}_1$, the scattering is directed along $\f{k} = 2 \f{k}_1 + \f{k}_2$ in order to fulfill the temporal phase matching condition $\f{k} - \f{\tilde{k}} = \f{k}_1 + \f{k}_2$. In that case, the atomic part of $\ket{\psi_{\rm conv}^{\rm s}}$ is given by $\ket{\psi_{\rm conv}^{\rm s}}
 \propto \Sigma^-\ket{s} = \sqrt{(S-s+1)s}\ket{s-1}$, and therefore the probability for this process also scales with $Ss$ in the limit $s \ll S$.
Fortunately, this process is not supported by the level scheme of the atom. In other words, the detuning for this process is comparable to optical energies. From Figure~\ref{fig:all_processes}(d) it can be seen that for our example, the detuning is $E_{23} + \omega_1$ respectively $- E_{12} - \omega_1$ (in the case where the control photon $\omega_2$ gets absorbed, it would just be $E_{23} + \omega_2$ and $- E_{12} - \omega_2$).
Therefore, this process does not scale with $1/\Delta^2$ as in (\ref{eq:2gamma}) and thus can be neglected for small detunings.
\subsection{Comparison and conclusion for many atoms}
\label{ssec:comparison_and_conclusion_for_many_atoms}
Summarizing, the two-photon absorption probability is enhanced by a factor $s$ compared to the one-photon scattering probability
\bea
\frac{P_{2\gamma}^{\rm s}}{P_{1\gamma}^{\rm s}} \approx s \; \frac{P_{2\gamma}}{P_{1\gamma}}
\,.
\ea
Concluding this section, we see that the requirement (\ref{eq:constraint}) could be achieved 
by sufficiently large $s$
\bea
\label{eq:ratio-n}
\kappa
=
\frac{\xi_{2\gamma}}{\xi_{1\gamma}}
=
\frac{s}{\Or (\pi\omega^2_{1,2}A )}
\,.
\ea
\subsection{Sustaining the excited state}
\label{ssec:sustaining_the_excited_state}
As discussed above, the state of coherent excitation of s atoms, $\ket{s}$, decays very rapidly by directed, spontaneous two-photon emission. In order to sustain the state $\ket{s}$, we propose to apply two pump lasers with wave numbers $\f{k_1}',\f{k_2}'$ which satisfy the same spatial and temporal phase matching conditions $\f{k_1}'+\f{k_2}'=\f{k_1}+\f{k_2}$ and $\omega_1'+\omega_2'=\omega_1+\omega_2$ as the control and target photon.
The quotient of the atoms which can be kept excited, $s/S$, should be primarily depending on the strength of the pump laser fields, which are naturally limited. To obtain a relation between the quotient $s/S$ and the intensities of the pump lasers, $I_1$ and $I_2$, we can start from a Hamiltonian similar to (\ref{Dicke_short}), but regarding the pump laser fields classically. Moreover we write the full Hamiltonian, including the energy due to excitations, and not only the interaction part
\bea
\label{eq:LaserHamiltonian}
\hat{H}_{\rm Laser}
=
E_{13} \left( \hat{\Sigma}^z + S/2 \right) + \left( g \hat{\Sigma}^+ + {\rm H.c.} \right)
\,.
\ea
In this case, the coupling constant $g$ reads (see \ref{eq:effective_coupling})
\bea
\label{eq:laser_coupling}
|g|
=
\frac{|\f{g}^{12}| |\f{g}^{23}|}{\Delta'} A_1 A_2
=
\frac{4 \pi \alpha_{\rm QED} E_{12} E_{23} \ell_{\rm atom}^2}{\Delta'} A_1 A_2
\,,
\ea
where $\Delta' = E_{12} - \omega_1'$ denotes the detuning of the pump laser field.
We can then treat the quasispin-$S$ system as a harmonic oscillator by applying a Holstein-Primakoff \cite{Holstein:1940uq} transformation,
\bea
\label{HolsteinPrimakoff}
\hat{\Sigma}^+ 
= 
\hat{a}^\dagger \sqrt{S - \hat{a}^\dagger \hat{a}^{\phantom{\dagger}}} 
= 
\left( \hat{\Sigma}^- \right)^\dagger,\quad
\hat{\Sigma}^z 
= 
\hat{a}^\dagger \hat{a}^{\phantom{\dagger}} - S/2
\,,
\ea
on (\ref{eq:LaserHamiltonian}). This yields
\bea
\label{LaserHamiltonianHP}
\hat{H}_{\rm Laser}
=
E_{13} \hat{a}^\dagger \hat{a}^{\phantom{\dagger}} + \left( g \hat{a}^\dagger \sqrt{S - \hat{a}^\dagger \hat{a}^{\phantom{\dagger}}} + {\rm H.c.} \right)
\,.
\ea
In our envisaged limit $S \gg s$ the Hamiltonian (\ref{LaserHamiltonianHP}) looks just like a harmonic oscillator coupled to the classical laser fields
\bea
\label{LaserHamiltonian2}
\hat{H}_{\rm Laser}
=
E_{13} \hat{a}^\dagger \hat{a}^{\phantom{\dagger}} + \sqrt{S} \left( g \hat{a}^\dagger + {\rm H.c.} \right)
\,.
\ea
Given a certain coupling constant $g$, which particularly depends on the laser-field intensities and the detuning, there is an associated coherent, steady state $\hat{a}^{\phantom{\dagger}} \ket{\alpha_{\rm g}} = \alpha_{\rm g} \ket{\alpha_{\rm g}}$, which can be found by solving the characteristic equation $\hat{H}_{\rm Laser} \ket{\alpha_{\rm g}} = E_\alpha \ket{\alpha_{\rm g}}$
\bea
\hat{H}_{\rm Laser} \ket{\alpha_{\rm g}} &=& \left[ \left( E_{13} \alpha_{\rm g} + \sqrt{S} g \right) \hat{a}^\dagger + \sqrt{S}  g^* \alpha_{\rm g} \right] \ket{\alpha_{\rm g}}
= E_\alpha \ket{\alpha_{\rm g}}
\,.
\ea
Thus we found the eigenstate of the $S$-atom system, it is the coherent state with
\bea
\label{CoherentState}
\alpha_{\rm g}
=
-g \sqrt{S} / E_{13}
\,,
\ea
and 
\bea
E_\alpha = - E_{13} | \alpha_{\rm g} |^2
\,.
\ea
The average exciton number in a coherent state is given by $\langle \hat{n} \rangle =\langle \hat{a}^\dagger \hat{a} \rangle = | \alpha_{\rm g} |^2$. Thus, by setting $| \alpha_{\rm g} | = \sqrt{s}$, we get an expression for the quotient $s/S$
\bea
\label{CoherentState}
\frac{s}{S}
=
\left( \frac{|g|}{E_{13}} \right)^2 = \left( \frac{4 \pi \alpha_{\rm QED} E_{12} E_{23} \ell_{\rm atom}^2}{E_{13} \Delta'} \right)^2 A_1^2 A_2^2
\,.
\ea
The vector potential is related to the intensity by $I = A^2 \omega^2$ (remember $c = \epsilon_0 = 1$) and we may express $E_{13}$ by the two frequencies $E_{13} = \omega_1' + \omega_2'$. Thus the intensities of the two pump lasers, $I_1,I_2$, are related to the ratio $s/S$ via
\bea
\label{eq:intensity}
\frac{s}{S} =  
\left(
\frac{4\pi\alpha_{\rm QED}E_{12}E_{23}\ell_{\rm atom}^2}
{\omega_1'\omega_2'\left(\omega_1'+\omega_2'\right)\Delta'}
\right)^2
I_1I_2
\,.
\ea
The detuning $\Delta'$ of the pump laser field cannot be chosen too small, because this would lead to non-negligible population of the middle (2p) level which we integrated out in our Hamiltonian (\ref{eq:LaserHamiltonian}).
The population of the intermediate level is governed by 
\bea
\label{eq:psi_2}
\psi_{\rm 2p} = -\frac{1}{\Delta'} \left( {g^{12}}^* A_1 \psi_{\rm 1s} + g^{23} A_2^* \psi_{\rm 3s}\right)
\,.
\ea
In other words, the detuning should be big enough such that ${g^{12}}^* A_1 / \Delta' \ll 1$ as well as $g ^{23} A_2^* / \Delta' \ll 1$, i.e. nearly no population of the middle (2p) level. In our Hamiltonian notation this corresponds to $\left| \f{g}^{12} \right| A_1 \ll \Delta'$ and $\left| \f{g}^{23} \right| A_2 \ll \Delta'$. Inserting $\left| \f{g}^{12} \right|$, $\left| \f{g}^{23} \right|$ and again $I = A^2 \omega^2$ gives
\bea
&&\sqrt{4 \pi \alpha_{\rm QED} I_1} \left| \f{\ell}_{\rm atom}^{12} \right| E_{12} / \omega_1 \ll \Delta'\,,\nonumber\\
&&\sqrt{4 \pi \alpha_{\rm QED} I_2} \left| \f{\ell}_{\rm atom}^{23} \right| E_{23} / \omega_2 \ll \Delta'
\,.
\ea
Assuming that $E_{12} / \omega_1 \approx E_{23} / \omega_2 \approx 1$ and that the laser intensities for both pump lasers are the same, $I_1 = I_2 = I$, we find that the detuning $\Delta'$ of the pump laser field should satisfy the condition
\bea
\label{eq:det_cond}
\Delta'\gg\sqrt{4\pi\alpha_{\rm QED}I}\ell_{\rm atom} 
\,,
\ea
in order to avoid unwanted excitations of the middle (2p) level. With a typical dipole length of six Bohr radii $\ell_{\rm atom}=6a_{\rm B}$ and for a large but possibly realistic intensity of 
$I = 10^{10}\,{\rm W}/{\rm cm}^2$, this translates into $\Delta'>10^{14}\,\rm Hz$ or $\Delta'>0.06\,\rm eV$. 
\section{Example values}
\label{sec:example_values}
Let us insert some typical parameters.
We assume that target and control photon are in the optical regime
(say around 500~nm) and adjust the detuning of the target photon to be 
$\Delta=3\cdot10^{12}\,\rm Hz$, which is several orders of magnitude 
larger than the typical natural line width of the middle (2p) level 
(in the range of GHz). 
Choosing the detuning of the control photon to be an order of magnitude 
larger, i.e., $3\cdot10^{13}\,\rm Hz$, the loss rate of the control 
photon can be neglected.
We consider three different values of the error probability $P_{\rm error}$, namely $P_{\rm error} = 0.5$, $P_{\rm error} = 0.25$ and $P_{\rm error} = 0.1$. At first, we determine the required ratio of the two-photon absorption and the one-photon loss rate $\kappa = \xi_{2\gamma} / \xi_{1\gamma}$ and the needed number of segments of the gate. Recall that in section \ref{ssec:three_branch_gate}, we already derived the relationship
\bea
\kappa
=
\frac{\xi_{2\gamma}}{\xi_{1\gamma}}
=
\frac{\pi^2}{2P_{\rm error}^2}
\gg1
\,,
\ea
which, however, is only valid for $N \gg 1$ and $\xi_{2\gamma} \gg \xi_{1\gamma}$. For experimental realization, small values for $N$ and $\kappa$ are desirable. Thus we calculate some possible values for $N$ and $\kappa$ numerically. As there is some margin for trade-off between $N$ and $\kappa$, we present three different possible choices of $N$ and $\kappa$, a choice with small $N$, a balanced one and a choice with small $\kappa$.
\begin{table}[h]
\centering
\begin{tabular}{|c|c|c||c|c|c||c|c|c|c|}
\hline
$\,P_{\rm error}\,$ & $\;N\;$ & 
$\,\kappa\,$ & $\,P_{\rm error}\,$ & $\;N\;$ & 
$\,\kappa\,$ & $\,P_{\rm error}\,$ & $\;N\;$ & 
$\,\kappa\,$\\ 
\hline
 50\% & 8 & 22 & 25\% & 20 & 120 & 10\% & 50 & 1~430\\
  & 10 & 12 & & 25 & 76 & & 60 & 760\\
  & 40 & 8 & & 70 & 55 & & 160 & 440\\
\hline
\end{tabular}
\caption{Example values of the number $N$ of segments and the corresponding ratio $\kappa$. For each value of the error probability $P_{\rm error}$, three possible choices of $N$ and $\kappa$ are given. For instance, to reach an error threshold of $P_{\rm error}=50\%$, it would be possible to build a gate with $N = 10$ and $\kappa = 12$. However, when $\kappa = 12$ poses too hard a challenge, it is e.g. also possible to go with $N = 40$  and $\kappa = 8$.}
\label{tab:ex_values_nkappa}
\end{table}
\noindent
For each possible realization of the set-up in Figure~\ref{fig:3branch}, we also specify a two-photon absorption probability $P_{2\gamma}^{{\rm segm}}$ and a one-photon loss probability $P_{1\gamma}^{{\rm segm}}$ \emph{per segment}, which is related to the ratio $\kappa$ and the number $N$ of segments by (see section \ref{ssec:three_branch_gate})
\bea
\label{eq:probs_segment}
P_{2\gamma}^{{\rm segm}} &=& 1 - (\rme^{-\xi_{2\gamma}})^2 = 1 - \exp \left( -2 \sqrt{\kappa} \frac{\sqrt{2} \pi}{N} \right)\,,\nonumber\\
P_{1\gamma}^{{\rm segm}} &=& 1 - (\rme^{-\xi_{1\gamma}})^2 = 1 - \exp \left( -2 \frac{1}{\sqrt{\kappa}} \frac{\sqrt{2} \pi}{N} \right)
\,.
\ea
\noindent
Furthermore, we state what is the number $n$ of repetitions necessary to reach the respective ratio $\kappa$ via the mechanism described in section \ref{sec:repeated_inducing} (sketched in Figure~\ref{fig:repeated}). The required amplification $n$ is simply given by (see above, section \ref{ssec:comparison_and_conclusion})
\bea
\label{repetitions_n}
n \frac{P_{2\gamma}}{P_{1\gamma}} \stackrel{!}{=} \kappa
\,.
\ea
This expression was evaluated using (\ref{eq:2gamma}) and (\ref{eq:scat_prob}), inserting the values given in this section and assuming a focus at the diffraction limit, $A = \left(\lambda/2\right)^2$.
As the required amplification of the two-photon absorption probability can also be achieved via coherent excitation of a large number $s$ of atoms/molecules (see section \ref{sec:coherent_interaction_with_Sgg1_atoms}), the obtained value in (\ref{repetitions_n}) can also correspond to $s$ or even to $ns$, when both enhancement mechanisms are combined.
The discussed values are given below in Table~\ref{tab:example_values_smallN} for the choice of small $N$, in Table~\ref{tab:example_values_sensible} for the balanced choice and in Table~\ref{tab:example_values_smallKappa} for the choice of small $\kappa$.
\begin{table}[h]
\centering
\begin{tabular}{|c|c|c|c|c|c|c|}
\hline
$\,P_{\rm error}\,$ & $\;N\;$ & 
$\,P_{2\gamma}^{\rm segm}\,$ & $\,P_{1\gamma}^{\rm segm}\,$ & 
$\,\kappa=\xi_{2\gamma}/\xi_{1\gamma}\,$ & $\,n$, $s$ or $ns\,$ \\ 
\hline
 50\% & 8 & 99\% & 21\% & 22 & 471 \\
 25\% & 20 & 99\% & 4\%  & 120 & 2~567 \\
 10\% & 50 & 99.9\% & 0.5\% & 1~430 & 30~588 \\
\hline
\end{tabular}
\caption{Example values for the choice of small $N$. Listed is the number of segments $N$, the two-photon absorption probability $P_{2\gamma}^{{\rm segm}}$ and 
the one-photon loss probability $P_{1\gamma}^{{\rm segm}}$ \emph{per segment}, 
the corresponding ratio $\kappa$, as well as the enhancement factor 
$n$, $s$ or $ns$ necessary for reaching this ratio.}
\label{tab:example_values_smallN}
\end{table}
\begin{table}[h]
\centering
\begin{tabular}{|c|c|c|c|c|c|c|}
\hline
$\,P_{\rm error}\,$ & $\;N\;$ & 
$\,P_{2\gamma}^{\rm segm}\,$ & $\,P_{1\gamma}^{\rm segm}\,$ & 
$\,\kappa=\xi_{2\gamma}/\xi_{1\gamma}\,$ & $\,n$, $s$ or $ns\,$ \\ 
\hline
 50\% & 10 & 95\% & 23\% & 12 & 257 \\
 25\% & 25 & 95\% & 4\%  & 76 & 1~626 \\
 10\% & 60 & 98\% & 0.5\% & 760 & 16~256 \\
\hline
\end{tabular}
\caption{Example values for the balanced choice. Listed is the number of segments $N$, the two-photon absorption probability $P_{2\gamma}^{{\rm segm}}$ and 
the one-photon loss probability $P_{1\gamma}^{{\rm segm}}$ \emph{per segment}, 
the corresponding ratio $\kappa$, as well as the enhancement factor 
$n$, $s$ or $ns$ necessary for reaching this ratio.}
\label{tab:example_values_sensible}
\end{table}
\begin{table}[h]
\centering
\begin{tabular}{|c|c|c|c|c|c|c|}
\hline
$\,P_{\rm error}\,$ & $\;N\;$ & 
$\,P_{2\gamma}^{\rm segm}\,$ & $\,P_{1\gamma}^{\rm segm}\,$ & 
$\,\kappa=\xi_{2\gamma}/\xi_{1\gamma}\,$ & $\,n$, $s$ or $ns\,$ \\ 
\hline
 50\% & 40 & 47\% & 8\% & 8 & 171 \\
 25\% & 70 & 61\% & 2\%  & 55 & 1~176 \\
 10\% & 160 & 69\% & 0.3\% & 440 & 9~412 \\
\hline
\end{tabular}
\caption{Example values for the choice of small $\kappa$. Listed is the number of segments $N$, the two-photon absorption probability $P_{2\gamma}^{{\rm segm}}$ and 
the one-photon loss probability $P_{1\gamma}^{{\rm segm}}$ \emph{per segment}, 
the corresponding ratio $\kappa$, as well as the enhancement factor 
$n$, $s$ or $ns$ necessary for reaching this ratio.}
\label{tab:example_values_smallKappa}
\end{table}
\noindent
To see which values of $s$ might be realistic, let us insert a detuning 
of the pump beam of $\Delta'=3\cdot10^{14}\,\rm Hz$ into 
expression~(\ref{eq:intensity}), where we get an excitation ratio 
of $s/S=1.7\cdot10^{-5}$. 
In order to obtain a reasonable value for $S$, we imagine a glass plate of $d = 10~\mu$m thickness, for example, and a cross section area of $A = \left(\lambda/2\right)^2$, where we assume $\lambda = 500~{\rm nm}$. With a density of $SiO_2$ of $\rho \approx 2.5~{\rm g}/{\rm cm}^3$ and a molar mass of $M = 60.1~{\rm g}/{\rm mol}$ \cite{Chemistry:2009fk}, we find
\bea
N_{\rm molecules} \approx N_A \frac{\rho A d}{M} \approx 1.6 \cdot 10^{10}
\,,
\ea
where $N_A$ is the Avogadro constant.
If one percent of these molecules is optically active, we get 
$S=1.6\cdot10^{8}$ and thus $s=2~720$.
Inserting the aforementioned values for the detuning of the 
target photon ($\Delta=3\cdot10^{12}\,\rm Hz$) and the 
control photon ($3\cdot10^{13}\,\rm Hz$), the two-photon absorption 
probability in (\ref{eq:2gamma}) would be around $P_{2\gamma}=\Or(10^{-11})$, 
which roughly fits to the value of $S=1.6\cdot10^{8}$ discussed above.

Going to the limit, we may imagine increasing these two detunings to 
$\Delta=3\cdot10^{13}\,\rm Hz$ for the target photon and 
for the control photon $3\cdot10^{14}\,\rm Hz$ 
(which is in the infra-red region).
In this case, the two-photon absorption probability in (\ref{eq:2gamma}) 
is two orders of magnitude lower and we could use a glass plate of 
1~mm thickness, which increases the maximum number $s$ of excited 
atoms/molecules by two orders of magnitude. 
The values in the three Tables remain the same with the exception of the 
last column, where the required numbers for $n$, $s$ or $ns$ increase by 25\%. 

For the amplification mechanism sketched in 
Figure~\ref{fig:repeated}, a very tight focus is desirable.
For the other enhancement mechanism based on (\ref{eq:Dicke}), however, 
the spatial phase matching conditions become problematic if we focus down 
to the diffraction limit $A = \left(\lambda/2\right)^2$ due to the 
uncertainty in $\f{k}$.
Therefore, let us consider increasing the cross section area $A$. 
On the one hand, this would decrease the ratio (\ref{eq:ratio}) even 
further -- but, on the other hand, the number $S$ of atoms/molecules
within this area $A$ grows by the same factor.
If we keep the pump laser intensity constant, the enhancement factor 
$s$ compensates the shrinking ratio (\ref{eq:ratio}), i.e., a weaker 
focus with smaller uncertainty in $\f{k}$ is also feasible.

\section{Alternative level scheme}
\label{sec:alternative_level_scheme}

\begin{figure}[h]
\begin{center}
\psfrag{e1}{$E_1$}
\psfrag{e2}{$E_2$}
\psfrag{e3}{$E_3$}
\psfrag{1s}{$\ket{\psi_1}$}
\psfrag{2s}{$\ket{\psi_2}$}
\psfrag{3p}{$\ket{\psi_3}$}
\psfrag{delta}{$\Delta$}
\psfrag{omega_1}{$\omega_1$}
\psfrag{omega_2}{$\omega_2$}
\includegraphics[width=0.3\columnwidth]{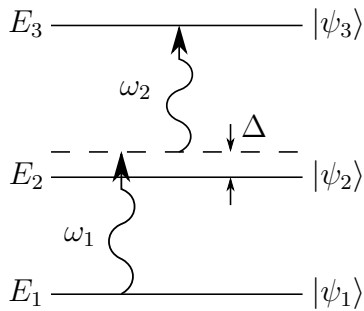}
\caption{Sketch (not to scale) of the level scheme. 
Both photons together are in resonance with the transition between 
$\ket{\psi_1}$ and $\ket{\psi_3}$, $E_3-E_1=\omega_1+\omega_2$, 
but one-photon absorption is suppressed by the detuning $\Delta$, 
where $E_2-E_1=\omega_1-\Delta$. 
}
\label{fig:new_niveau}
\end{center}
\end{figure}

As another idea for suppressing one-photon loss in comparison to two-photon 
absorption, let us replace the level scheme discussed in section 
\ref{sec:single_three_level_atom} (see Figure~\ref{fig:niveau}) by a 
three-level system as in Figure~\ref{fig:new_niveau} where the middle level 
$\ket{\psi_2}$ lives much longer than the upper level $\ket{\psi_3}$.  
Such a level scheme is often referred to as $\Lambda$-system. 

In this case, the coupling strength $g_{12}$ of the 
$\ket{\psi_1}\to\ket{\psi_2}$ transition is much smaller than the other 
coupling strength $g_{23}$ of the $\ket{\psi_2}\to\ket{\psi_3}$ transition. 
According to Eq.~(\ref{eq:2gamme_pre}), the two-photon absorption probability 
$P_{2\gamma}$ scales with $g_{12}^2$.   
In contrast, as shown in Eq.~(\ref{eq:p1gamma_app}), the one-photon 
scattering probability $P_{1\gamma}$ behaves as $g_{12}^4$.
Thus, if we denote the ratio of the two coupling strengths  by 
$f=g_{12}/g_{23}\ll1$, we see that 
$\kappa = P_{2\gamma}/P_{1\gamma}$ scales with $1/f^2$, 
and consequently can be enhanced strongly.

Let us insert some example values. 
We assume that the life-time of the upper $\ket{\psi_3}$ level is 
of the order $10^{-9}$s, which is a usual value for fast optical 
transitions. 
The life-time of the middle $\ket{\psi_2}$ level is supposed to be 
much longer, about $10^{-6}$s.
This corresponds to a ratio of $f = g_{12}/g_{23} = 0.03$.
Note that such differences are not unusual in quantum optics.
For example, the coupling constants of an electric dipole (E1) transition
(with $\Delta l = \pm 1$, cf.~\cite{Cohen-Tannoudji:1977fk}) 
and an electric quadrupole (E2) transition (with $\Delta l = 0, \pm 2$) 
differ by a factor of about $f = \ell_{\rm atom} / \lambda$ 
(cf.~\cite{Sakurai:1987fk}) which is 
even smaller than the value $f = g_{12}/g_{23} = 0.03$ we have chosen.
Inserting this value, we get at the diffraction limit 
\bea
\kappa = \frac{P_{2\gamma}}{P_{1\gamma}} 
\approx \frac{3}{2 \pi^3} \cdot \frac{1}{f^2} \approx 54
\,,
\ea
i.e., an enhancement by a factor of $10^3$ in comparison to 
(\ref{eq:ratio_example}).
This value of $\kappa$ corresponds to an error probability 
as low as $P_{\rm error} \approx 25\;\%$ 
(see Table~\ref{tab:ex_values_nkappa}).

Of course one needs to keep in mind that the two-photon absorption 
probability (\ref{eq:2gamma}) is also reduced by a factor of 
$f^2 = 9\cdot10^{-4}$ in comparison with the example values in 
section \ref{sec:example_values}. 
This poses constraints on the number of $\Lambda$-systems required 
for achieving a large enough two-photon absorption probability.
In principle, this probability can be increased by reducing the 
detuning $\Delta$, but this reduction is limited by the line-width 
of the middle $\ket{\psi_2}$ level. 
Unfortunately, this is not the natural line-width given by the 
inverse of the life-time, but gets broadened, 
e.g., by inhomogeneities and thermal effects.
Here, we assume a detuning of $\Delta=3\cdot10^{9}\,\rm Hz$ 
(for the target-photon). 
This value is a bit lower than in that of the gain media in typical 
solid-state (e.g., ruby) lasers, but could be achieved with gas lasers 
(see, e.g., \cite{Duarte:1995jh}) or ions in crystals \cite{Wrigge:2008fk,Sun:2005rh,Macfarlane:1987ie}.
Inserting this detuning, the two-photon absorption probability is about 
$P_{2\gamma}=\Or(10^{-7})$ per $\Lambda$-system.
An absorption probability of order unity can then be achieved by packing 
$10^7$ or more of these $\Lambda$-systems in the focal point with a volume 
of order $1\;\mu{\rm m}^3$.  
This corresponds to a density of $10^{19}\rm cm^{-3}$ which is quite feasible.  

Since it is probably hard to increase the spatial density of 
$\Lambda$-systems much more, higher values of $f$ would require 
smaller detunings $\Delta<3\cdot10^{9}\,\rm Hz$ in order to maintain 
$P_{2\gamma}=\Or(1)$.
As one possibility, one could imagine using ultra-cold atoms or molecules
as absorbed media. 
Although their spatial density is also limited, the line-widths are 
typically much sharper. 
For example, lowering the detuning one additional order of 
magnitude ($\Delta=3\cdot10^{8}\,\rm Hz$) would allow for 
$f = 3\cdot10^{-3}$ and thus for $\kappa=5.4\cdot10^3$ which 
would correspond to an error probability far below 10\%.

\section{Summary and conclusion}
\label{sec:summary_and_conclusion}

In summary, we presented a scheme for entangling photons via the quantum 
Zeno effect which might be realizable with present-day technology.
Compared to the previously proposed Franson gate 
\cite{Franson:2004fj,Leung:2006jo,Franson:2007yo,Myers:2007fc,Leung:2007dw}, 
our three-branch gate (section \ref{ssec:three_branch_gate}) 
exhibits a significantly reduced error probability 
(section \ref{sssec:comparison_with_franson_gate}) 
due to its improved design.
Our scheme could be physically implemented using strong two-photon 
absorption in optical fibres, as already envisaged for the Franson-gate  
\cite{Franson:2006mw,You:2008kn,You:2009jw,Hendrickson:2010pi}. 
In this case, the benefit of our scheme is given simply by the reduced 
error probability.

Moreover, we explored three mechanisms to enhance the two-photon 
absorption rate compared to the one-photon loss rate. 
Because wave-guides, resonators or optical fibres also have their drawbacks
(as they, for example, can induce additional losses or decoherence), we
mainly focus on a free-space realization of our entangling gate.

First (section \ref{sec:repeated_inducing}), we presented an apparatus 
(Figure~\ref{fig:repeated}) which allows to enhance the two-photon absorption 
probability compared to the one-photon scattering probability by a factor 
$n$ by sending the two photons $n$ times through the same, given absorber. 
The whole apparatus then replaces the absorbers in our three-branch gate 
(brown circles in Figure~\ref{fig:3branch}).
Second (section \ref{sec:coherent_interaction_with_Sgg1_atoms}), 
we proposed to employ coherent excitation of a large number of 
atoms/molecules, as known as ``Dicke super-radiance'' 
\cite{Scully:2007fk,Scully:2006fk,Sete:2010fu}. 
It was shown that the two-photon absorption probability is enhanced by 
a factor $S s$ where $S$ is the number of atoms and $s$ is the number 
of excitations. 
Furthermore it was shown that any important one-photon process only 
scales with the number of atoms $S$. 
Thus we achieve the desired effect, the two-photon absorption is enhanced 
by a factor $s$ compared to one-photon processes.
Third, we consider a special level scheme ($\Lambda$-system)  
in section \ref{sec:alternative_level_scheme}, which possesses a 
strongly reduced one-photon scattering rate.

Finally, we inserted potentially realistic example values in sections 
\ref{sec:example_values} and \ref{sec:alternative_level_scheme}, 
showing that the presented scheme is not out of reach experimentally. 
Even though the example parameters indicate that it might be hard 
to directly reach the error threshold of $10^{-4}$ or $10^{-5}$ 
required for universal quantum computation \cite{DiVincenzo:2000fk}, 
the achievable success probabilities for entangling photons are already 
comparably high.
To prevent any misunderstanding, it should at this point be stressed 
that in this work, the (pseudo) deterministic creation of entanglement 
between two {\em given} photons was considered, which is very different 
from the {\em spontaneous} creation of entangled photon pairs via 
parametric down conversion. 

As an outlook for future developments it should be stated that the design 
of our set-up in Figure~\ref{fig:3branch} does not raise the claim 
to be optimal. 
It is possible that improved set-ups can be developed which yield an 
even lower error probability, given the same ratio of two-photon 
absorption compared to one-photon loss.
Another key point where improvements could ensue is by providing 
new ways to enhance two-photon absorption compared to one-photon loss. 
With regard to the 
mechanisms proposed in this paper, better mirrors 
(i.e. more repetitions possible, see section \ref{sec:repeated_inducing}),  
stronger lasers [to improve the ratio $s/S$, see (\ref{eq:intensity})], 
or absorber media with sharper line-widths (in order to reduce the 
detuning, cf.~section \ref{sec:alternative_level_scheme}) are desirable.  

In order to explore another direction in the multi-dimensional
parameter space, one could consider realizing the three-level systems 
via Rydberg atoms or artificial atoms such as quantum dots, 
which generate electronic bound states with a level structure similar to 
Figure~\ref{fig:niveau}. 
In this case, the dipole length $\ell_{\rm atom}$ could be increased 
by several orders of magnitude, see (\ref{eq:intensity}), but the
realistic energy and intensity scales would also have to be modified. 
As another point, by placing photon detectors around the apparatus, 
one can detect occurring errors with a certain probability and 
thus apply error correcting schemes \cite{Myers:2007fc,Leung:2007dw}. 
For example, Myers and Gilchrist demonstrated that the the performance 
of a quantum Zeno gate can be greatly enhanced using photon loss codes 
\cite{Myers:2007fc}.
\ack

The authors thank 
Uwe Bovensiepen, Alexander Tarasevitch (Universit\"at Duisburg-Essen),  
Vahid Sandoghdar (and his group members at the Max Planck Institute 
for the Science of Light) for valuable conversations. 
R.S.\ acknowledges  fruitful discussions during the 
First NASA Quantum Future Technologies Conference (January 2012), 
especially with Mark Saffman (University of Wisconsin), 
where the idea for set-up in section~\ref{sec:alternative_level_scheme}
was born. 
This work was supported by the DFG (SFB-TR12).

\appendix
\section{Success probabilities of the optical set-ups}
\label{sec:appendix_a}
\subsection{Success probabilities of the two-branch gate}
\label{ap:ssec:success_probabilities_of_the_two_branch_gate}
The error probabilities for both operating modes of the two-branch gate, as defined in section \ref{sssec:error_probabilites_2branch}
\bea
P_{\rm error}^{1\gamma} &=& 1 - \left|(0,1) \cdot \vec\psi_{\rm out}^{1\gamma}\right|^2 \,, \nonumber\\
P_{\rm error}^{2\gamma} &=& 1 - \left|(1,0) \cdot \vec\psi_{\rm out}^{2\gamma}\right|^2
\,,
\ea
are analytically given by (\ref{eq:outputstate_general}), and were expanded for the limit $N \gg 1$ in zeroth order, see (\ref{eq:error_nocontrol}) and (\ref{eq:error_control}). When expanding them until first non-vanishing order in $1/N$ while keeping the assumption $N \xi_{2\gamma} \gg 1 \gg N \xi_{1\gamma}$, they read
\bea
\label{ap:eq:error_nocontrol_2branch}
P_{\rm error}^{1\gamma} = N \xi_{1\gamma} + \xi_{1\gamma} + \Or \left( N^{-2} \right)
\,,
\ea
and
\bea
\label{ap:eq:error_control_2branch}
P_{\rm error}^{2\gamma} = &&\frac{\pi^2}{2 N \xi_{2\gamma}} + \frac{2 \pi^2 - \pi^4}{48 N^2} + \frac{4 \pi^4 + \pi^6}{192 N^3 \xi_{2\gamma}}+\nonumber\\
&&+ \frac{\pi^2 \left( \xi_{1\gamma} + \xi_{2\gamma} \right)}{24 N} + \Or \left( N^{-3} \right)
\,,
\ea
where in the two-photon case the first order in $1/N$ vanishes.
\subsection{Success probabilities of the three-branch gate}
\label{ap:ssec:success_probabilities_of_the_three_branch_gate}
The error probabilities for both operating modes of the three-branch gate, as defined in section \ref{sssec:error_probabilites_3branch}
\bea
P_{\rm error}^{1\gamma} &=& 1 - \left|(0,0,1) \cdot \vec\psi_{\rm out}^{1\gamma}\right|^2\,, \nonumber\\
P_{\rm error}^{2\gamma} &=& 1 - \left|(1,0,0) \cdot \vec\psi_{\rm out}^{2\gamma}\right|^2
\,,
\ea
are analytically given by (\ref{eq:3branch_outputstate}), and were expanded for the limit $N \gg 1$ in zeroth order, see (\ref{eq:3branch_error_prob}). Using \texttt{Mathematica}, they can also be expanded until first non-vanishing order in $1/N$ while keeping the assumption $N \xi_{2\gamma} \gg 1 \gg N \xi_{1\gamma}$. Then they read
\bea
\label{ap:eq:error_nocontrol_3branch}
P_{\rm error}^{1\gamma} = \frac{N \xi_{1\gamma}}{2} + \xi_{1\gamma} + \Or \left( N^{-2} \right)
\,,
\ea
and
\bea
\label{ap:eq:error_control_3branch}
P_{\rm error}^{2\gamma} = &&\frac{\pi^2}{N \xi_{2\gamma}} + \frac{4 \pi^2 - 3 \pi^4}{48 N^2} + \frac{2 \pi^4 + \pi^6}{24 N^3 \xi_{2\gamma}}\nonumber+\\
&&+ \frac{\pi^2 \left( \xi_{1\gamma} + \xi_{2\gamma} \right)}{12 N} + \Or \left( N^{-3} \right)
\,,
\ea
where again in the two-photon case the first order in $1/N$ vanishes.
\section{Details on the coupling constants}
\label{sec:appendix_b}
\subsection{Derivation of the coupling constants}
\label{ap:ssec:derivation_of_the_coupling_constants}
In order to examine the corresponding coupling constants, we need to calculate the atomic transition matrix elements \cite{Sakurai:1987fk}. For transitions which are given by the mixed term $\propto \hat{\f{p}}\cdot\hat{\f{A}\,}( \f{r}_0 + \f{\delta}\hat{\f{r}}, t )$, dipole approximation is applied, i.e. $\hat{\f{p}}\cdot\hat{\f{A}\,}( \f{r}_0 + \f{\delta}\hat{\f{r}}, t ) \approx \hat{\f{p}}\cdot\hat{\f{A}\,}(\f{r}_0, t )$, and thus only the $\hat{\f{p}}$-operator works on the electron states. For the transition between 1s and 2p, for example, the transition matrix element $\f{g}^{12}$ reads
\bea
\f{g}^{12} &=& -\frac{q}{m} \bra{\rm 1s} \hat{\f{p}} \ket{\rm 2p} = -\frac{q}{m} \bra{\rm 1s} \rmi m \left[ \hat{H}_0, \hat{\f{r}} \right] \ket{\rm 2p}\nonumber\\
&=& \rmi q E_{12} \bra{\rm 1s} \hat{\f{r}} \ket{\rm 2p} = \rmi q E_{12} \f{\ell}_{\rm atom}^{12}
\,,
\ea
where the commutator relation $\left[ \hat{\f{p}}^2, \hat{\f{r}} \right] = - 2 \rmi \hat{\f{p}}$ was used and $\f{\ell}_{\rm atom}^{12} = \bra{\rm 1s} \hat{\f{r}} \ket{\rm 2p}$ is a vector with the typical dipole (coupling) length of about six Bohr radii $\ell_{\rm atom}^{12} = 6 a_0$ and a direction depending on the orientation of the electronic orbitals 1s and 2p.
Taking also the amplitude ${\f{g}^{\lambda}_{\fk{k}}}^*$ from the vector potential into account we arrive at a coupling constant 
\bea
g_{\vphantom{\fk{\tilde{k}}}\fk{k}, \lambda}^{12}
=
{\f{g}^{\lambda}_{\fk{k}}}^* \cdot \f{g}^{12} \approx E_{12} \sqrt{\frac{\alpha_{\rm QED}}{\left( 2 \pi \right)^2 \omega_k}} \left( \hat{\f{\epsilon}}^{\lambda}_{\fk{k}} \cdot \f{\ell}_{\rm atom}^{12} \right)
\,,
\ea
and analogously for $g_{\vphantom{\fk{\tilde{k}}}\fk{k}, \lambda}^{23}$
\bea
g_{\vphantom{\fk{\tilde{k}}}\fk{k}, \lambda}^{23}
=
{\f{g}^{\lambda}_{\fk{k}}}^* \cdot \f{g}^{23} \approx E_{23} \sqrt{\frac{\alpha_{\rm QED}}{\left( 2 \pi \right)^2 \omega_k}} \left( \hat{\f{\epsilon}}^{\lambda}_{\fk{k}} \cdot \f{\ell}_{\rm atom}^{23} \right)
\,,
\ea
where $\alpha_{\rm QED}$ is the fine-structure constant.
The transition matrix elements for the direct two-photon absorption (or emission) process, which is possible via the quadratic term $\propto \hat{\f{A}\,}^2(\f{r}_0 + \f{\delta}\hat{\f{r}}, t )$, are zero when using the dipole approximation, as there would be no (momentum) operator left working on the electron states. Therefore it is necessary to look at higher orders of $\rme^{\rm\rmi\fk{k}\cdot\fk{\delta}\hat{\fk{r}}}$
\bea
\label{eq:higher_than_dipole}
\bra{\rm 1s} \rme^{\rm\rmi\fk{k}\cdot\fk{\delta}\hat{\fk{r}}} \ket{\rm 3s} 
&=& 
\bra{\rm 1s} \left[ 1 + \rmi \f{k} \cdot\f{\delta}\hat{\f{r}} - \left(\f{k} \cdot \f{\delta}\hat{\f{r}}\right)^2/2 + \Or \left( \left( \f{k}\cdot\f{\delta}\hat{\f{r}}\right)^3 \right) \right] \ket{\rm 3s}\nonumber\\
&=&
- \frac{1}{2} \bra{\rm 1s} \left( \f{k} \cdot \f{\delta}\hat{\f{r}}\right)^2 \ket{\rm 3s} + \Or \left( \left( \f{k} \cdot \f{\delta}\hat{\f{r}}\right)^3 \right)
\,.
\ea
As for the hydrogen atom, we assumed $\bra{\rm 1s} \f{\delta}\hat{\f{r}} \ket{\rm 3s} = 0$ for parity reasons (1s and 3s are both spherically symmetric). Therefore we will take account of the quadratic order as the first non-vanishing order, while neglecting higher orders which should be much smaller. This results in the following transition matrix element
\bea
g^{13} 
&=&
\frac{q^2}{2m} \bra{1s} \rme^{\rmi (\fk{k} + \fk{\tilde{k}})\cdot \fk{\delta}\hat{\fk{r}}} \ket{3s}\nonumber\\
&\approx&
- \frac{q^2}{4m} \bra{1s} \left[ ( \f{k} + \f{\tilde{k}} ) \cdot \f{\delta}\hat{\f{r}} \right]^2 \ket{3s}\nonumber\\
&\lesssim&
- \frac{q^2}{4m} ( \f{k} + \f{\tilde{k}} )^2 \big(\ell_{\rm atom}^{13}\big)^2
\,,
\ea
and the following coupling constant (note that we have quadratic order in ${\f{g}^{\lambda}_{\fk{k}}}^*$ as well)
\bea
g_{\vphantom{\fk{\tilde{k}}}\fk{k} \fk{\tilde{k}}, \lambda \tilde{\lambda}}^{13}
&=&
\left( {\f{g}^{\lambda}_{\vphantom{\fk{\tilde{k}}}\fk{k}}}^* \cdot {\f{g}^{\tilde{\lambda}}_{\fk{\tilde{k}}}}^* \right) g^{13}\nonumber\\
&\approx& \frac{\alpha_{\rm QED}}{4 \left( 2 \pi \right)^2 m} \frac{1}{\sqrt{\omega_k \omega_{\tilde{k}}}} ( \f{k} + \f{\tilde{k}} )^2 \big(\ell_{\rm atom}^{13}\big)^2 \left( \hat{\f{\epsilon}}^{\lambda}_{\fk{k}} \cdot \hat{\f{\epsilon}}^{\tilde{\lambda}}_{\fk{\tilde{k}}} \right)
\,.
\ea
For $g_{\vphantom{\fk{\tilde{k}}}\fk{k} \fk{\tilde{k}}, \lambda \tilde{\lambda}}^{11}$, the atomic part of the coupling constant simply reads $g^{11} = q^2 / \left( 2 m \right)$ because there is no atomic transition occurring in the case of direct $\propto \hat{\f{A}\,}^2(\f{r}_0 + \f{\delta}\hat{\f{r}}, t )$-scattering. There is an additional factor $2$ because the process arises as a mixed term in $\hat{\f{A}\,}^2(\f{r}_0 + \f{\delta}\hat{\f{r}}, t )$. Thus
\bea
g_{\vphantom{\fk{\tilde{k}}}\fk{k} \fk{\tilde{k}}, \lambda \tilde{\lambda}}^{11}
=
2 \left( {\f{g}^{\lambda}_{\vphantom{\fk{\tilde{k}}}\fk{k}}}^* \cdot \f{g}^{\tilde{\lambda}}_{\fk{\tilde{k}}} \right) \frac{q^2}{2m} 
= 
\frac{\alpha_{\rm QED}}{\left( 2 \pi \right)^2 m} \frac{1}{\sqrt{\omega_k \omega_{\tilde{k}}}} \left( \hat{\f{\epsilon}}^{\lambda}_{\fk{k}} \cdot \hat{\f{\epsilon}}^{\tilde{\lambda}}_{\fk{\tilde{k}}} \right)
\,.
\ea
Using the recently derived coupling constants, the Hamiltonian (\ref{eq:Hamiltonian_interaction}) can be rewritten in terms of atomic transition operators like $\hat{\sigma}^{12} = \ket{\rm 1s}\bra{\rm 2p}$ and photonic annihilation/creation operators $\hat{a}_{\vphantom{\fk{\tilde{k}}}\fk{k}, \lambda}^{\phantom{\dagger}} / \hat{a}_{\vphantom{\fk{\tilde{k}}}\fk{k}, \lambda}^\dagger$ in order to prepare the subsequent perturbation theory calculations. There, (\ref{eq:ham_inter_mod}), we did not quote the full result, which reads
\bea
\label{eq:ham_inter_mod_rem_complete}
\hat{H}_{1, {\rm I}} ( \f{r}_0, t )
&=& \sum_{\fk{k}, \lambda} \left( g_{\vphantom{\fk{\tilde{k}}}\fk{k}, \lambda}^{12} \hat{\sigma}^{12} \rme^{-\rmi \Delta_k^{12} t} + g_{\vphantom{\fk{\tilde{k}}}\fk{k}, \lambda}^{23} \hat{\sigma}^{23} \rme^{-\rmi \Delta_k^{23} t} \right) \hat{a}_{\vphantom{\fk{\tilde{k}}}\fk{k}, \lambda}^\dagger \rme^{-\rmi\fk{k}\cdot\fk{r}_0} + \rm{H.c.}\nonumber\\
&+& \sum_{\fk{k}, \fk{\tilde{k}}, \lambda, \tilde{\lambda}} g_{\vphantom{\fk{\tilde{k}}}\fk{k} \fk{\tilde{k}}, \lambda \tilde{\lambda}}^{13} \rme^{\rmi(\omega_k + \omega_{\tilde{k}} - E_{13})t} \hat{a}_{\vphantom{\fk{\tilde{k}}}\fk{k}, \lambda}^\dagger \hat{a}_{\fk{\tilde{k}}, \tilde{\lambda}}^\dagger \hat{\sigma}^{13} \rme^{-\rmi(\fk{k} + \fk{\tilde{k}})\cdot\fk{r}_0} + \rm{H.c.}\nonumber\\
&+& \sum_{\fk{k}, \fk{\tilde{k}}, \lambda, \tilde{\lambda}} g_{\vphantom{\fk{\tilde{k}}}\fk{k} \fk{\tilde{k}}, \lambda \tilde{\lambda}}^{11} \rme^{\rmi(\omega_k - \omega_{\tilde{k}})t} \hat{a}_{\vphantom{\fk{\tilde{k}}}\fk{k}, \lambda}^\dagger \hat{a}_{\fk{\tilde{k}}, \tilde{\lambda}}^{\phantom{\dagger}} \rme^{\rmi(\fk{\tilde{k}} - \fk{k})\cdot\fk{r}_0}\nonumber\\
&-& \sum_{\fk{k}, \lambda} \left( g_{\vphantom{\fk{\tilde{k}}}\fk{k}, \lambda}^{12} \hat{\sigma}^{12} \rme^{-\rmi(E_{12} + \omega_k)t} + g_{\vphantom{\fk{\tilde{k}}}\fk{k}, \lambda}^{23} \hat{\sigma}^{23} \rme^{-\rmi(E_{23} + \omega_k)t} \right) \hat{a}_{\vphantom{\fk{\tilde{k}}}\fk{k}, \lambda}^{\phantom{\dagger}} \rme^{\rmi\fk{k}\cdot\fk{r}_0} + \rm{H.c.}\nonumber\\
&+& \sum_{\fk{k}, \fk{\tilde{k}}, \lambda, \tilde{\lambda}} g_{\vphantom{\fk{\tilde{k}}}\fk{k} \fk{\tilde{k}}, \lambda \tilde{\lambda}}^{13} \rme^{-\rmi(\omega_k + \omega_{\tilde{k}} + E_{13})t} \hat{a}_{\vphantom{\fk{\tilde{k}}}\fk{k}, \lambda}^{\phantom{\dagger}} \hat{a}_{\fk{\tilde{k}}, \tilde{\lambda}}^{\phantom{\dagger}} \hat{\sigma}^{13} \rme^{\rmi(\fk{k} + \fk{\tilde{k}})\cdot\fk{r}_0} + \rm{H.c.}\nonumber\\
&+& \sum_{\fk{k}, \fk{\tilde{k}}, \lambda, \tilde{\lambda}} g_{\vphantom{\fk{\tilde{k}}}\fk{k} \fk{\tilde{k}}, \lambda \tilde{\lambda}}^{11} \frac{g^{13}}{g^{11}} \rme^{\rmi(\omega_k - \omega_{\tilde{k}} - E_{13})t} \hat{a}_{\vphantom{\fk{\tilde{k}}}\fk{k}, \lambda}^\dagger \hat{a}_{\fk{\tilde{k}}, \tilde{\lambda}}^{\phantom{\dagger}} \hat{\sigma}^{13} \rme^{\rmi(\fk{\tilde{k}} - \fk{k})\cdot\fk{r}_0}\nonumber\\
&+& \sum_{\fk{k}, \fk{\tilde{k}}, \lambda, \tilde{\lambda}} g_{\vphantom{\fk{\tilde{k}}}\fk{k} \fk{\tilde{k}}, \lambda \tilde{\lambda}}^{11} \frac{{g^{13}}^*}{g^{11}} \rme^{\rmi(\omega_k - \omega_{\tilde{k}} + E_{13})t} \hat{a}_{\vphantom{\fk{\tilde{k}}}\fk{k}, \lambda}^\dagger \hat{a}_{\fk{\tilde{k}}, \tilde{\lambda}}^{\phantom{\dagger}} \hat{\sigma}^{13\dagger} \rme^{\rmi(\fk{\tilde{k}} - \fk{k})\cdot\fk{r}_0}\nonumber\\
&-& \sum_{\fk{k}, \fk{\tilde{k}}, \lambda, \tilde{\lambda}} \frac{g_{\vphantom{\fk{\tilde{k}}}\fk{k} \fk{\tilde{k}}, \lambda \tilde{\lambda}}^{11}}{2} \rme^{-\rmi(\omega_k + \omega_{\tilde{k}})t} \hat{a}_{\vphantom{\fk{\tilde{k}}}\fk{k}, \lambda}^{\phantom{\dagger}} \hat{a}_{\fk{\tilde{k}}, \tilde{\lambda}}^{\phantom{\dagger}} \rme^{\rmi(\fk{k} + \fk{\tilde{k}})\cdot\fk{r}_0} + \rm{H.c.}
\,.
\ea
In section \ref{ssec:hamiltonian_and_initial_state}, the remaining terms $\hat{H}_{1, {\rm I}}^{\rm rem} ( \f{r}_0, t )$ were suppressed because they are unimportant for the subsequent analysis
\bea
\label{eq:ham_inter_mod_rem}
\hat{H}_{1, {\rm I}}^{\rm rem} ( \f{r}_0, t )
&=& \sum_{\fk{k}, \fk{\tilde{k}}, \lambda, \tilde{\lambda}} g_{\vphantom{\fk{\tilde{k}}}\fk{k} \fk{\tilde{k}}, \lambda \tilde{\lambda}}^{13} \rme^{-\rmi(\omega_k + \omega_{\tilde{k}} + E_{13})t} \hat{a}_{\vphantom{\fk{\tilde{k}}}\fk{k}, \lambda}^{\phantom{\dagger}} \hat{a}_{\fk{\tilde{k}}, \tilde{\lambda}}^{\phantom{\dagger}} \hat{\sigma}^{13} \rme^{\rmi(\fk{k} + \fk{\tilde{k}})\cdot\fk{r}_0} + \rm{H.c.}\nonumber\\
&+& \sum_{\fk{k}, \fk{\tilde{k}}, \lambda, \tilde{\lambda}} g_{\vphantom{\fk{\tilde{k}}}\fk{k} \fk{\tilde{k}}, \lambda \tilde{\lambda}}^{11} \frac{g^{13}}{g^{11}} \rme^{\rmi(\omega_k - \omega_{\tilde{k}} - E_{13})t} \hat{a}_{\vphantom{\fk{\tilde{k}}}\fk{k}, \lambda}^\dagger \hat{a}_{\fk{\tilde{k}}, \tilde{\lambda}}^{\phantom{\dagger}} \hat{\sigma}^{13} \rme^{\rmi(\fk{\tilde{k}} - \fk{k})\cdot\fk{r}_0}\nonumber\\
&+& \sum_{\fk{k}, \fk{\tilde{k}}, \lambda, \tilde{\lambda}} g_{\vphantom{\fk{\tilde{k}}}\fk{k} \fk{\tilde{k}}, \lambda \tilde{\lambda}}^{11} \frac{{g^{13}}^*}{g^{11}} \rme^{\rmi(\omega_k - \omega_{\tilde{k}} + E_{13})t} \hat{a}_{\vphantom{\fk{\tilde{k}}}\fk{k}, \lambda}^\dagger \hat{a}_{\fk{\tilde{k}}, \tilde{\lambda}}^{\phantom{\dagger}} \hat{\sigma}^{13\dagger} \rme^{\rmi(\fk{\tilde{k}} - \fk{k})\cdot\fk{r}_0}\nonumber\\
&-& \sum_{\fk{k}, \fk{\tilde{k}}, \lambda, \tilde{\lambda}} \frac{g_{\vphantom{\fk{\tilde{k}}}\fk{k} \fk{\tilde{k}}, \lambda \tilde{\lambda}}^{11}}{2} \rme^{-\rmi(\omega_k + \omega_{\tilde{k}})t} \hat{a}_{\vphantom{\fk{\tilde{k}}}\fk{k}, \lambda}^{\phantom{\dagger}} \hat{a}_{\fk{\tilde{k}}, \tilde{\lambda}}^{\phantom{\dagger}} \rme^{\rmi(\fk{k} + \fk{\tilde{k}})\cdot\fk{r}_0} + \rm{H.c.}
\,.
\ea
In the perturbation theory calculations of the Hamiltonian (\ref{eq:ham_inter_mod_rem_complete}), all terms of order $\hat{\f{A}\,}^3$ or higher are ignored. Thus all the remaining terms (\ref{eq:ham_inter_mod_rem}) can only contribute to the first-order amplitude. However, the first, third and fourth term all have a non-vanishing temporal phase (when $\omega_{1/2} < E_{13}$) and thus vanish for infinite interaction time. The second term is not important as it starts from the 3s-level where in our initial state the atom occupies the 1s-level.
\subsection{Integrating out the 2p-level}
\label{ap:ssec:integrating_out_the_(2p)_level}
The two-photon absorption probability of (\ref{eq:Dicke}) in standard first-order perturbation theory in case of $S = 1$ is just $P_{2\gamma} = T^2 \left| g \right|^2$, where $T$ is the total interaction time. Comparing this result with the two-photon absorption probability of the original Hamiltonian, (\ref{eq:2gamma}), one can extract the effective coupling constant $g$ to
\bea
\label{eq:coupling_constant}
\left| g \right|
=
\frac{8 \pi}{T} \frac{\left|g_{\fk{k_1},\lambda_1}^{12}\right| \left|g_{\fk{k_2},\lambda_2}^{23}\right|}{\Delta} \frac{1}{A}
=
\frac{2 \alpha_{\rm QED}}{\pi \sqrt{\omega_1 \omega_2} T} \frac{E_{12} E_{23}}{\Delta} \frac{\ell_{\rm atom}^2}{A} 
\,.
\ea
\section{Miscellaneous}
\label{sec:appendix_c}
\subsection{Commutation relations}
\label{ap:ssec:commutation_relations}
In order to simplify the results of the perturbation theory calculations, throughout the whole paper the following commutation relations (\ref{eq:comm_rel_1}) and (\ref{eq:comm_rel_2}) were applied. They can be derived using the definition of $\hat{a}_{1/2}^{\phantom{\dagger}}$,
\bea
\hat{a}_{1/2}^{\phantom{\dagger}} = \sum_{\fk{k}} \, f_{1/2}\left( \f{k} \right) \hat{a}_{\vphantom{\fk{\tilde{k}}}\fk{k}, \lambda_{1/2}}^{\phantom{\dagger}}
\,,
\ea
and the fundamental commutation relation for annihilation/creation operators
\bea
\left[ \hat{a}_{\vphantom{\fk{\tilde{k}}}\fk{k}, \lambda}^{\phantom{\dagger}}, \hat{a}_{\fk{\tilde{k}}, \tilde{\lambda}}^\dagger \right] = \delta^3 \left( \f{k} - \f{\tilde{k}} \right) \delta_{\vphantom{\tilde{\lambda}}\lambda \tilde{\lambda}}
\,.
\ea
From these two definitions, we can easily calculate the following commutation rule
\bea
\label{eq:comm_tool}
\hat{a}_{\vphantom{\fk{\tilde{k}}}\fk{k}, \lambda}^{\phantom{\dagger}} \hat{a}_{1/2}^\dagger = \hat{a}_{1/2}^\dagger \hat{a}_{\vphantom{\fk{\tilde{k}}}\fk{k}, \lambda}^{\phantom{\dagger}} + f_{1/2}^* ( \f{k} ) \delta_{\vphantom{\tilde{\lambda}}\lambda \lambda_{1/2}}
\,.
\ea
Applying (\ref{eq:comm_tool}) repeatedly yields the commutation relations employed in section \ref{ssec:perturbation_theory}
\bea
\label{eq:comm_rel_1}
\hat{a}_{\vphantom{\fk{\tilde{k}}}\fk{k}, \lambda}^\dagger \hat{a}_{\fk{\tilde{k}}, \tilde{\lambda}}^{\phantom{\dagger}} \hat{a}_1^\dagger \hat{a}_2^\dagger = &&\hat{a}_{\vphantom{\fk{\tilde{k}}}\fk{k}, \lambda}^\dagger \hat{a}_1^\dagger \hat{a}_2^\dagger \hat{a}_{\fk{\tilde{k}}, \tilde{\lambda}}^{\phantom{\dagger}}+\nonumber\\
 &+& f_2^* ( \f{\tilde{k}} ) \delta_{\vphantom{\tilde{\lambda}}\lambda_2 \tilde{\lambda}} \hat{a}_{\vphantom{\fk{\tilde{k}}}\fk{k}, \lambda}^\dagger \hat{a}_1^\dagger+\nonumber\\
 &+& f_1^* ( \f{\tilde{k}} ) \delta_{\vphantom{\tilde{\lambda}}\lambda_1 \tilde{\lambda}} \hat{a}_{\vphantom{\fk{\tilde{k}}}\fk{k}, \lambda}^\dagger \hat{a}_2^\dagger
\,,
\ea
and
\bea
\label{eq:comm_rel_2}
\hat{a}_{\vphantom{\fk{\tilde{k}}}\fk{k}, \lambda}^{\phantom{\dagger}} \hat{a}_{\fk{\tilde{k}}, \tilde{\lambda}}^{\phantom{\dagger}} \hat{a}_1^\dagger \hat{a}_2^\dagger = &&\hat{a}_1^\dagger \hat{a}_2^\dagger \hat{a}_{\vphantom{\fk{\tilde{k}}}\fk{k}, \lambda}^{\phantom{\dagger}} \hat{a}_{\fk{\tilde{k}}, \tilde{\lambda}}^{\phantom{\dagger}}+\nonumber\\
&+& f_2^* \left( \f{k} \right) \delta_{\vphantom{\tilde{\lambda}}\lambda_2 \lambda} \hat{a}_1^\dagger \hat{a}_{\fk{\tilde{k}}, \tilde{\lambda}}^{\phantom{\dagger}} + f_1^* \left( \f{k} \right) \delta_{\vphantom{\tilde{\lambda}}\lambda_1 \lambda} \hat{a}_2^\dagger \hat{a}_{\fk{\tilde{k}}, \tilde{\lambda}}^{\phantom{\dagger}}+\nonumber\\
&+& f_2^* ( \f{\tilde{k}} ) \delta_{\vphantom{\tilde{\lambda}}\lambda_2 \tilde{\lambda}} \left( \hat{a}_1^\dagger \hat{a}_{\vphantom{\fk{\tilde{k}}}\fk{k}, \lambda}^{\phantom{\dagger}} + f_1^* ( \f{k} ) \delta_{\vphantom{\tilde{\lambda}}\lambda_1 \lambda} \right)+\nonumber\\
&+& f_1^* ( \f{\tilde{k}} ) \delta_{\vphantom{\tilde{\lambda}}\lambda_1 \tilde{\lambda}} \left( \hat{a}_2^\dagger \hat{a}_{\vphantom{\fk{\tilde{k}}}\fk{k}, \lambda}^{\phantom{\dagger}} + f_2^* ( \f{k} ) \delta_{\vphantom{\tilde{\lambda}}\lambda_2 \lambda} \right)
\,.
\ea
\subsection{Scattering probability integration}
\label{ap:ssec:scattering_probability_integration}
Here the derivation of the scattering probability integration is given in full detail. We start with expression (\ref{eq:p1gamma_app}) from section \ref{ssec:one_photon_scattering_probability}, where we already neglected the $f^4$-terms
\bea
P_{1\gamma} &=& ( 2 \pi )^2 \sum_{\fk{k}, \lambda, \zeta = 1,2} \left| \sum_{\fk{\tilde{k}}} \left( \frac{g_{\vphantom{\fk{\tilde{k}}}\fk{k}, \lambda}^{12} 
\big(g_{\fk{\tilde{k}}, \lambda_\zeta}^{12}\big)^*}{\Delta_{\tilde{k}}^{12}} - g_{\vphantom{\fk{\tilde{k}}}\fk{k} \fk{\tilde{k}}, \lambda \lambda_\zeta}^{11} \right)\times\right.\nonumber\\
&&\hspace{1.0cm}\left.\times\,\delta \left( \omega_k - \omega_{\tilde{k}} \right) f_\zeta^* ( \f{\tilde{k}} ) \rme^{\rmi(\fk{\tilde{k}} - \fk{k})\cdot\fk{r}_0} \vphantom{\sum_{\fk{\tilde{k}}}}\right|^2
\,.
\ea
We then carry out the inner integration assuming constant integrand $\f{\tilde{k}} \approx \f{k_\zeta}$, which yields
\bea
P_{1\gamma} &=& 2 ( 2 \pi )^2 \sum_{\fk{k}, \lambda, \zeta = 1,2} \left| \frac{g_{\vphantom{\fk{\tilde{k}}}\fk{k}, \lambda}^{12} \big(g_{\vphantom{\fk{\tilde{k}}}\fk{k_\zeta}, \lambda_\zeta}^{12}\big)^*}{\Delta_{k_\zeta}^{12}} - g_{\vphantom{\fk{\tilde{k}}}\fk{k} \fk{k_\zeta}, \lambda \lambda_\zeta}^{11} \right|^2 \frac{\Delta k_x \Delta k_y}{\Delta k_z}
\,.
\ea
Now we write out the outer integration over $\f{k}$ explicitly in spherical coordinates such that
\bea
\f{k} = k \left( \sin \theta \cos \phi, \sin \theta \sin \phi, \cos \theta \right)
\,,
\ea
bearing in mind that the integral over the radius is restricted to a small interval due to the Dirac delta function of the inner integration
\bea
P_{1\gamma} = 2 \left( 2 \pi \right)^2 \sum_{\lambda, \zeta = 1,2} &&\int_{k_\zeta - \Delta k_z}^{k_\zeta + \Delta k_z} k^2 \,dk \int_{0}^{\pi} \sin\theta \,d\theta \int_{0}^{2 \pi} \,d\phi \times\nonumber\\
&&\times\left| \frac{g_{\vphantom{\fk{\tilde{k}}}\fk{k}, \lambda}^{12} \big(g_{\vphantom{\fk{\tilde{k}}}\fk{k_\zeta}, \lambda_\zeta}^{12}\big)^* }{\Delta_{k_\zeta}^{12}} - g_{\vphantom{\fk{\tilde{k}}}\fk{k} \fk{k_\zeta}, \lambda \lambda_\zeta}^{11} \right|^2 \frac{\Delta k_x \Delta k_y}{\Delta k_z}
\,.
\ea
\noindent
The radius-integration is quickly resolved by again assuming a constant integrand, and what remains to be calculated is (remember that $\hat{\f{\epsilon}}^{\lambda_\zeta}_{\fk{k_\zeta}} \cdot \f{\ell}_{\rm atom}^{12} = \ell_{\rm atom}$)
\bea
P_{1\gamma} &=& 4 \left( 2 \pi \right)^2 \sum_{\lambda, \zeta = 1,2} \int_{0}^{\pi} \sin\theta \,d\theta \int_{0}^{2 \pi} \,d\phi\times\nonumber\\
&&\hspace{1.0cm}\times \left| k_\zeta \frac{g_{\vphantom{\fk{\tilde{k}}}\fk{k}, \lambda}^{12} \big(g_{\vphantom{\fk{\tilde{k}}}\fk{k_\zeta}, \lambda_\zeta}^{12}\big)^* }{\Delta_{k_\zeta}^{12}} - k_\zeta g_{\vphantom{\fk{\tilde{k}}}\fk{k} \fk{k_\zeta}, \lambda \lambda_\zeta}^{11} \right|^2 \Delta k_x \Delta k_y \nonumber\\
&=& \frac{\alpha_{\rm QED}^2 \ell_{\rm atom}^4}{\pi^2} \sum_{\lambda, \zeta = 1,2} \int_{0}^{\pi} \sin\theta \,d\theta \int_{0}^{2 \pi} \,d\phi\times\nonumber\\
&&\hspace{1.0cm}\times \left|\left( \frac{E_{12}^2}{\Delta_{k_\zeta}^{12}} \frac{\f{\ell}_{\rm atom}^{12}}{\ell_{\rm atom}} - \frac{1}{m \ell_{\rm atom}^2} \hat{\f{\epsilon}}^{\lambda_\zeta}_{\fk{k_\zeta}} \right) \cdot \hat{\f{\epsilon}}^{\lambda}_{\fk{k}}\right|^2 \Delta k_x \Delta k_y
\,.
\ea
The two orthogonal, transverse polarization vectors can be written as
\bea
\hat{\f{\epsilon}}^{1}_{\fk{k}} &=& \left( \sin \theta, - \cos \phi, 0 \right),\nonumber\\
\hat{\f{\epsilon}}^{2}_{\fk{k}} &=& \left( \cos \theta \cos \phi, \cos \theta \sin \phi, -\sin \theta \right)
\,.
\ea
Using the following integral equation which can easily be calculated and is valid for any vector $\f{v}$ 
\bea
\sum_{\lambda} \int_{0}^{\pi} \sin\theta \,d\theta \int_{0}^{2 \pi} \,d\phi \left| \f{v} \cdot \hat{\f{\epsilon}}^{\lambda}_{\fk{k}} \right|^2 = \frac{8}{3} \pi \left|\f{v}\right|^2
\,,
\ea
we arrive at the result already stated in section \ref{ssec:one_photon_scattering_probability}
(remember $\f{\ell}_{\rm atom}^{12} \parallel \hat{\f{\epsilon}}^{\lambda_\zeta}_{\fk{k_\zeta}}$)
\bea
P_{1\gamma} = \frac{8 \alpha_{\rm QED}^2}{3 \pi} \sum_{\zeta = 1,2} \left[\frac{E_{12}^2}{\Delta_{k_\zeta}^{12}} - \frac{1}{m \ell_{\rm atom}^2}\right]^2 \frac{\ell_{\rm atom}^4}{A}
\,.
\ea

\section*{References}
\bibliographystyle{iopart-num}
\bibliography{l6}

\providecommand{\newblock}{}
\begin{thebibliography}{10}
\expandafter\ifx\csname url\endcsname\relax
  \def\url#1{{\tt #1}}\fi
\expandafter\ifx\csname urlprefix\endcsname\relax\def\urlprefix{URL }\fi
\providecommand{\eprint}[2][]{\url{#2}}

\bibitem{Benioff:1980kx}
Benioff P 1980 {\em J. Stat. Phys.\/} {\bf 22} 563

\bibitem{Feynman:1982fk}
Feynman R~P 1982 {\em Int. J. Theor. Phys.\/} {\bf 21} 467

\bibitem{Feynman:1986uq}
Feynman R~P 1986 {\em Found. Phys.\/} {\bf 16} 507

\bibitem{Shor:1997vn}
Shor P~W 1997 {\em SIAM J. Comput.\/} {\bf 26} 1484

\bibitem{Grover:1997ys}
Grover L~K 1997 {\em Phys. Rev. Lett.\/} {\bf 79} 325

\bibitem{Lloyd:1996vn}
Lloyd S 1996 {\em Science\/} {\bf 273} 1073

\bibitem{Exp:2000kx}
Braunstein S~L, Lo H~K and Kok P 2000 {\em Fortschr. Phys.\/} {\bf 48} 767

\bibitem{DiVincenzo:2000fk}
DiVincenzo D~P 2000 {\em Fortschr. Phys.\/} {\bf 48} 771

\bibitem{Alleaume:2004uq}
All{\'e}aume R, Treussart F, Messin G, Dumeige Y, Roch J~F, Beveratos A,
  Brouri-Tualle R, Poizat J~P and Grangier P 2004 {\em New J. Phys.\/} {\bf 6}
  92

\bibitem{Chen:2006kx}
Chen S, Chen Y~A, Strassel T, Yuan Z~S, Zhao B, Schmiedmayer J and Pan J~W 2006
  {\em Phys. Rev. Lett.\/} {\bf 97} 173004

\bibitem{Varnava:2008fk}
Varnava M, Browne D~E and Rudolph T 2008 {\em Phys. Rev. Lett.\/} {\bf 100}
  060502

\bibitem{Nielsen:2000uq}
Nielsen M~A and Chuang I~L 2000 {\em {Quantum Computation and Quantum
  Information}\/} (Cambridge: Cambridge University Press)

\bibitem{Knill:2001vn}
Knill E, Laflamme R and Milburn G~J 2001 {\em Nature\/} {\bf 409} 46

\bibitem{Raussendorf:2001ve}
Raussendorf R and Briegel H~J 2001 {\em Phys. Rev. Lett.\/} {\bf 86} 5188

\bibitem{Pittman:2001zr}
Pittman T~B, Jacobs B~C and Franson J~D 2001 {\em Phys. Rev.\/} A {\bf 64}
  062311

\bibitem{Ralph:2001fk}
Ralph T~C, White A~G, Munro W~J and Milburn G~J 2001 {\em Phys. Rev.\/} A {\bf
  65} 012314

\bibitem{Franson:2002kx}
Franson J~D, Donegan M~M, Fitch M~J, Jacobs B~C and Pittman T~B 2002 {\em Phys.
  Rev. Lett.\/} {\bf 89} 137901

\bibitem{Knill:2002uq}
Knill E 2002 {\em Phys. Rev.\/} A {\bf 66} 052306

\bibitem{Yoran:2003fk}
Yoran N and Reznik B 2003 {\em Phys. Rev. Lett.\/} {\bf 91} 037903

\bibitem{Nielsen:2004uq}
Nielsen M~A 2004 {\em Phys. Rev. Lett.\/} {\bf 93} 040503

\bibitem{al:2004kx}
Hayes A~J~F, Gilchrist A, Myers C~R and Ralph T~C 2004 {\em J. Opt. B: Quantum
  Semiclassical Opt.\/} {\bf 6} 533

\bibitem{Browne:2005vn}
Browne D~E and Rudolph T 2005 {\em Phys. Rev. Lett.\/} {\bf 95} 010501

\bibitem{Kok:2007nx}
Kok P, Munro W~J, Nemoto K, Ralph T~C, Dowling J~P and Milburn G~J 2007 {\em
  Rev. Mod. Phys.\/} {\bf 79} 135

\bibitem{Gilchrist:2007bs}
Gilchrist A, Hayes A~J~F and Ralph T~C 2007 {\em Phys. Rev.\/} A {\bf 75}
  052328

\bibitem{Hayes:2008bh}
Hayes A~J~F, Gilchrist A and Ralph T~C 2008 {\em Phys. Rev.\/} A {\bf 77}
  012310

\bibitem{Gong:2010ly}
Gong Y~X, Zou X~B, Ralph T~C, Zhu S~N and Guo G~C 2010 {\em Phys. Rev.\/} A
  {\bf 81} 052303

\bibitem{Hayes:2010dq}
Hayes A~J~F, Haselgrove H~L, Gilchrist A and Ralph T~C 2010 {\em Phys. Rev.\/}
  A {\bf 82} 022323

\bibitem{Jennewein:2011vn}
Jennewein T, Barbieri M and White A~G 2011 {\em J. Mod. Opt.\/} {\bf 58} 276

\bibitem{Berry:2011fk}
Berry D~W and Lvovsky A~I 2011 {\em Phys. Rev.\/} A {\bf 84} 042304

\bibitem{Pittman:2002fk}
Pittman T~B, Jacobs B~C and Franson J~D 2002 {\em Phys. Rev. Lett.\/} {\bf 88}
  257902

\bibitem{Pittman:2003uq}
Pittman T~B, Fitch M~J, Jacobs B~C and Franson J~D 2003 {\em Phys. Rev.\/} A
  {\bf 68} 032316

\bibitem{OBrien:2003kx}
O'Brien J~L, Pryde G~J, White A~G, Ralph T~C and Branning D 2003 {\em Nature\/}
  {\bf 426} 264

\bibitem{Sanaka:2004ys}
Sanaka K, Jennewein T, Pan J~W, Resch K and Zeilinger A 2004 {\em Phys. Rev.
  Lett.\/} {\bf 92} 017902

\bibitem{Gasparoni:2004vn}
Gasparoni S, Pan J~W, Walther P, Rudolph T and Zeilinger A 2004 {\em Phys. Rev.
  Lett.\/} {\bf 93} 020504

\bibitem{Zhao:2005zr}
Zhao Z, Zhang A~N, Chen Y~A, Zhang H, Du J~F, Yang T and Pan J~W 2005 {\em
  Phys. Rev. Lett.\/} {\bf 94} 030501

\bibitem{Pittman:2005ly}
Pittman T~B, Jacobs B~C and Franson J~D 2005 {\em Phys. Rev.\/} A {\bf 71}
  032307

\bibitem{Chen:2008ys}
Chen J, Altepeter J~B, Medic M, Lee K~F, Gokden B, Hadfield R~H, Nam S~W and
  Kumar P 2008 {\em Phys. Rev. Lett.\/} {\bf 100} 133603

\bibitem{Schmid:2009qf}
Schmid C, Kiesel N, Weber U~K, Ursin R, Zeilinger A and Weinfurter H 2009 {\em
  New J. Phys.\/} {\bf 11} 033008

\bibitem{Lemr:2010ly}
Lemr K, {\v C}ernoch A, Soubusta J and Fiur{\'a}{\v s}ek J 2010 {\em Phys.
  Rev.\/} A {\bf 81} 012321

\bibitem{Barz:2010ve}
Barz S, Cronenberg G, Zeilinger A and Walther P 2010 {\em Nat. Photonics\/}
  {\bf 4} 553

\bibitem{Mikova:2012zr}
Mikov{\'a} M, Fikerov{\'a} H, Straka I, Mi{\v c}uda M, Fiur{\'a}{\v s}ek J,
  Je{\v z}ek M and Du{\v s}ek M 2012 {\em Phys. Rev.\/} A {\bf 85} 012305

\bibitem{Misra:1977ul}
Misra B and Sudarshan E~C~G 1977 {\em J. Math. Phys.\/} {\bf 18} 756

\bibitem{Itano:1990pd}
Itano W~M, Heinzen D~J, Bollinger J~J and Wineland D~J 1990 {\em Phys. Rev.\/}
  A {\bf 41} 2295

\bibitem{Kofman:1996cr}
Kofman A~G and Kurizki G 1996 {\em Phys. Rev.\/} A {\bf 54} R3750

\bibitem{Schulman:1998nx}
Schulman L~S 1998 {\em Phys. Rev.\/} A {\bf 57} 1509

\bibitem{Kwiat:1998tg}
Kwiat P~G 1998 {\em Phys. Scr.\/} {\bf 1998} 115

\bibitem{Kofman:2000kl}
Kofman A~G and Kurizki G 2000 {\em Nature\/} {\bf 405} 546

\bibitem{Kofman:2001oq}
Kofman A~G, Kurizki G and Opatrn{\'y} T 2001 {\em Phys. Rev.\/} A {\bf 63}
  042108

\bibitem{Horodecki:2001fk}
Horodecki P 2001 {\em Phys. Rev.\/} A {\bf 63} 022108

\bibitem{Franson:2004fj}
Franson J~D, Jacobs B~C and Pittman T~B 2004 {\em Phys. Rev.\/} A {\bf 70}
  062302

\bibitem{Leung:2006jo}
Leung P~M and Ralph T~C 2006 {\em Phys. Rev.\/} A {\bf 74} 062325

\bibitem{Franson:2007yo}
Franson J~D, Pittman T~B and Jacobs B~C 2007 {\em J. Opt. Soc. Am. B: Opt.
  Phys.\/} {\bf 24} 209

\bibitem{Myers:2007fc}
Myers C~R and Gilchrist A 2007 {\em Phys. Rev.\/} A {\bf 75} 052339

\bibitem{Leung:2007dw}
Leung P~M and Ralph T~C 2007 {\em New J. Phys.\/} {\bf 9} 224

\bibitem{Huang:2008xq}
Huang Y~P and Moore M~G 2008 {\em Phys. Rev.\/} A {\bf 77} 062332

\bibitem{Sakurai:1994uq}
Sakurai J~J and Tuan S~F 1994 {\em {Modern Quantum Mechanics}\/} (Reading:
  Addison-Wesley Publishing Company)

\bibitem{Franson:2006mw}
Franson J~D and Hendrickson S~M 2006 {\em Phys. Rev.\/} A {\bf 74} 053817

\bibitem{You:2008kn}
You H, Hendrickson S~M and Franson J~D 2008 {\em Phys. Rev.\/} A {\bf 78}
  053803

\bibitem{You:2009jw}
You H, Hendrickson S~M and Franson J~D 2009 {\em Phys. Rev.\/} A {\bf 80}
  043823

\bibitem{Hendrickson:2010pi}
Hendrickson S~M, Lai M~M, Pittman T~B and Franson J~D 2010 {\em Phys. Rev.
  Lett.\/} {\bf 105} 173602

\bibitem{Scully:1997fk}
Scully M~O and Zubairy M~S 1997 {\em {Quantum Optics}\/} (Cambridge: Cambridge
  University Press)

\bibitem{Ryder:1996fk}
Ryder L~H 1996 {\em {Quantum Field Theory}\/} (Cambridge: Cambridge University
  Press)

\bibitem{Dicke:1954fk}
Dicke R~H 1954 {\em Phys. Rev.\/} {\bf 93} 99

\bibitem{Rehler:1971uq}
Rehler N~E and Eberly J~H 1971 {\em Phys. Rev.\/} A {\bf 3} 1735

\bibitem{Scully:2006fk}
Scully M~O, Fry E~S, Ooi C~H~R and W\'odkiewicz K 2006 {\em Phys. Rev. Lett.\/}
  {\bf 96} 010501

\bibitem{Eberly:2006ly}
Eberly J~H 2006 {\em J. Phys. B: At. Mol. Opt. Phys.\/} {\bf 39} S599

\bibitem{Scully:2007fk}
Scully M~O 2007 {\em Laser Phys.\/} {\bf 17} 635

\bibitem{Svidzinsky:2008qf}
Svidzinsky A~A, Chang J~T and Scully M~O 2008 {\em Phys. Rev. Lett.\/} {\bf
  100} 160504

\bibitem{Porras:2008kx}
Porras D and Cirac J~I 2008 {\em Phys. Rev.\/} A {\bf 78} 053816

\bibitem{Scully:2009fk}
Scully M~O and Svidzinsky A~A 2009 {\em Science\/} {\bf 325} 1510

\bibitem{Sete:2010fu}
Sete E~A, Svidzinsky A~A, Eleuch H, Yang Z, Nevels R~D and Scully M~O 2010 {\em
  J. Mod. Opt.\/} {\bf 57} 1311

\bibitem{Wiegner:2011zr}
Wiegner R, von Zanthier J and Agarwal G~S 2011 {\em Phys. Rev.\/} A {\bf 84}
  023805

\bibitem{Lipkin:2002fk}
Lipkin H~J 2002 {\em Multiple Facets of Quantization and Supersymmetry\/} {\bf
  Michael Marinov Memorial Volume} 128

\bibitem{Holstein:1940uq}
Holstein T and Primakoff H 1940 {\em Phys. Rev.\/} {\bf 58} 1098

\bibitem{Chemistry:2009fk}
{Thieme Chemistry} 2009 {\em {R{\"O}MPP Online - Version 3.5}\/} (Stuttgart:
  Georg Thieme Verlag KG)

\bibitem{Cohen-Tannoudji:1977fk}
Cohen-Tannoudji C, Diu B and Lalo\"e F 1977 {\em Quantum Mechanics Volume 2\/}
  (New York: Wiley)

\bibitem{Sakurai:1987fk}
Sakurai J~J 1987 {\em {Advanced Quantum Mechanics}\/} (Redwood City:
  Addison-Wesley Publishing Company)

\bibitem{Duarte:1995jh}
Duarte F~J 1995 {\em {Tunable Lasers Handbook}\/} (San Diego: Academic Press)

\bibitem{Wrigge:2008fk}
Sandoghdar V {2012 Private communication, see also Wrigge G, Hwang J, Gehardt
  I, Zumofen G and Sandoghdar V 2008 {\it Opt. Express} {\bf 16} 17358}

\bibitem{Sun:2005rh}
Sun Y~C {2005 "Rare Earth Materials in Optical Storage and Data Processing
  Applications", in: {\it Spectroscopic Properties of Rare Earths in Optical
  Materials} (Springer)}

\bibitem{Macfarlane:1987ie}
Macfarlane R~M and Shelby R~M {1987 "Coherent Transient and Holeburning
  Spectroscopy of Rare Earth Ions in Solids", in: {\it Spectroscopy of Solids
  containing Rare Earth Ions} (North-Holland)}

\end{thebibliography}

\end{document}